\documentclass[useAMS,usenatbib]{mn2e}
\usepackage{float}
\input{boxed1.sty}
\floatstyle{simplerule}

\usepackage{amssymb}
\usepackage{graphicx}
\usepackage{longtable}
\usepackage[table]{xcolor}
\usepackage[dvipdfm]{hyperref}
\hypersetup{
   colorlinks=true,             
   linkcolor=blue,              
   urlcolor=blue,               
   anchorcolor=blue,            
   citecolor=blue,              
}

\definecolor{grey80}{rgb}{0.90,0.90,0.90}
\usepackage{amsmath}

\newcounter{mycount}

\title[Dynamics in the vicinity of (101955) Bennu]{\centering Dynamics in the vicinity of (101955) Bennu: Solar radiation pressure effects in equatorial orbits}
\author[T. G. G. Chanut, S. Aljbaae,  A. F. B. A. Prado and V. Carruba]
{T. G. G. Chanut$^{1}$\thanks{\href{thierry@feg.unesp.br}{thierry@feg.unesp.br} (TGGC), \href{safwan.aljbaae@gmail.com}{safwan.aljbaae@gmail.com} (SA), \href{antonio.prado@inpe.br}{antonio.prado@inpe.br} (AFBAP), \href{vcarruba@feg.unesp.br} {vcarruba@feg.unesp.br} (VC)}, S. Aljbaae$^{1\star}$,  A. F. B. A. Prado$^{2\star}$ and V. Carruba$^{1\star}$\\
$^{1}$S\~{a}o Paulo State University (UNESP), School of Natural Sciences and Engineering, Guaratinguet\'{a}, SP, 12516-410, Brazil \\
$^{2}$Division of Space Mechanics and Control, INPE, C.P. 515, 12227-310 S\~ao Jos\'e dos Campos, SP, Brazil\\
}

\begin{document}
\date{Accepted 2017 May 10. Received 2017 April 29; in original form 2017 January 13}
\pagerange{\pageref{firstpage}--\pageref{lastpage}} \pubyear{2017}
\maketitle
\label{firstpage}
\begin{abstract}
Here we study the dynamical effects of the solar radiation pressure (SRP) on a spacecraft that will survey the near-Earth rotating asteroid (101955) Bennu when the projected shadow is accounted for. The spacecraft's motion near (101955) Bennu is modelled in the rotating frame fixed at the centre of the asteroid, neglecting the sun gravity effects. We calculate the solar radiation pressure at the perihelion, semi-major axis and aphelion distances of the asteroid from the Sun. The goals of this work are to analyse the stability for both homogeneous and inhomogeneous mass distribution and study the effects of the solar radiation pressure in equatorial orbits close to the asteroid (101955) Bennu.  As results, we find that the mascon model divided into ten equal layers seems to be the most suitable for this problem. We can highlight that the centre point $E$8, which was linearly stable in the case of the homogeneous mass distribution, becomes unstable in this new model changing its topological structure. For a Sun initial longitude $\psi_0 = -180^o$, starting with the spacecraft longitude $\lambda = 0$, the orbits suffer fewer impacts and some (between 0.4 and 0.5 km), remaining unwavering even if the maximum solar radiation is considered. When we change the initial longitude of the Sun to $\psi_0 = -135^o$, the orbits with initial longitude $\lambda = 90^0 $ appear to be more stable. Finally, when the passage of the spacecraft in the shadow is accounted for, the effects of solar radiation pressure are softened, and we find more stable orbits.
\end{abstract}
\section{Introduction}
Discovered by the $LINEAR$\footnote{\href{http://neo.jpl.nasa.gov/missions/linear.html}{http://neo.jpl.nasa.gov/missions/linear.html}} Project in September 1999 \citep{Williams_1999}, asteroid (101955) Bennu (formerly designated 1999 RQ36) is an Apollo Near-Earth Object ($NEO$). As noted by \citet{Lauretta_2015}, in 2135, Bennu will pass inside the orbit of the Moon (0.002 $AU$ over the surface of the Earth). This asteroid is the target of the $OSIRIS-REx$\footnote{\href{http://science.nasa.gov/missions/osiris-rex/}{http://science.nasa.gov/missions/osiris-rex/}} asteroid sample return mission, that $NASA$ launched in 2016 to collect a sample from the space rock and return it to Earth by 2023. Bennu is an exciting target for an asteroid sample return mission. It is different from all other near-Earth asteroids previously visited by spacecraft. Asteroid (433) Eros, target of the NEAR-Shoemaker mission, and (25143) Itokawa, target of the Hayabusa mission, are both S-type asteroids with irregular shapes. In contrast, Bennu is a spectral B-type asteroid, and has a distinct spheroidal shape. While Eros and Itokawa are similar
to ordinary chondrite meteorites, Bennu is likely to be formed by carbonaceous chondrites, which are meteorites that record the organic compound history of the early solar System. An important parameter for the mission design is the shape model that provides its gravity field. The polyhedron shape of asteroid (101955) Bennu was created by \citet{Nolan_2013}, based on radar images and optical light-curves collected in 1999 and 2005. Recent computational tools from the polyhedral shape model were created to predict and control the dynamical evolution of an orbiter around irregular small bodies in their complex gravity fields. \citet{Tsoulis_2012} refined the approach of \citet{Werner&Scheeres_1997} by presenting the derivation of certain singularity terms, which emerge for special locations of the computation point with respect to the attracting polyhedral source. This method has been applied to the investigation of the actual dynamical environments around (433) Eros and (216) Kleopatra \citep{Chanut_2014,Chanut_2015b}. Although very accurate, the method requires a large computational effort for integrating the orbits around these bodies. 

One of the solutions for this problem was to adapt the method of mascons to polyhedra \citep{Venditti_2013}. In \citet{Chanut_2015a}, it was shown that this solution is more accurate than the classical mascon method \citep{Geissler_1997}, and faster than the Tsoulis method.  In all of the above-cited cases, the solar radiation pressure effects are not taken into account for the dynamics. Actually, the effect of this force become more pronounced as the spacecraft flies farther away from the asteroid, since the orbital motion close to the body is generally dominated by its own gravity field \citep{Scheeres_2000}. Nevertheless, \citet{Scheeres_2006} have shown that, for a body smaller than few kilometres, like asteroid (25143) Itokawa, the solar radiation pressure parameter become relevant even in close proximity to the body. Moreover, in the Hill problem, the influence of solar radiation pressure can lead to both unstable and stable spacecraft orbits near bodies with a diameter smaller than few kilometres \citep{Hussmann_2012}. However, the solar radiation pressure model, as well as the eclipse model around rotating asteroids, were only recently presented in \citet{Xin_2016}. When SRP perturbations are considered, the equilibrium points usually do not exist anymore, since the equations of motion are no longer time independent. However, forced motions called "dynamical substitutes" appear around the geometrical equilibrium points. This issue has been thoroughly discussed in \citet{Xin_2016} and will not be part of the scope of this work. 

In this paper, we analyze the stability of both homogeneous and inhomogeneous mass distribution and the effects of solar radiation pressure in equatorial orbits in the proximity of the asteroid (101955) Bennu. This will be attended using the mascon gravitation model and the solar radiation pressure force defined by \citet{Scheeres_2002} and \citet{Xin_2016} for the dynamical model. 
First, we show the general properties of (101955) Bennu in Section 2. In Section 3, the methodology of mascon gravity tensor of the polyhedral model is briefly presented. Then, we discuss the equations of motion and the conserved quantity in Section 4, as well as the equilibria when the solar radiation pressure is not accounted for. We find the exact location of the eight equilibrium points and compare their stability when the mass distribution occurs in different density layers. The dynamical model of initially equatorial orbits close to (101955) Bennu, when the solar radiation pressure is accounted for, and the results of numerical simulations are presented in Section 5. Finally, we discuss and conclude in Section 6.

\section[]{Computed Physical Features\\* from the shape of (101955) Bennu}
The three-dimensional shape of near-Earth asteroid (101955) Bennu (provisional designation 1999 RQ36) is based on radar images and optical lightcurves \citep{Nolan_2013}. Bennu was observed both in 1999, at its discovery apparition, and in 2005, using the 12.6-cm radar at the Arecibo Observatory and the 3.5-cm radar at the Goldstone tracking station. From the data set of EAR-A-I0037-5-BENNUSHAPE-V1.0. of NASA Planetary Data System (2013), we build a polyhedral model with 1348 vertices and 2692 faces shown in Fig. 1. Radar astrometry combined with infrared astronomy provides an estimate of asteroid mass of $7.8(\pm 0.9) \times  10^{10} kg $. When linked with the shape model, the (101955) Bennu bulk density is $1.26 \pm 0.7 \, g/cm^3$  \citep{Chesley_2014}.
 The shape model provides a total volume of $0.062\, km^3$ with an equivalent diameter of $0.492\, km$. We fit the body centered at its center of mass with the rotation pole of the model lying along the z axis (e.g. Table 1). So, the overall dimensions of the asteroid shape model in the principal directions are: $ -0.2783 \leq  x  \leq  0.2881$ km, $-0.2661 \leq  y  \leq  \,\, 0.2698 $ km, $ -0.2457 \,\leq  z  \leq  \,\, 0.2631 $ km. 

 \begin{table}
  \centering
   \begin{minipage}{1\linewidth}
   \caption{Coordinates shift to the center of mass and rotation matrix to the principal axes of inertia using the algorithm of \citet{Mirtich_1996} }
   \label{symbols}
   \resizebox{1.0\textwidth}{!}{
   \begin{tabular}{cccc}
   \hline
    Center of mass ($m$) & +0.04355522 & -0.00089925 & +0.00624339 \\
    Eigenvectors (in columns) & +0.99999950   &+0.00100076  & -0.00004066 \\
                              & -0.00100076   &+0.99999950   & +0.00000585 \\
                              & +0.00004067   &-0.00000581   & +0.99999999 \\
     \hline   
\end{tabular}}
 \end{minipage}
\end{table}

\begin{figure}
   \centering
   \includegraphics[width=0.48\linewidth]{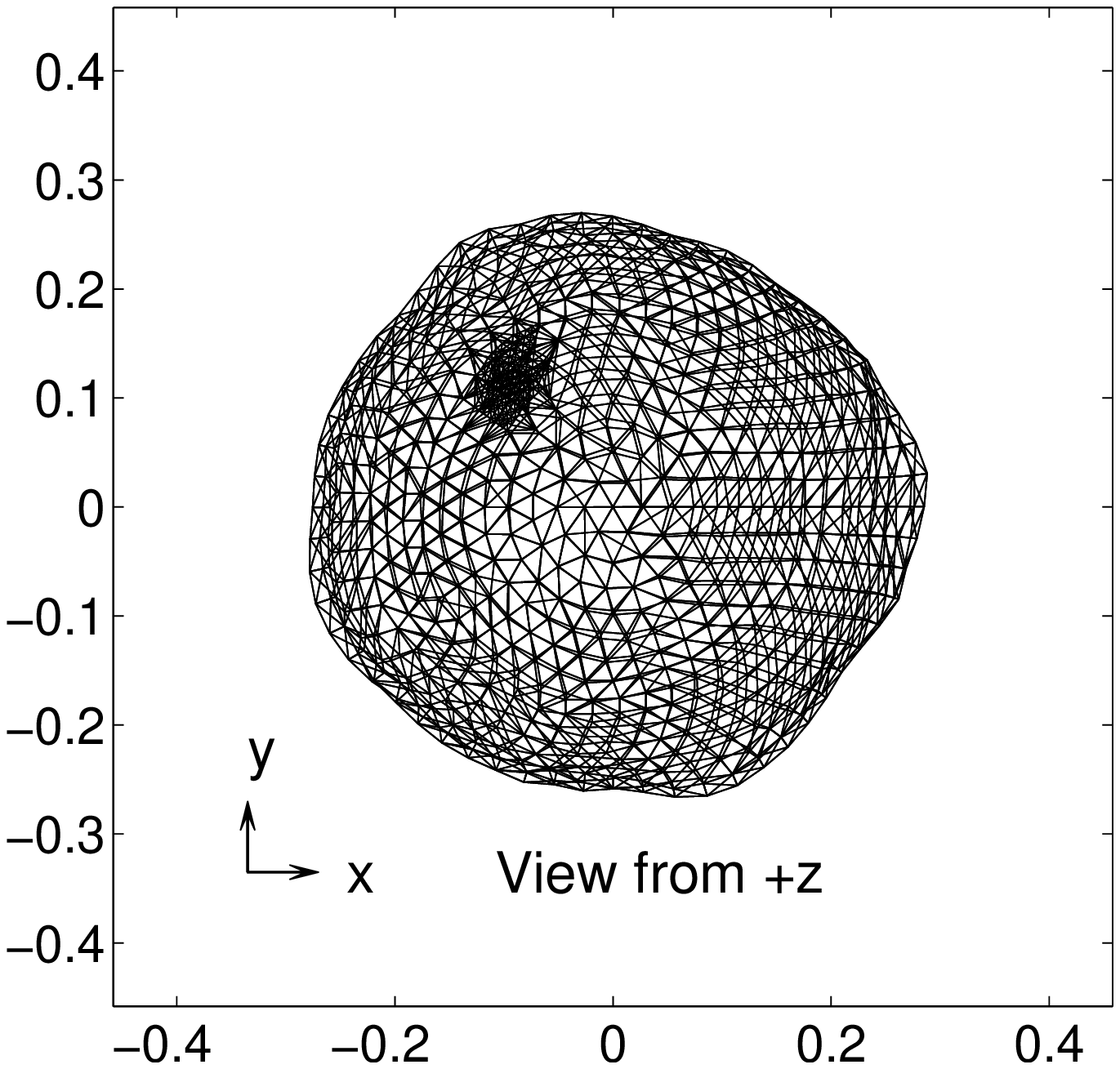}
   \includegraphics[width=0.48\linewidth]{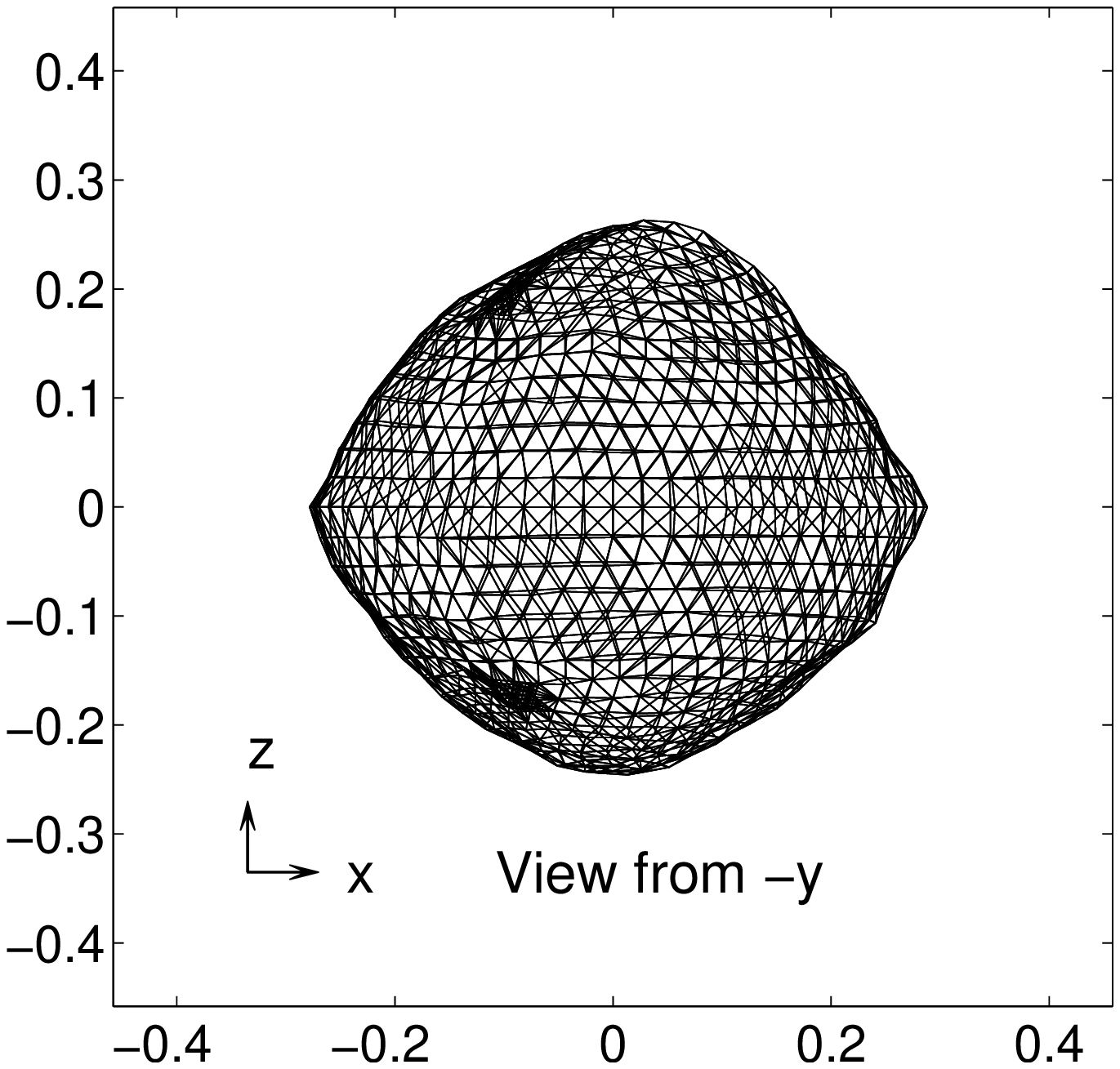}\\
   \includegraphics[width=0.48\linewidth]{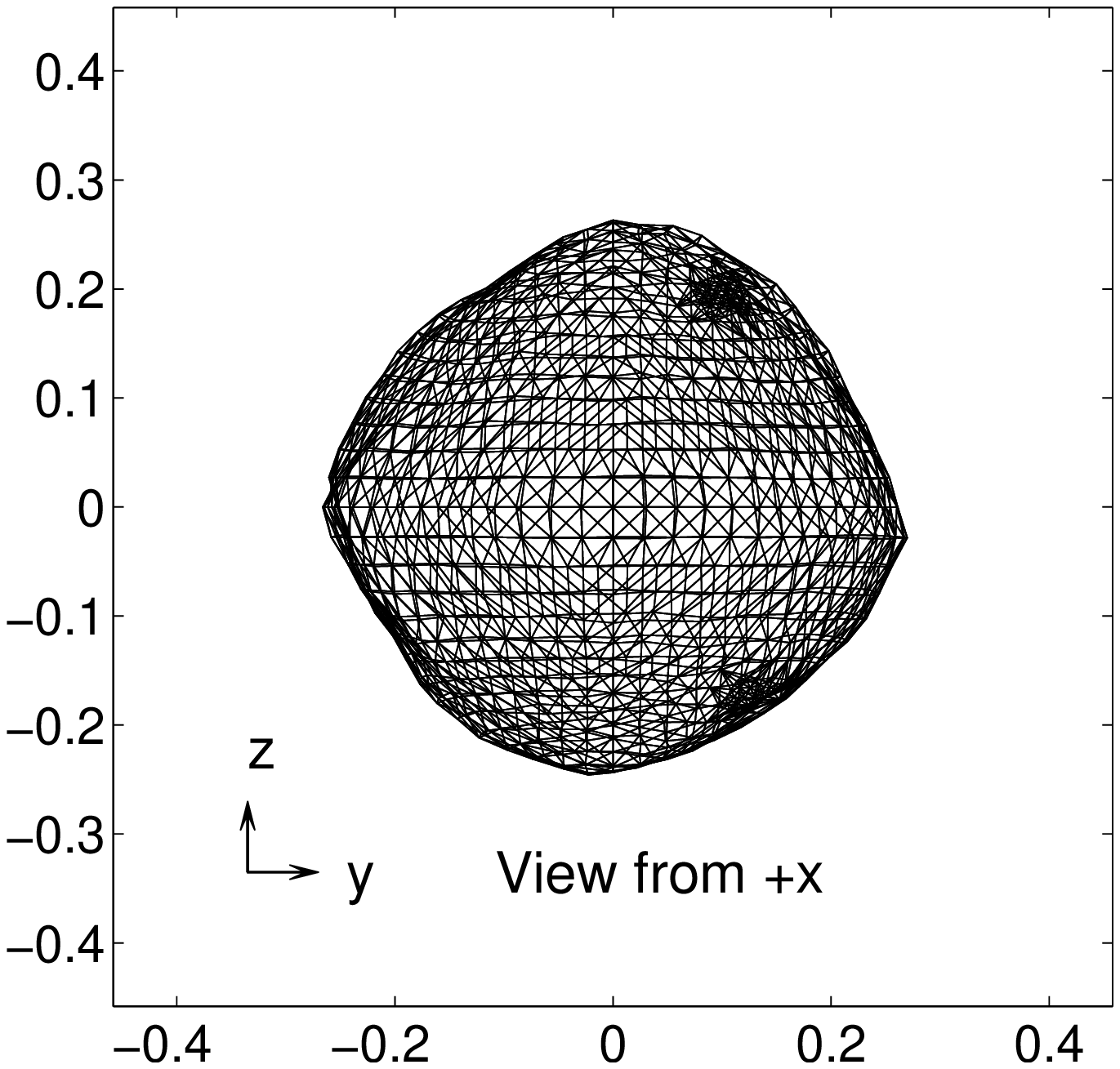}   
   \includegraphics[width=0.48\linewidth]{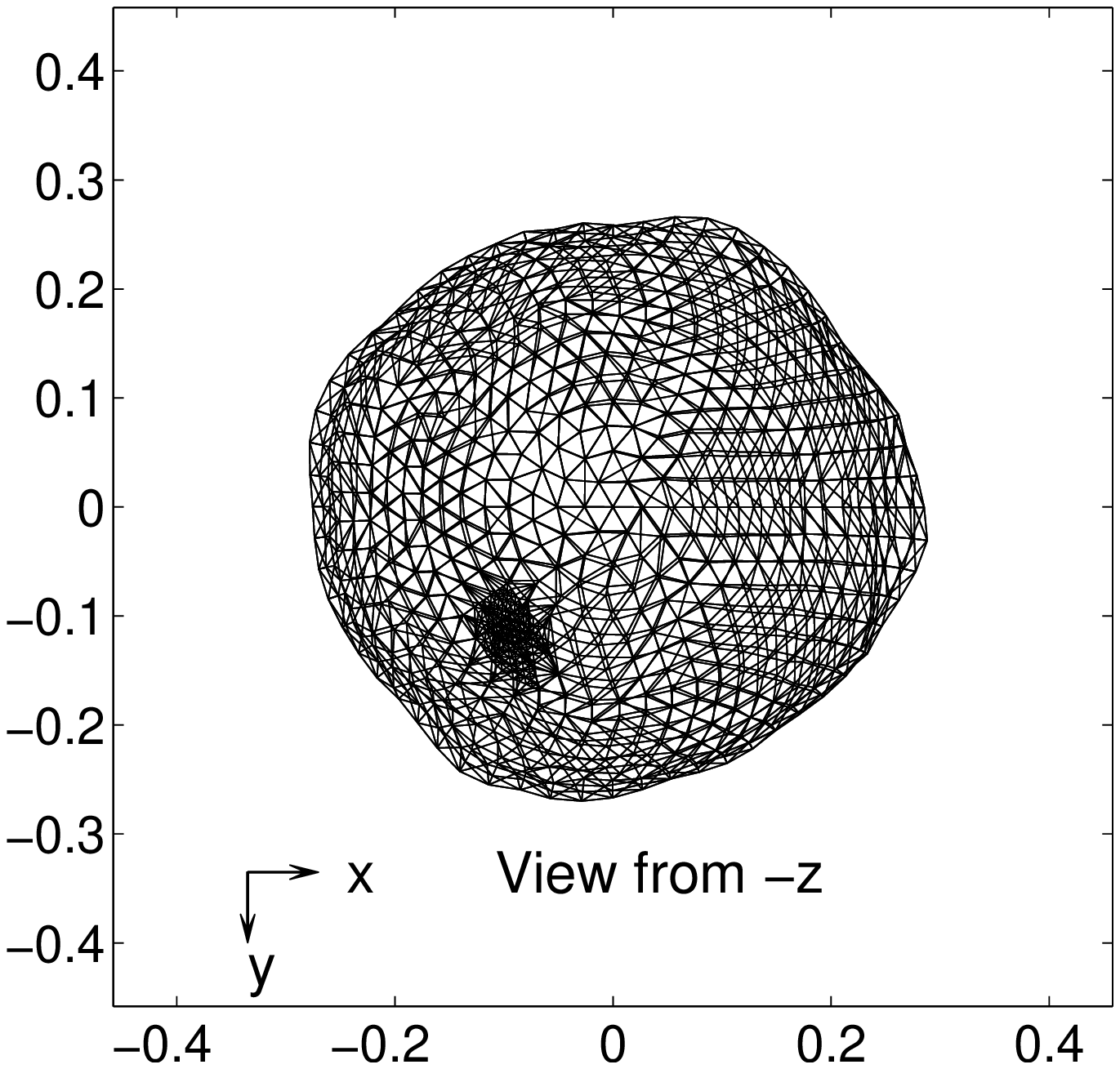}\\
   \includegraphics[width=0.48\linewidth]{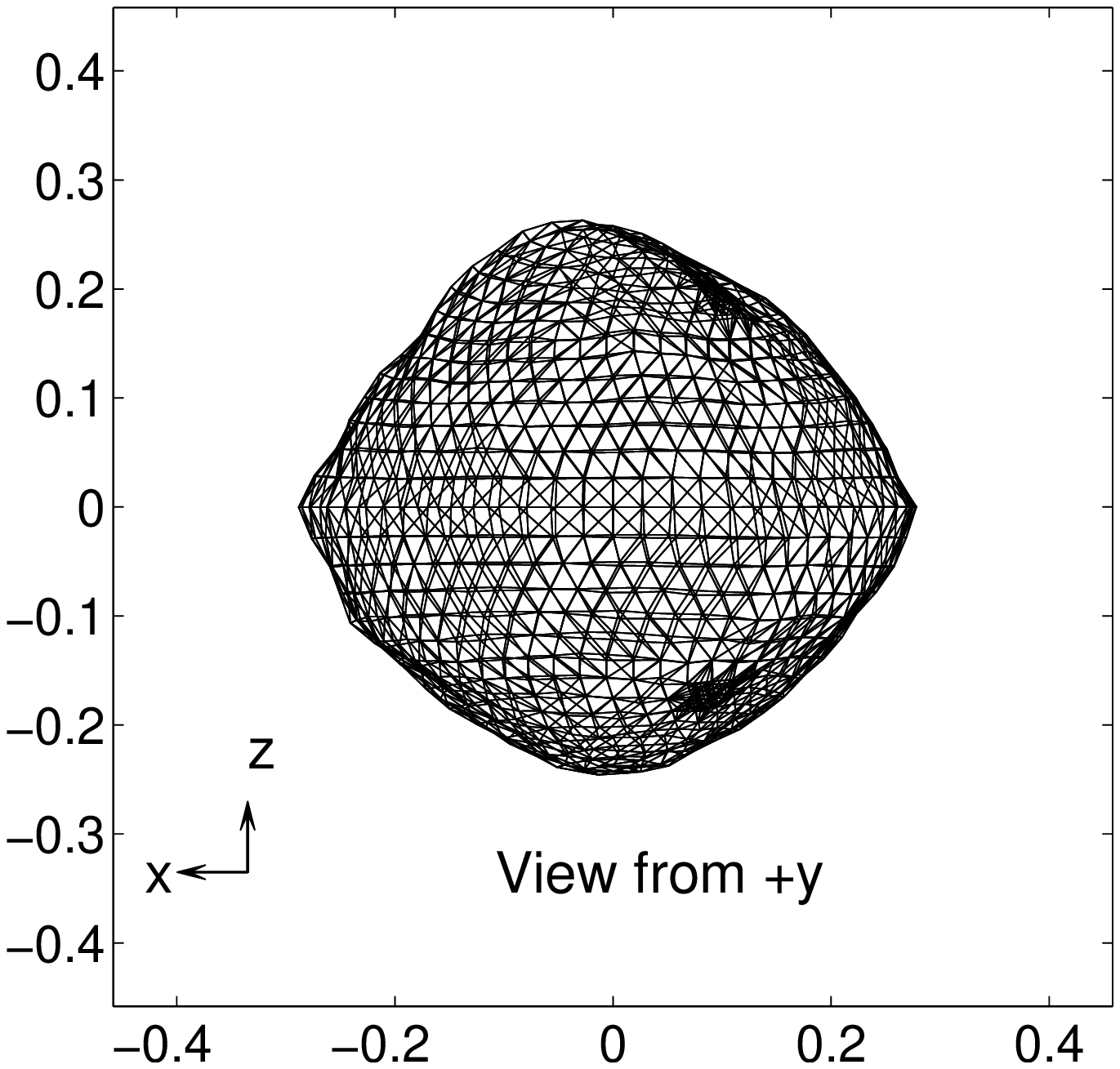}
   \includegraphics[width=0.48\linewidth]{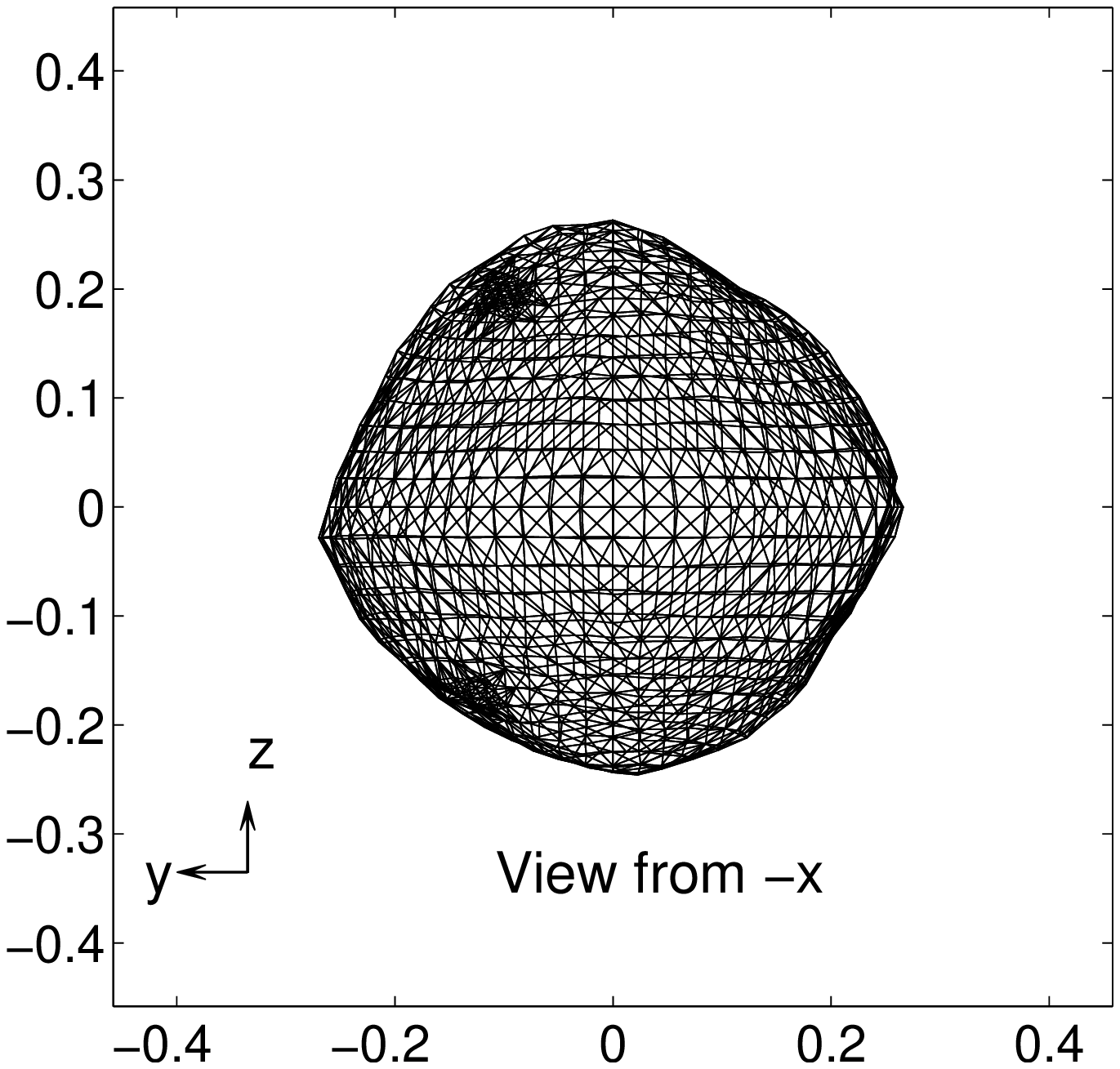}\\
   \caption{Polyhedral shape model in 3D of asteroid (101955) Bennu shown in 6 perspective views ($\pm$ x, $\pm$ y, and $\pm$ z) with a scale-size of 0.9981 relative to the original shape. The shape was built with 2692 triangular faces.}
   \label{Fig1}
\end{figure} 

 We use the algorithm of \citet{Mirtich_1996} to obtain the values of the moments of the principal axes of inertia, respectively, $I_{xx}$, $I_{yy}$, and $I_{zz}$, which are:
\begin{eqnarray}
    \nonumber I_{xx} = 1.8130 \times 10^{9} \,kg\,km^2,\\  
      I_{yy} = 1.8836  \times 10^{9} \,kg\,km^2,\\  
    \nonumber I_{zz} = 2.0334 \times 10^{9} \,kg\,km^2.  
 \end{eqnarray}

From the moments of inertia \citep{Dobrovolskis_1996}, we find an equivalent ellipsoid with semi-major axes of $260 \times  251 \times  231 \, m$, which is close to the dynamically equivalent equal volume ellipsoid (DEEVE) shown in \citet{Lauretta_2015}. 
We can also derive from the moments of inertia the most important terms of the harmonic expansion that correspond to the second-degree and -order gravity coefficients, and these are equal to \citep{Werner_1997}:
 \begin{eqnarray}
    \nonumber C_{20}R^{2}_0 = -2.3734\times 10^{-3} \,km^2,\\\nonumber\\  
      C_{22}R^{2}_0 =\,\,\,\,2.2646\times 10^{-4} \,\,km^2,   
 \end{eqnarray}
 where the normalisation radius $R_0$ is arbitrary chosen.
 Differently from the gravity field coefficients up to order and degree 4 shown in \citet{Nolan_2013}, we find more useful to present them unnormalized in Table 2. In fact, the coefficients need to be fully normalised from order 10 onwards to avoid divergence, because of the order of their magnitudes according to the formula shown in \citep{Kaula_1966}. As we can see, and differently from the Earth, the zonal gravity terms $C_{20}$, $C_{30}$ and $C_{40}$ have a closer order of magnitude. This reveals an irregular gravity field with a shape somewhat more pointed and cylindrical.

\begin{table}
\centering
\begin{minipage}[h]{0.95\linewidth}
\caption{  Bennu Unnormalized Gravity Field Coefficients up to degree and order 4 for the constant density of 1.26 $g.cm^{-3}$ and the total mass of $7.80 \times 10^{10} kg$. They are computed for a reference distance of $0.2459$ $km$. The frame is centered at the center of mass and aligned with the principal moments of inertia. } \label{Table_20_Harmonics_Bennu}
\resizebox{1.0\textwidth}{!}{
\begin{tabular}{cccc}
    \hline
Order & Degree & $ C_{nm}$ & $ S_{nm}$\\
\hline

 0 & 0 & 1.0000000000                      & - \\
 1 & 0 & 0                     & - \\
 1 & 1 & 0                     & 0  \\
 2 & 0 & -3.9257534110 $ \times 10^{-2} $                      &- \\
 2 & 1 & 0  & 0  \\
 2 & 2 & 3.7458971411 $ \times 10^{-3} $  & 0  \\
 3 & 0 & 1.4711698072 $ \times 10^{-2} $  & - \\
 3 & 1 & 1.6900954267 $ \times 10^{-3} $  & 1.6633040165 $ \times 10^{-3} $  \\
 3 & 2 & 3.5430315751 $ \times 10^{-5} $  & 3.7683848394 $ \times 10^{-5} $  \\
 3 & 3 & 3.6694725097 $ \times 10^{-4} $  & -1.2605818113 $ \times 10^{-4} $  \\
 4 & 0 & 3.0760445246 $ \times 10^{-2} $  & - \\
 4 & 1 & 3.8222964288 $ \times 10^{-4} $  & 1.7720543174 $ \times 10^{-3} $  \\
 4 & 2 & -4.8369718690 $ \times 10^{-4} $ & 1.6740004235 $ \times 10^{-4} $  \\
 4 & 3 & -6.4236840214 $ \times 10^{-5} $ & 5.4053957662 $ \times 10^{-6} $  \\
 4 & 4 & 4.5899589934 $ \times 10^{-5} $  & 6.6026679964 $ \times 10^{-5} $  \\

\hline

\end{tabular}}
\end{minipage}
\end{table}

 Another parameter ($\sigma$) of the asteroid's shape from the gravity field is defined in \citet{Hu_2004}. If $\sigma = 1$, the body has a prolate inertia matrix, while one with $\sigma = 0$ corresponds to an oblate matrix. For (101955) Bennu  $\sigma = 0.3205$. Thus, we can affirm that asteroid (101955) Bennu is closer to an oblate shape value, which is compatible with Fig.1.
 
 \section[]{Mascon gravity gradient of \\* the polyhedral model}
   The first attempt to evaluate the potential of three-dimensional bodies by the polyhedron method was developed by \citet{Werner_1994}. The polyhedron is divided into a collection of simple tetrahedra with one of the vertices at the origin and the opposite face represented by a trinomial with predefined orientation. The polyhedral approach allows us to calculate the total volume of a constant density polyhedron and evaluate its gravitational field with a good accuracy. However, the computational cost is high, depending on the number of tetrahedra that form the polyhedron. In order to reduce this cost, \citet{Chanut_2015a} developed a mathematical approach of the mascon gravity tensor with respect to a shaped polyhedral source. 
Thus, the gravitational potential suffered by an external point P from the tetrahedron is: 
\begin{equation}
U_T=\frac{\mu}{r},
\end{equation}
where $ r=(\xi ^2+\eta ^2+\zeta ^2)^{1/2}$ is the distance between the center of mass of the tetrahedron and the external point P, as represented in Fig. 2. We take $\mu=GM_T$, where the gravitational parameter is $G=6.67259 \times 10^{-20}\,km^3/kg/s^2$ and $M_T$ represents the tetrahedron's mass. Therefore, the potential and the first order derivatives of the shaped polyhedral source are 
\begin{equation}
U=\sum\limits_{i=1}^n \frac{\mu_i}{r_i},
\end{equation}
and

\begin{equation}
U_{\chi}=\sum\limits_{i=1}^n \frac{\partial U}{\partial \chi_i}= \sum\limits_{i=1}^n \left( \frac{\partial U}{\partial r_i} \right) 
\left( \frac{\partial r_i}{\partial \chi_i} \right)= \sum\limits_{i=1}^n -\frac{\mu_i\chi_i}{r^3_i},
\end{equation}
where $\chi = (\xi, \eta, \zeta)$ and
\begin{figure}
   \centering
   \includegraphics[width=1.2\linewidth]{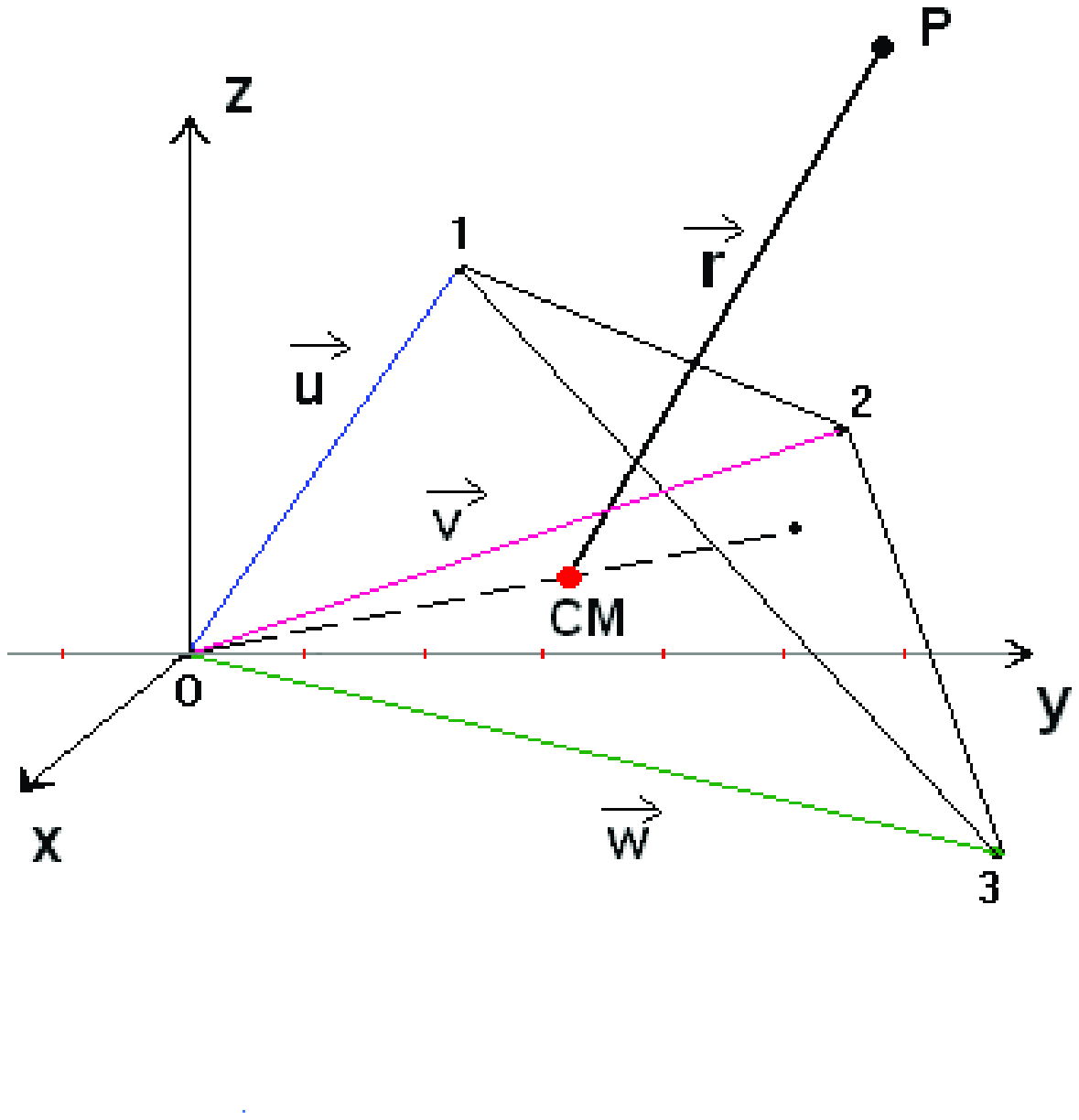}
   \vspace{-2cm}{}
         \caption{ Representation of a tetrahedron with vertex 0 at the origin and
the vectors $\bmath{u}$, $\bmath{v}$ and $\bmath{w}$ coming out from this vertex \citep{Chanut_2015a}.
}
         \label{Fig2}
   \end{figure} 
the sum represents the total quantity of tetrahedra that compound the shaped polyhedral source with $n$ the number of faces and $i$ the index of each face. 
\section[]{Equations of motion and \\* potential energy}
\subsection[]{Equations of motion}
As discussed by \citet{Scheeres_2000}, the solar gravity becomes relevant when the spacecraft flies away farther from the asteroid, which allows us to neglect any solar gravitational effect in the vicinity of (101955) Bennu. The sidereal rotation period of (101955) Bennu determined from both the lightcurve and radar data is 4.297 h \citep{Nolan_2013}. The data reveal a spheroidal asteroid undergoing retrograde rotation. The spin rate of the asteroid is denoted as $\omega$. We use a rotating reference frame that is centered on the asteroid  \citep{Szebehely_1967}. Thus, in the body-fixed reference frame, the equations of motion are:
 \begin{eqnarray}
    \ddot{x}-2\omega \dot{y} = \omega^2 x +U_x,\\  
    \ddot{y}+2\omega \dot{x} = \omega^2 y +U_y,\\ 
    \ddot{z}= U_z.
 \end{eqnarray}
 where $U_x$, $U_y$ and $U_z$ denote the first-order partial derivatives of the potential. Equations (6-8) admit an integral of motion, the Jacobi function, defined as:
    \begin{equation}
   J= \frac{1}{2} (\dot{x}^2+\dot{y}^2+\dot{z}^2) -\frac{1}{2} \omega^2(x^2+y^2)-U(x,y,z)   
    \end{equation}
 where
 \begin{equation}
    \frac{1}{2} \omega^2(x^2+y^2)+U(x,y,z) =  V(x,y,z)
 \end{equation}
 is the modified potential energy and
  \begin{equation}
    \frac{1}{2} (\dot{x}^2+\dot{y}^2+\dot{z}^2) = T_E
 \end{equation}
 represents the kinetic energy of the particle regarding the rotating asteroid.
\subsection[]{Zero velocity curves and equilibria}
 We can provide concrete informations of the possible motion of a particle analyzing the zero-velocity surfaces defining the Jacobi function as constant $C$ where $J+C=0$. 
 Note that $V(x,y,z)\geq 0$ over the entire space, and because $T_E\geq 0$, let us constrain the study to the inequality
 \begin{equation}
   V(x,y,z)\geq C,
 \end{equation}
 that divides the $x, y, z $ space into regions where the motion of the particle is allowed and where it is not, given a specific value for $C$. The general situation was discussed in more detail by \citet{Scheeres_1994}. Setting $T_E=0$ on the $x, y, z $ space, the equation
\begin{equation}
   V(x,y,z)=C,
 \end{equation}
 represents the zero-velocity surfaces and
 \begin{figure*}
   \centering
      Tsoulis     \hspace{7.5cm}           Mascon 4\\
   \includegraphics[width=0.48\linewidth]{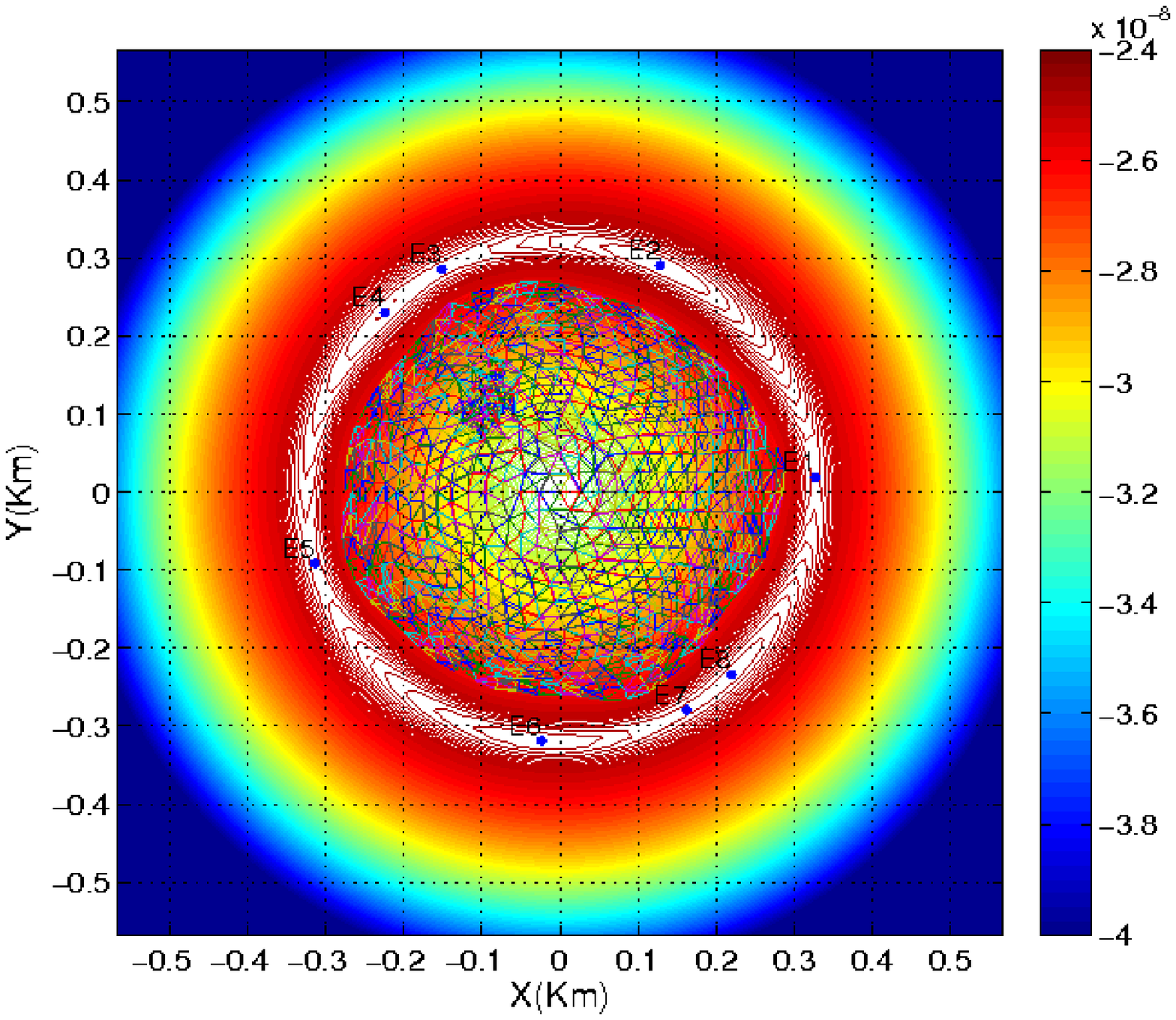}
   \includegraphics[width=0.48\linewidth]{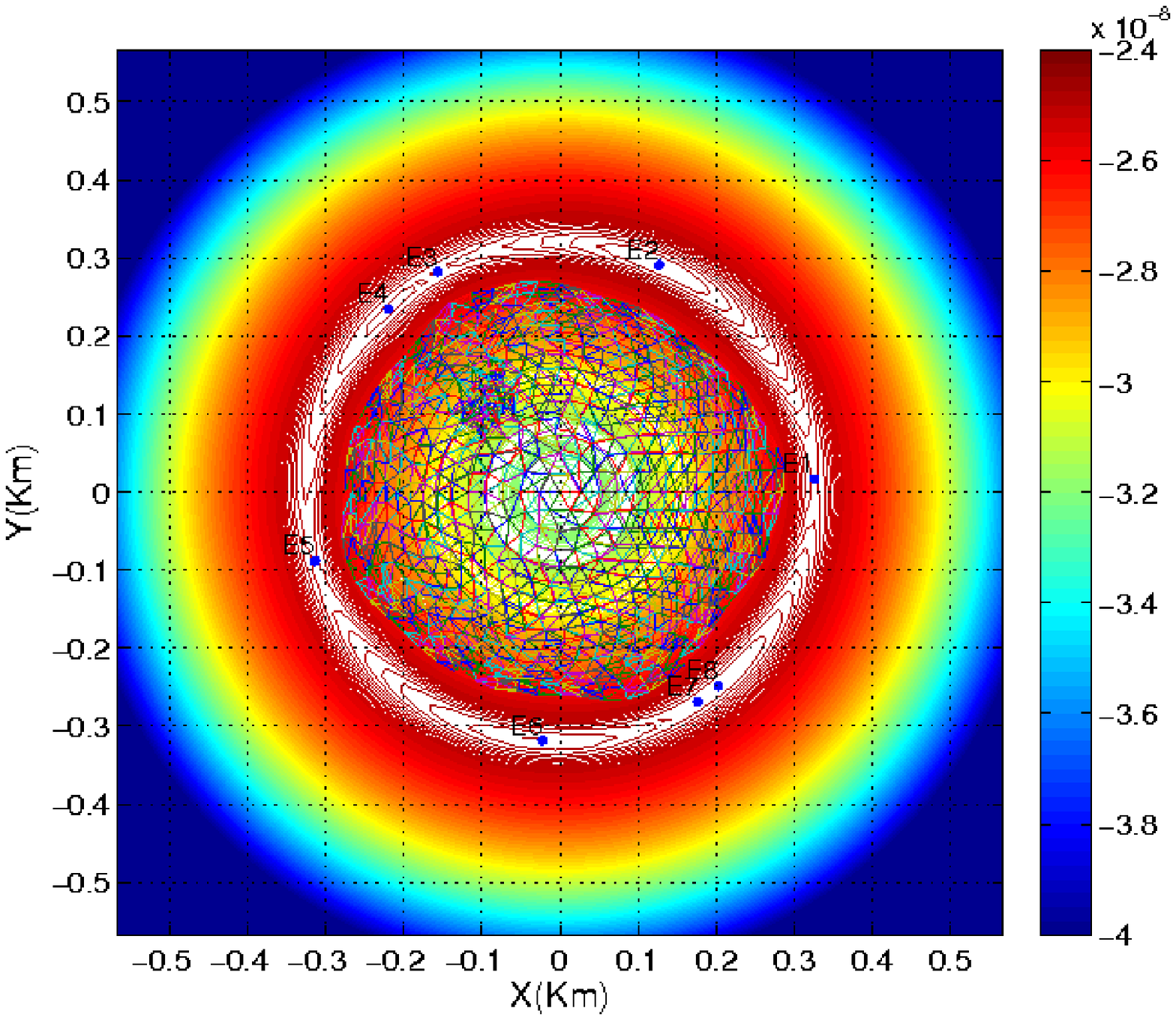}
   
      Mascon 6     \hspace{7.5cm}           Mascon 10\\
   \includegraphics[width=0.48\linewidth]{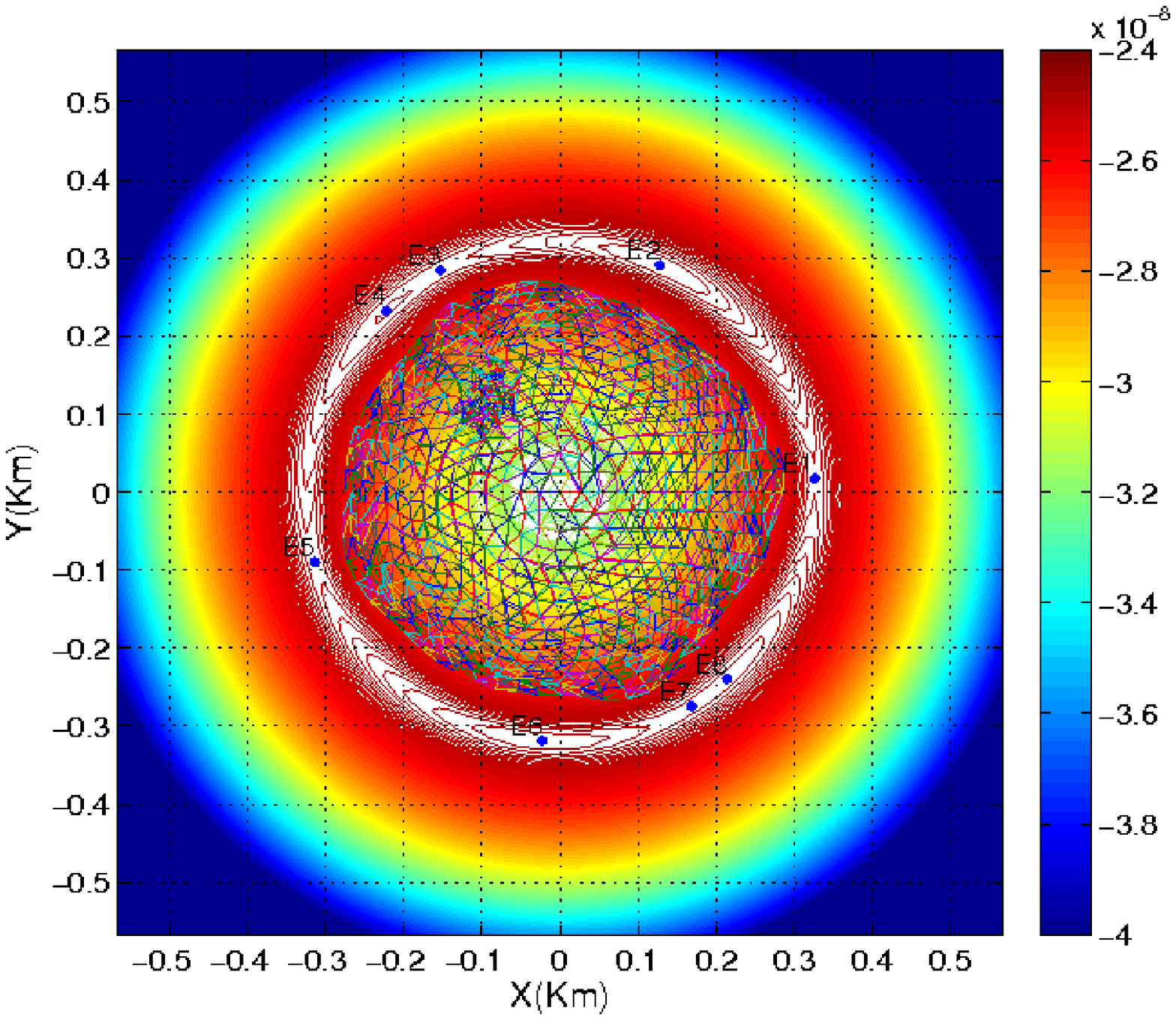}
   \includegraphics[width=0.48\linewidth]{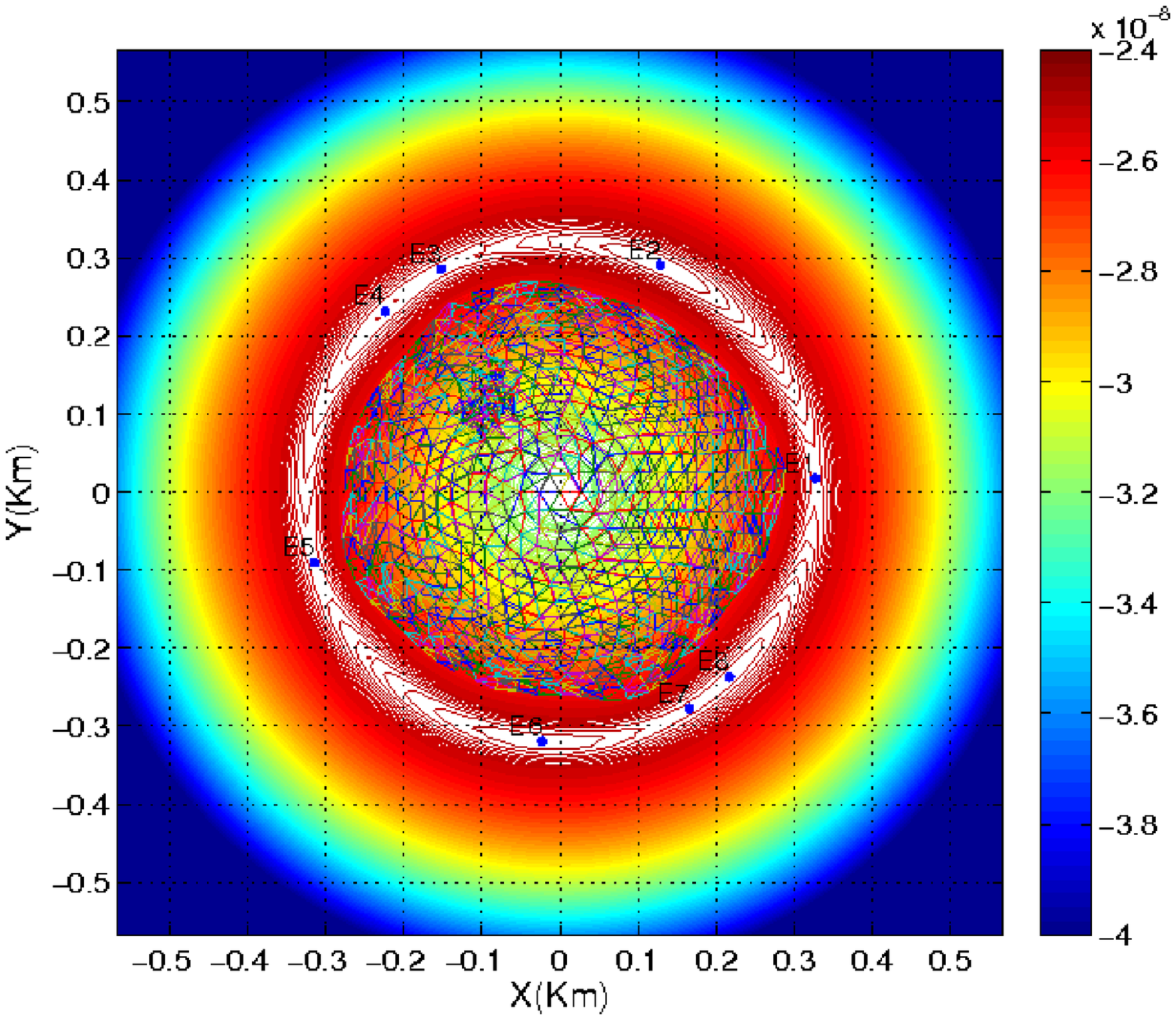}\\
    \caption{Zero-velocity curves and equilibrium points of asteroid (101955) Bennu when the SRP is not accounted for. The color code gives the intensity of the Jacobi constant in $km^2$ $s^{-2}$. The equilibrium points outside the body, indicated by $E$1, $E$2, $E$3 ,$E$4, $E$5, $E$6, $E$7 and $E$8, are displayed using several gravity models highlighted on the top of each figure .
}
         \label{Fig3}
\end{figure*}
 Fig. 3 shows their projections onto the  $z = 0$ plane for different gravity models. 
The first model improved by \citet{Tsoulis_2012}, called polyhedral model, is the more accurate, but needs a large computational effort. The other models are divisions in several layers of tetrahedrons and were developed by \citet{Chanut_2015a}. In the present case, we share each tetrahedron up to 10 layers of equal thickness. When the value of the Jacobi constant $C$ is varied, the surfaces change. For some C values, the surfaces intersect or close on themselves at points in the $ x $ - $ y $ - $ z $ space, ordinarily called equilibrium points. Due to its oblate spherical shape, there are eight equilibrium points in the potential field close to the asteroid (101955) Bennu. As the surfaces are evaluated close to the critical values of $C$, Fig. 3 indicates the location of the eight equilibrium points projected onto the equatorial plane.  

In their analysis of the Potential Field and Equilibrium Points of Irregular-shaped Minor Celestial Bodies, \citet{Wang_2014} found the location of the equilibrium points of asteroid Bennu with different values of the bulk density ($0.97 \, g/cm^3$) and sidereal rotation period ($4.288 \, h$). As we can see, in the above-mentioned figure, the equilibrium points seem to be farther from the body with the present density and rotation period. Furthermore, from Fig. 3, the location of equilibrium points with the Mascon 10 model are closer if compared to the classic method.

\begin{table}
  \centering
   \begin{minipage}{1\linewidth}
   \caption{CPU time needed to compute the $10^6$ points of the grid on a Pentium 3.10 GHz computer}
   \label{symbols}
   \resizebox{1.0\textwidth}{!}{
   \begin{tabular}{|c|c|c|c|c|}
   \hline
    Asteroid &Mascon 4 & Mascon 6 & Mascon 10 & Tsulis \\
    \hline
    (101955) Bennu & 11m32s & 13m25s & 21m15s & 121m39s  \\
     \hline   
\end{tabular}}
 \end{minipage}
\end{table}

On the other hand, each additional division in the Mascon model somewhat increases the computational cost, as shown in Table 3.
We have, for example, a remarkable difference between the Mascon 4 and Mascon 10 models in Fig. 3, with only an accretion of ten minutes in execution. The computational cost is very important to be taken into account, because it increases considerably in numerical integrations. The Mascon 10 seems to be convenient here, as we tested higher divisions and the computational time greatly increases with a little difference in accuracy.

The exact location of the equilibrium points can be found solving
 \begin{equation}
   \nabla  V(x,y,z)=0,
 \end{equation}
 and we show the results in Table 4 for the classical polyhedral model and for the Mascon 10 model with the corresponding energy value $C$ for each point.

  \begin{table*}
\centering
\begin{minipage}{0.9\linewidth}
\caption{Locations of equilibrium points and their related Jacobi constant $C$ values generated by the Tsoulis and Mascon gravity tensor method.}

\resizebox{1.0\textwidth}{!}{

\begin{tabular}{crrrrrrrr}
\hline
  & \textcolor{white}{00000000} & $x (km)$ & \textcolor{white}{00000000} & $y (km)$  & \textcolor{white}{00000000} & $z (km)$  & \textcolor{white}{00000000} & $C(km^2 s^{-2})$  \\
  \hline
  \multicolumn{9}{c}{Tsoulis (uniform density)}\\
E1   & \textcolor{white}{00000000} &     0.32712780   & \textcolor{white}{00000000} &    0.01881705   & \textcolor{white}{00000000} &   -0.00323254 & \textcolor{white}{00000000} & -2.5185392 $ \times 10^{-8} $ \\
E2   & \textcolor{white}{00000000} &     0.12835825   & \textcolor{white}{00000000} &    0.29049337   & \textcolor{white}{00000000} &   -0.00248251 & \textcolor{white}{00000000} & -2.4786276  $ \times 10^{-8} $ \\
E3   & \textcolor{white}{00000000} &    -0.15087024   & \textcolor{white}{00000000} &    0.28572902   & \textcolor{white}{00000000} &   -0.00799664 & \textcolor{white}{00000000} & -2.4977207 $ \times 10^{-8} $  \\
E4   & \textcolor{white}{00000000} &    -0.22440718   & \textcolor{white}{00000000} &    0.22998957   & \textcolor{white}{00000000} &   -0.00734351 & \textcolor{white}{00000000} & -2.4960632  $ \times 10^{-8} $ \\
E5   & \textcolor{white}{00000000} &    -0.31365215   & \textcolor{white}{00000000} &   -0.09138746   & \textcolor{white}{00000000} &   -0.00230723 & \textcolor{white}{00000000} & -2.5148952  $ \times 10^{-8} $ \\
E6   & \textcolor{white}{00000000} &    -0.02359439   & \textcolor{white}{00000000} &   -0.31856713   & \textcolor{white}{00000000} &    0.00026125 & \textcolor{white}{00000000} & -2.4857406  $ \times 10^{-8} $ \\
E7   & \textcolor{white}{00000000} &     0.16235990   & \textcolor{white}{00000000} &   -0.27883894   & \textcolor{white}{00000000} &   -0.00196089 & \textcolor{white}{00000000} & -2.4944932  $ \times 10^{-8} $ \\
E8   & \textcolor{white}{00000000} &     0.22021204   & \textcolor{white}{00000000} &   -0.23383764   & \textcolor{white}{00000000} &   -0.00285934 & \textcolor{white}{00000000} & -2.4938440  $ \times 10^{-8} $ \\
 
  \multicolumn{9}{c}{Mascon10 (uniform density)}\\
E1   & \textcolor{white}{00000000} &     0.32753073    & \textcolor{white}{00000000} &   0.01761348   & \textcolor{white}{00000000} &   -0.00322919  & \textcolor{white}{00000000} & -2.527486 $ \times 10^{-8} $  \\
E2   & \textcolor{white}{00000000} &     0.12836337   & \textcolor{white}{00000000} &    0.29110750   & \textcolor{white}{00000000} &   -0.00247424  & \textcolor{white}{00000000} & -2.488105 $ \times 10^{-8} $  \\
E3   & \textcolor{white}{00000000} &    -0.15201320   & \textcolor{white}{00000000} &    0.28556798   & \textcolor{white}{00000000} &   -0.00771526  & \textcolor{white}{00000000} & -2.506817 $ \times 10^{-8} $  \\
E4   & \textcolor{white}{00000000} &    -0.22366530   & \textcolor{white}{00000000} &    0.23150318   & \textcolor{white}{00000000} &   -0.00713717  & \textcolor{white}{00000000} & -2.505418 $ \times 10^{-8} $  \\
E5   & \textcolor{white}{00000000} &    -0.31424051   & \textcolor{white}{00000000} &   -0.09053520   & \textcolor{white}{00000000} &   -0.00227728  & \textcolor{white}{00000000} & -2.523795 $ \times 10^{-8} $  \\
E6   & \textcolor{white}{00000000} &    -0.02294123   & \textcolor{white}{00000000} &   -0.31917790   & \textcolor{white}{00000000} &    0.00015956  & \textcolor{white}{00000000} & -2.495140 $ \times 10^{-8} $  \\
E7   & \textcolor{white}{00000000} &     0.16595331   & \textcolor{white}{00000000} &   -0.27711388   & \textcolor{white}{00000000} &   -0.00201768  & \textcolor{white}{00000000} & -2.503559 $ \times 10^{-8} $  \\
E8   & \textcolor{white}{00000000} &     0.21747342   & \textcolor{white}{00000000} &   -0.23717355   & \textcolor{white}{00000000} &   -0.00278987  & \textcolor{white}{00000000} & -2.503119 $ \times 10^{-8} $  \\

 \multicolumn{9}{c}{Tsoulis (Two-layered structure)}\\

E1   & \textcolor{white}{00000000} &     0.32816567   & \textcolor{white}{00000000} &    0.01909258   & \textcolor{white}{00000000} &   -0.00344706 & \textcolor{white}{00000000} & -2.5227172 $ \times 10^{-8} $  \\
E2   & \textcolor{white}{00000000} &     0.12952321   & \textcolor{white}{00000000} &    0.29012470   & \textcolor{white}{00000000} &   -0.00265287 & \textcolor{white}{00000000} & -2.4791869 $ \times 10^{-8} $  \\
E3   & \textcolor{white}{00000000} &    -0.14925987   & \textcolor{white}{00000000} &    0.28742942   & \textcolor{white}{00000000} &   -0.00872964 & \textcolor{white}{00000000} & -2.5003119 $ \times 10^{-8} $ \\
E4   & \textcolor{white}{00000000} &    -0.22653434   & \textcolor{white}{00000000} &    0.22850336   & \textcolor{white}{00000000} &   -0.00803714 & \textcolor{white}{00000000} & -2.4981428 $ \times 10^{-8} $ \\
E5   & \textcolor{white}{00000000} &    -0.31437975   & \textcolor{white}{00000000} &   -0.09236478   & \textcolor{white}{00000000} &   -0.00247585 & \textcolor{white}{00000000} & -2.5187497 $ \times 10^{-8} $  \\
E6   & \textcolor{white}{00000000} &    -0.02444105   & \textcolor{white}{00000000} &   -0.31881361   & \textcolor{white}{00000000} &    0.00039831 & \textcolor{white}{00000000} & -2.4870682 $ \times 10^{-8} $  \\
E7   & \textcolor{white}{00000000} &     0.15938233   & \textcolor{white}{00000000} &   -0.28143337   & \textcolor{white}{00000000} &   -0.00204463 & \textcolor{white}{00000000} & -2.4968379 $ \times 10^{-8} $  \\
E8   & \textcolor{white}{00000000} &     0.22463361   & \textcolor{white}{00000000} &   -0.23007960   & \textcolor{white}{00000000} &   -0.00316965 & \textcolor{white}{00000000} & -2.4957693 $ \times 10^{-8} $  \\

  \multicolumn{9}{c}{Mascon10 (Two-layered structure)}\\

E1   & \textcolor{white}{00000000} &     0.32799244   & \textcolor{white}{00000000} &    0.01789287   & \textcolor{white}{00000000} &   -0.00345690  & \textcolor{white}{00000000} & -2.522122 $ \times 10^{-8} $ \\
E2   & \textcolor{white}{00000000} &     0.12936662   & \textcolor{white}{00000000} &    0.29014265   & \textcolor{white}{00000000} &   -0.00265115  & \textcolor{white}{00000000} & -2.479068 $ \times 10^{-8} $ \\
E3   & \textcolor{white}{00000000} &    -0.14967103   & \textcolor{white}{00000000} &    0.28699429   & \textcolor{white}{00000000} &   -0.00847789  & \textcolor{white}{00000000} & -2.499839 $ \times 10^{-8} $ \\
E4   & \textcolor{white}{00000000} &    -0.22573166   & \textcolor{white}{00000000} &    0.22924887   & \textcolor{white}{00000000} &   -0.00785274  & \textcolor{white}{00000000} & -2.497922 $ \times 10^{-8} $ \\
E5   & \textcolor{white}{00000000} &    -0.31437138   & \textcolor{white}{00000000} &   -0.09148043   & \textcolor{white}{00000000} &   -0.00245738  & \textcolor{white}{00000000} & -2.518088 $ \times 10^{-8} $ \\
E6   & \textcolor{white}{00000000} &    -0.02383261   & \textcolor{white}{00000000} &   -0.31881751   & \textcolor{white}{00000000} &    0.00030015  & \textcolor{white}{00000000} & -2.486889 $ \times 10^{-8} $ \\
E7   & \textcolor{white}{00000000} &     0.16163245   & \textcolor{white}{00000000} &   -0.27987105   & \textcolor{white}{00000000} &   -0.00209692  & \textcolor{white}{00000000} & -2.496327 $ \times 10^{-8} $ \\
E8   & \textcolor{white}{00000000} &     0.22256375   & \textcolor{white}{00000000} &   -0.23203595   & \textcolor{white}{00000000} &   -0.00311998  & \textcolor{white}{00000000} & -2.495480 $ \times 10^{-8} $ \\

\hline
\end{tabular}}
 \end{minipage}
\end{table*}

The difference between the two models occurs in the third decimal digit and we consider acceptable a difference on the order of a metre. So, we can conclude that Mascon 10 has a good accuracy with respect to the classical polyhedral model for the calculation of equilibria. Due to (101955) Bennu irregular shape, there is no symmetry between the saddle points and the centre points. \citet{Wang_2014} showed that all the eight equilibrium points outside the asteroid Bennu ($E$1 $-$ $E$8) are unstable with their chosen bulk density ($0.97 \, g/cm^3$). The odd indices identify saddle points, while the even indices are associated with centre points. However, recently \citet{Wang_2016} have shown that within the limits of density and rotation defined by \citet{Chesley_2014}, the centre points can become linearly stable by changing the topological structure from case 5 to case 1 \citep{Jiang_2014}. In \citet{Aljbaae_2017}, we presented the corrected form of the second order derivatives of \citet{Chanut_2015a}.
We solved the linearized state equations in the neighbourhood of the equilibrium points \citep{Jiang_2014}, and we show the unnormalized eigenvalues and their stability type in Table 5. From the physical features chosen, we confirm that the equilibrium point $E$8 become linearly stable. According to \citet{Scheeres_2016}, there is one stable
centre equilibrium point at the nominal density, and three stable centre equilibrium points at the highest density. If there are stable equilibria in Bennu, a concentration of dust distributed on the surface of the asteroid can be expected in the vicinity of these regions, since an impact may cause the launched regolith to be captured in those regions and returns to the surface.

   \begin{table*}
   \centering
   \begin{minipage}{1\linewidth}
   \caption{Eigenvalues of Jacobi matrix of the eight external equilibrium points and their stability} 
   
   \resizebox{1.0\textwidth}{!}{

   \begin{tabular}{crrrrrrrr}
   \hline
   
   Eigenvalues $\times 10^ {-4}$ & \multicolumn{1}{c}{E1} & \multicolumn{1}{c}{E2} & \multicolumn{1}{c}{E3} & \multicolumn{1}{c}{E4} & \multicolumn{1}{c}{E5} & \multicolumn{1}{c}{E6} & \multicolumn{1}{c}{E7} & \multicolumn{1}{c}{E8} \\ \hline
   \multicolumn{9}{c}{Tsoulis (uniform density)}\\
 
$\lambda_{1}$ &           4.6265 $i$ &           4.3031 $i$ &           4.5318 $i$ &           4.4365 $i$ &           4.6124 $i$ &           4.3623 $i$ &           4.5252 $i$ &           4.4442 $i$ \\ 
$\lambda_{2}$ &          -4.6265 $i$ &          -4.3031 $i$ &          -4.5318 $i$ &          -4.4365 $i$ &          -4.6124 $i$ &          - 4.3623 $i$ &          -4.5252 $i$ &          -4.4442 $i$ \\ 
$\lambda_{3}$ &           4.2744 $i$ & -0.4427 +  2.7267 $i$ &           3.9935 $i$ & -0.4715 + 2.6227 $i$ &           4.0714 $i$ & -0.3781 +  2.6694 $i$ &           3.8816 $i$ &           3.0387 $i$ \\ 
$\lambda_{4}$ &          -4.2744 $i$ & -0.4427 - 2.7267 $i$ &          -3.9935 $i$ & -0.4715 - 2.6227 $i$ &          -4.0714 $i$ & -0.3781 - 2.6694 $i$ &          -3.8816 $i$ &          -3.0387 $i$ \\ 
$\lambda_{5}$ &           -2.5846    & 0.4427 +  2.7267 $i$ &           -1.8683    & 0.4715 + 2.6227 $i$ &           -2.2036    & 0.3781 + 2.6694 $i$ &           -1.5965    &           2.0025 $i$ \\ 
$\lambda_{6}$ &            2.5846    & 0.4427 - 2.7267 $i$ &            1.8683    & 0.4715 - 2.6227 $i$ &            2.2036    & 0.3781 - 2.6694 $i$ &            1.5965    &          -2.0025 $i$ \\ 
\multicolumn{9}{c}{Mascon10 (uniform density)}\\
$\lambda_{1}$ &           4.5929 $i$ &           4.2946 $i$ &           4.5063 $i$ &           4.4263 $i$ &           4.5791 $i$ &           4.3548 $i$ &           4.4983 $i$ &           4.4340 $i$ \\ 
$\lambda_{2}$ &          -4.5929 $i$ &          -4.2946 $i$ &          -4.5063 $i$ &          -4.4263 $i$ &          -4.5791 $i$ &          -4.3548 $i$ &          -4.4983 $i$ &          -4.4340 $i$ \\ 
$\lambda_{3}$ &           4.2676 $i$ & -0.4021 +  2.7271 $i$ &           3.9808 $i$ & -0.2122 +  2.5973 $i$ &           4.0710 $i$ & -0.2472 +  2.6602 $i$ &           3.8659 $i$ &           3.1951 $i$ \\ 
$\lambda_{4}$ &          -4.2676 $i$ & -0.4021 - 2.7271 $i$ &          -3.9808 $i$ & -0.2122 - 2.5973 $i$ &          -4.0710 $i$ & -0.2472 - 2.6602 $i$ &          -3.8659 $i$ &          -3.1951 $i$ \\ 
$\lambda_{5}$ &           -2.5124    & 0.4021 +  2.7271 $i$ &           -1.7771    & 0.2122 +  2.5973 $i$ &           -2.1322    & 0.2472 +  2.6602 $i$ &           -1.4782    &           1.7680 $i$ \\ 
$\lambda_{6}$ &            2.5124    & 0.4021 - 2.7271 $i$ &            1.7771    & 0.2122 - 2.5973 $i$ &            2.1322    & 0.2472 - 2.6602 $i$ &            1.4782    &          -1.7680 $i$ \\ 
\hline

  Case (Stability)   & \multicolumn{1}{c}{ 2 (U)} & \multicolumn{1}{c}{5 (U)} & \multicolumn{1}{c}{2 (U)} & \multicolumn{1}{c}{5 (U)} & \multicolumn{1}{c}{2 (U)} & \multicolumn{1}{c}{5 (U)} & \multicolumn{1}{c}{2 (U)} & \multicolumn{1}{c}{1 (LS)} \\ \hline
\multicolumn{9}{c}{Tsoulis (Two-layered structure)}\\

   $\lambda_{1}$ &     4.6789 $i$          &           4.3283 $i$ &           4.5817 $i$ &           4.4743 $i$ &           4.6635 $i$ &           4.3917 $i$ &           4.5767 $i$ &           4.4760 $i$ \\ 
$\lambda_{2}$ &      -4.6789 $i$       &          -4.3283 $i$ &          -4.5817 $i$ &          -4.4743 $i$ &          -4.6635 $i$ &          -4.3917 $i$ &          -4.5767 $i$ &          -4.4760 $i$ \\ 
$\lambda_{3}$ &          4.2968 $i$ & -0.6551 +  2.7495 $i$ &           4.0060 $i$ & -0.7725 +  2.6616 $i$ &           4.0822 $i$ & -0.6391 +  2.6949 $i$ &           3.9090 $i$ & -0.4496 +  2.5850 $i$ \\ 
$\lambda_{4}$ &           -4.2968 $i$ & -0.6551 - 2.7495 $i$ &          -4.0060 $i$ & -0.7725 - 2.6616 $i$ &          -4.0822 $i$ & -0.6391 - 2.6949 $i$ &          -3.9090 $i$ & -0.4496 - 2.5850 $i$ \\ 
$\lambda_{5}$ &          -2.7129  & 0.6551 +  2.7495 $i$ &           -2.0112    & 0.7725 +  2.6616 $i$ &           -2.3277    & 0.6391 +  2.6949 $i$ &           -1.7975    & 0.4496 +  2.5850 $i$ \\ 
$\lambda_{6}$ &            2.7129    & 0.6551 +  -2.7495 $i$ &            2.0112    & 0.7725 - 2.6616 $i$ &            2.3277    & 0.6391 - 2.6949 $i$ &            1.7975    & 0.4496 - 2.5850 $i$ \\ 
\multicolumn{9}{c}{Mascon10 (Two-layered structure)}\\

$\lambda_{1}$ &     4.6481 $i$          &           4.3209 $i$ &           4.5584 $i$ &           4.4658 $i$ &           4.6323 $i$ &           4.3856 $i$ &           4.5522 $i$ &           4.4669 $i$ \\ 
$\lambda_{2}$ &    -4.6481 $i$      &          -4.3209 $i$ &          -4.5584 $i$ &          -4.4658 $i$ &          -4.6323 $i$ &          -4.3856 $i$ &          -4.5522 $i$ &          -4.4669 $i$ \\ 
$\lambda_{3}$ &      4.2909 $i$      & -0.6373 +  2.7511 $i$ &           3.9954 $i$ & -0.6724 +  2.6416 $i$ &           4.0821 $i$ & -0.5864 +  2.6878 $i$ &           3.8992 $i$ & -0.1809 +  2.5600 $i$ \\ 
$\lambda_{4}$ &      -4.2909 $i$   & -0.6373 - 2.7511 $i$ &          -3.9954 $i$ & -0.6724 - 2.6416 $i$ &          -4.0821 $i$ & -0.5864 - 2.6878 $i$ &          -3.8992 $i$ & -0.1809 - 2.5600 $i$ \\ 
$\lambda_{5}$ &          -2.6499   & 0.6373 +  2.7511 $i$ &           -1.9358    & 0.6724 +  2.6416 $i$ &           -2.2643    & 0.5864 +  2.6878 $i$ &           -1.7123    & 0.1809 +  2.5600 $i$ \\ 
$\lambda_{6}$ &            2.6499    & 0.6373 - 2.7511 $i$ &            1.9358    & 0.6724  - 2.6416 $i$ &            2.2643    & 0.5864 - 2.6878 $i$ &            1.7123    & 0.1809 - 2.5600 $i$ \\
\hline

  Case (Stability)   & \multicolumn{1}{c}{ 2 (U)} & \multicolumn{1}{c}{5 (U)} & \multicolumn{1}{c}{2 (U)} & \multicolumn{1}{c}{5 (U)} & \multicolumn{1}{c}{2 (U)} & \multicolumn{1}{c}{5 (U)} & \multicolumn{1}{c}{2 (U)} & \multicolumn{1}{c}{5 (U)} \\ \hline

   \end{tabular}}
   \end{minipage}
   \end{table*}

\begin{figure}
   \centering
   
    \includegraphics[width=0.96\linewidth]{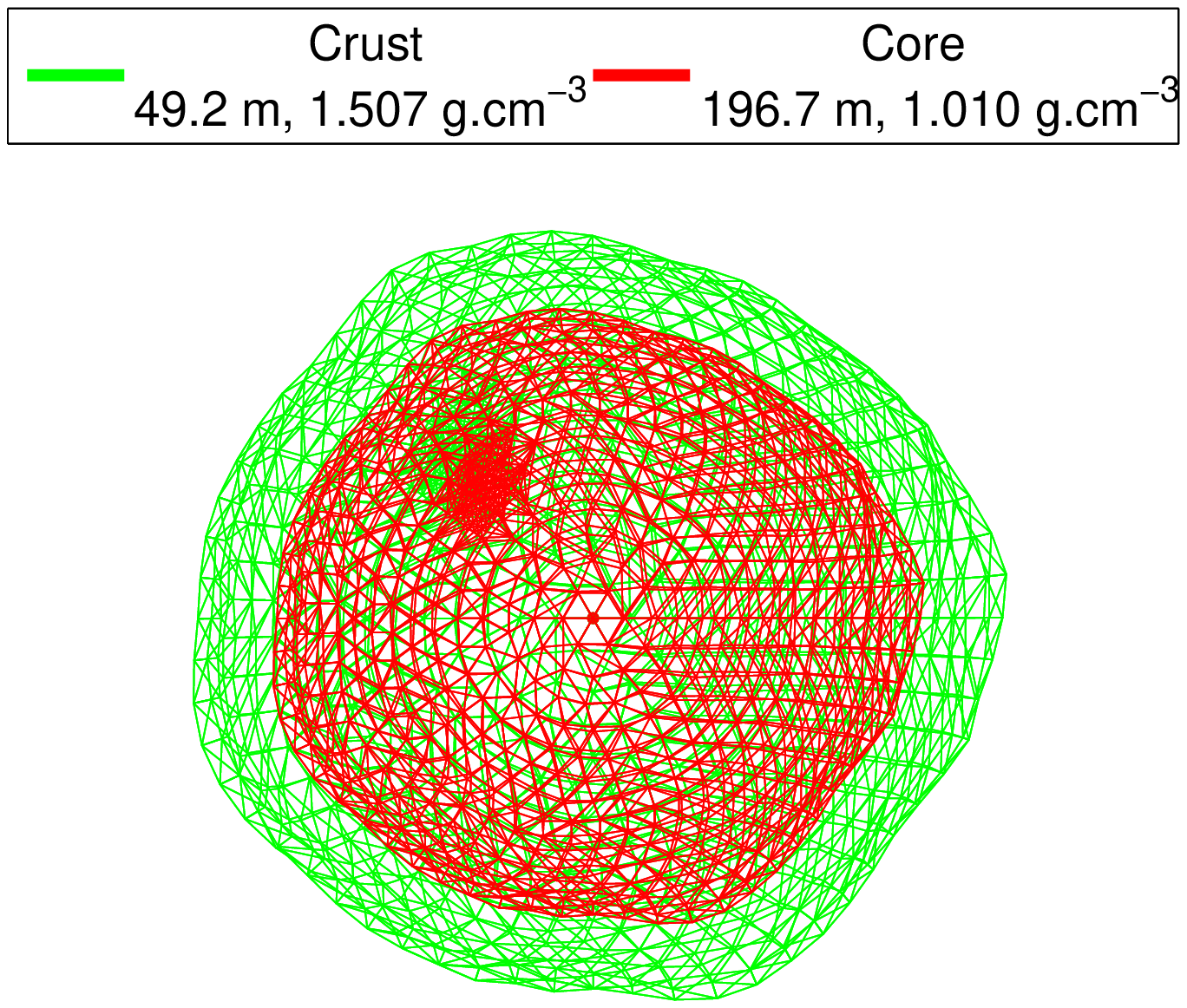}
    \caption{Two-layered structure of asteroid (101955) Bennu with a 50 m deep surface layer.
}
         \label{Fig4} 
\end{figure}
\begin{figure}
   \centering
    
    Mascon 10 (2 densities)\\
     \includegraphics[width=0.96\linewidth]{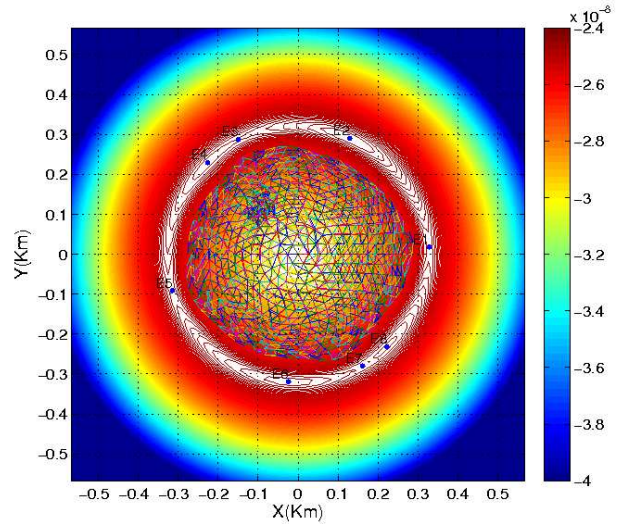}\\
     \caption{Zero-velocity curves and equilibrium points of asteroid (101955) Bennu. The color code gives the intensity of the Jacobi constant in $km^2$ $s^{-2}$. The equilibrium points outside the body are displayed using the Mascon 10 gravity model shown in Fig. 4.
}
         \label{Fig5} 
\end{figure}
The goal is now to investigate the variation in the gravity field due to the non-homogeneous mass distribution and to check the effects of the non-homogeneous distribution on the equilibria. In \citet{Aljbaae_2017}, we tested two different mass distributions for asteroid (21) Lutetia and we found that the two models of layered structures are not determinant for the stability of the equilibria. For (101955) Bennu, we take the two-layered model described as a surface model in \citet{Scheeres_2016}. This distribution is illustrated in Fig. 4 with the two volumes in the body adjusted for the total mass to remain constant. The model corresponds to a volume-equivalent diameter of 491.8 m, where the crust has a mean thickness of 49.2 m occupying 48.8\% of the total volume with a density of 1.507 $g.cm^{-3}$, that represents
58.72\% of the total mass. The core, based on rubble-pile characteristics, is considered with a density of 1.010 $g.cm^{-3}$. 
Figure 5 shows the projections of the zero-velocity surfaces onto the  $z = 0$ plane when the two-layered model of density distribution is considered. The equilibria do not fundamentally change unless a little displacement away from the body is considered. The equilibrium points $E$3 and $E$4 and also the points $E$7 and $E$8 seem to move away from each other. So, we can differentiate the topological structure of the equilibrium points better in the figure. It does not happen with the nominal constant density and for higher densities, these points mix and disappear \citep{Wang_2016, Scheeres_2016}. The eigenvalues and their stability type for the two gravity models are shown at the bottom of Table 5. What we can highlight is that the centre point $E$8, which was linearly stable in the case of the homogeneous mass distribution, returns to be unstable changing the topological structure from case 1 to case 5. This situation already existed with the low density, and it can make a previously stable trajectory about the centre point to become unstable, allowing the spacecraft or any particles to be ejected or briefly collide with the asteroid. This possibility must be taken into account by the $OSIRIS-REx$ mission. Other factors, such as the solar radiation pressure, can make periodic motions about those equilibrium points to be unstable \citep{Xin_2016}, but this point will not be investigated in the paper. Another important point is that the running time shown in Table 6 indicated that the classical polyhedron model of gravitation is not very well suited for the division into several different density layers, while the Mascon mathematical model described in \citet{Chanut_2015a} seems more appliable for this purpose. Unlike the tsoulis method, in the mascon model case, the calculations are done only one time for the total potential, independently of each layer density. However, certain distributions may favor the calculations in the execution. It is the case here, for the two density layers.

\begin{table}
  \centering
   \begin{minipage}{1\linewidth}
   \caption{CPU time needed to find the eight equilibrium points on a Pentium 3.10 GHz computer}
   \label{symbols}
   \resizebox{1.0\textwidth}{!}{
   \begin{tabular}{|c|c|c|c|c|}
   \hline
    Asteroid &Mascon 10 & Mascon 10(2) & Tsoulis & Tsoulis(2) \\
    \hline
    (101955) Bennu & 65m27s & 62m26s & 148m57s & 414m04s  \\
     \hline   
\end{tabular}}
 \end{minipage}
\end{table}

\section[]{Dynamics close to (101955) Bennu with solar radiation pressure (SRP)}
  \subsection[]{Solar radiation pressure dynamical model}
We consider the case of a spacecraft in the vicinity of an asteroid and significantly far from any other celestial
body. In fact, since the mean motion of the asteroid around the Sun is generally much smaller than its spin rate, the radiation pressure from the Sun has a constant direction and magnitude at a given distance. When studying the dynamics around it a short time interval, we can assume that during this time interval, the position of the Sun is ``frozen'' in the inertial space. The formulation of Hill's problem taking into account the effect of SRP was presented by \citet{Scheeres_2002}. However, in the body-fixed frame of the asteroid, as represented in Fig. 6, the Sun has a circular orbit perpendicular to the asteroid's spin axis, with a constant latitude $\theta$ and a time-varying longitude $\psi = - \omega t + \psi_0$, where $\psi_0$ is the initial longitude of the Sun.
\begin{figure}
   \centering
   \includegraphics[width=1.0\linewidth]{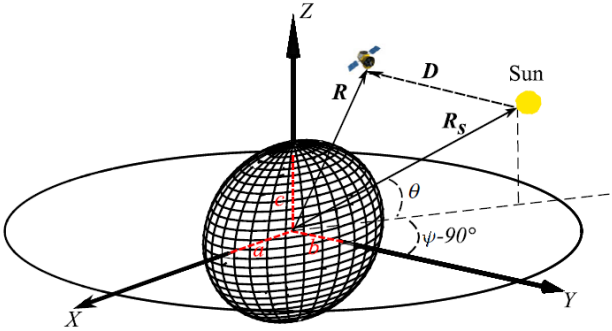}
   \vspace{-1cm}{}
         \caption{Schematic illustration of the SRP model \citep{Xin_2016}.
}
         \label{Fig6}
   \end{figure}  
  Therefore, the solar radiation pressure is acting in the anti-sunward direction. Consequently, the equations of motion taking SRP into account are:
 \begin{eqnarray}
    \ddot{x}-2\omega \dot{y} = \omega^2 x +U_x -\nu g \cos\theta \cos \psi ,\\  
    \ddot{y}+2\omega \dot{x} = \omega^2 y +U_y -\nu g \cos\theta \sin \psi,\\ 
    \ddot{z}= U_z -\nu g \sin\theta.
 \end{eqnarray}
 where $U_x$, $U_y$ and $U_z$ denote the first-order partial derivatives of the potential and g is the SRP magnitude computed as:
\begin{eqnarray}
        g = \frac{\beta }{D^2}.
 \end{eqnarray}
where $\beta = (1+\eta)G1/B$, $G1 = 1 \times 10^8\, kg km^3s^{-2}m^{-2} $ is a solar constant, $B$ is the spacecraft mass to area ratio in $(kg/m^2)$, $\eta $ is the reflectance of the spacecraft, and $D$ is the distance of the spacecraft from the Sun, in km. Here, we consider $D$ = $R_S$ (the heliocentric distance of the asteroid) while $R$ is negligible with respect to $R_S$ close to the body. For the spacecraft OSIRIS-REx mass-area ratio, we took the values of \citet{Lantoine_2013} that allow $B \simeq  96\, kg/m^2$ and a reflectance of the material $\eta=0.2$. $\nu$ represents the solar eclipse for the shadow projected by an ellipsoidal asteroid taking the values 1 or 0 as defined in \citep{Xin_2016}.

\subsection[]{Numerical simulations and initial conditions}
In this section, we numerically test and analyse the evolution of the dynamics of equatorial, direct orbits, considering the 3-D gravitational perturbation and the solar radiation pressure (SRP) close to the asteroid (101955) Bennu. As done in the previous section, we choose the Mascon gravity tensor implemented by \citet{Chanut_2015a}, with each tetrahedron divided into ten equal thickness layers (Mascon 10). The path of the spacecraft is calculated using the equations of motion (Eqs. 15 to 17) in the rotating body-fixed reference frame, and the equations are integrated by the Bulirsch-Stoer numerical algorithm. The initial orbits of the spacecraft are launched at the periapsis radius of the equatorial plane ($z = 0$) for four values of its longitude $\lambda \in (0,360^o)$. The interval between the initial values of $\lambda $ is $90^o$ and the initial periapsis radii are taken from 0.34 $km$ up to 0.7 $km$, with an interval of 10 $m$. Moreover, the initial eccentricities of the spacecraft's orbit start from 0 up to 0.4, with an interval of 0.01. The initial velocities are taken from the two body problem in the rotating body-fixed reference frame when there is no other perturbative force depending on time (Eqs. 6 to 8) or the initial position of the spacecraft is in the shadow. If we put the same initial velocities between cases with and without solar radiation pressure, this means that the initial position of the spacecraft is always in the shadow, which is only true when the Sun is in the oposite direction. Let $v0_x$, $v0_y$ and $v0_z$ be those initial velocities. However, integrating the equations 15 to 17 with respect to time and for $t = 0$, the term containing the time-varing longitude appears in the initial velocities where $v'0_x = v0_x + \frac{g}{\omega}\cos\theta\sin \psi_0$, $v'0_y = v0_y - \frac{g}{\omega}\cos\theta\cos \psi_0$ and $v'0_z = v0_z = 0$. 
The pole orientation in ecliptic coordinates was determined by \citet{Nolan_2013} as $(-88^o, 45^o)$. So, we take $\theta = 2^o$ and two values for the whole simulations with the Sun initial longitude  $\psi_0 = -180+45^o$ or $\psi_0 = -135^o$, and $\psi_0 = -180^o$ when the sun direction lies along the x-axis. To verify if it exists a certain symmetry in the behaviour around asteroid (101955) Bennu as pointed out by \citep{Xin_2016} in the case of the ellipsoid, we will test the intermediate longitudes of the Sun $\psi_0 = -90^o$ and $\psi_0 = -45^o$ at the initial spacecraft longitude $\lambda = 90^o$. 
\citet{Broschart_2014} have corrected some values of parameters of various mission configurations. However,they have computed the reflectance in their $B$ parameters while we compute the reflectance only in the $\beta$ parameter. As values of $B$, we take $B= 96\, kg/m^2$ for the Osiris-Rex probe. The reflectance $\eta=20\%$ is taken into account in our calculation of $\beta$. In order to evaluate the maximum, medium and minimum values of the $g$ parameter of Eq.18, we consider the motion of the spacecraft close to the asteroid Bennu when it is at the perihelion distance $D=0.8969\, AU$, semi-major axis distance $D=1.1264\, AU$ and aphelion distance $D=1.3559\, AU$ from the Sun. 
 When the path of the spacecraft crosses the boundaries of the ellipsoid approximation with semi-major axes ($0.29 \times  0.27 \times  0.26 \, km$), we consider that the spacecraft has impacted with the asteroid and the integration is stopped. The integration's time is $\approx  60$ days, adequate to determinate the final destiny of the orbits. This time allows us to consider acceptable not to change the distance of the asteroid from the Sun during the integration.

\begin{figure*}
   \centering
     Without SPR ($g=0$)\\ 
   \textcolor{white}{.}\\ \hspace{0.5cm} $0^{\circ}$ \hspace{8cm} $90^{\circ}$\\  
 \includegraphics[width=0.48\linewidth]{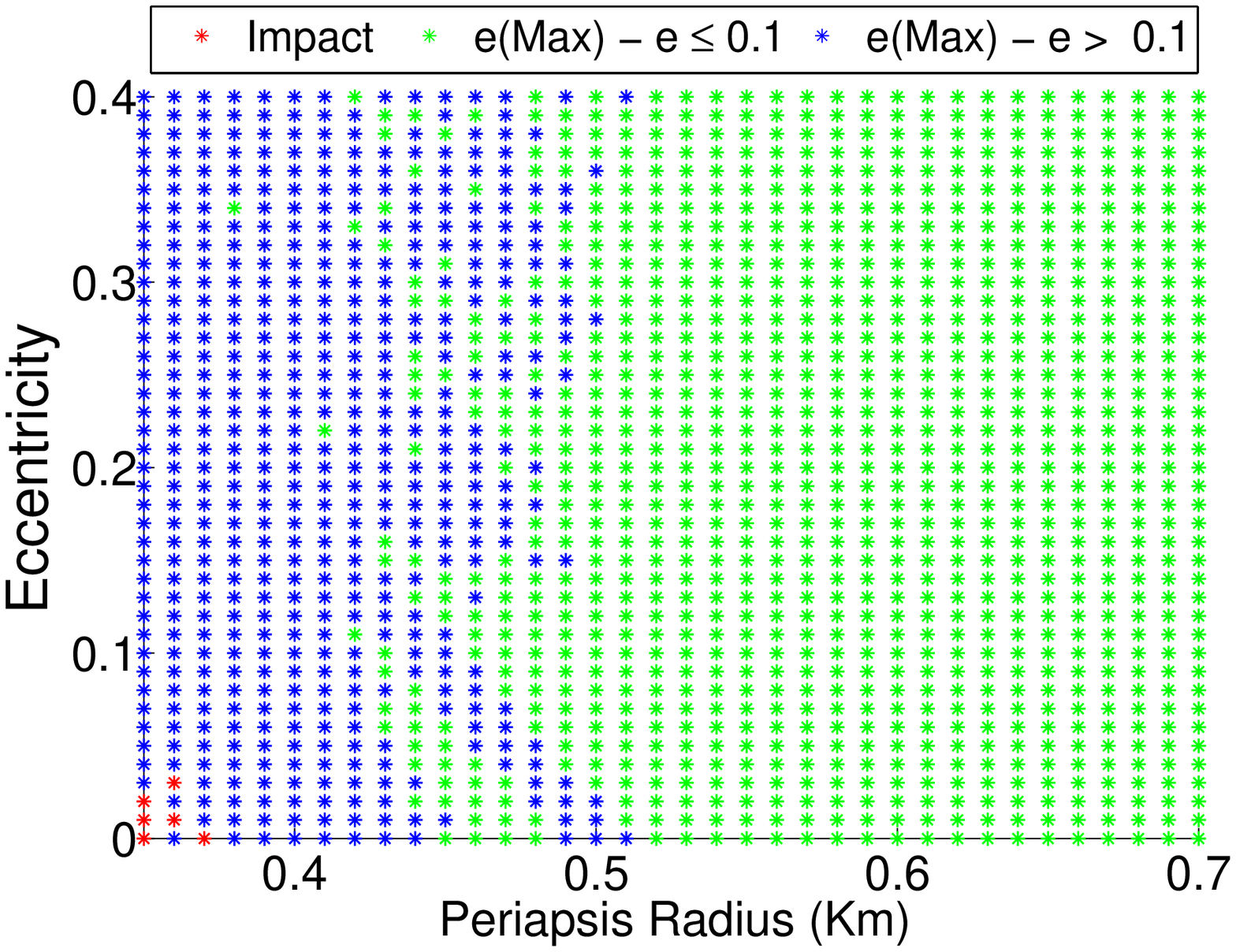}
 \includegraphics[width=0.48\linewidth]{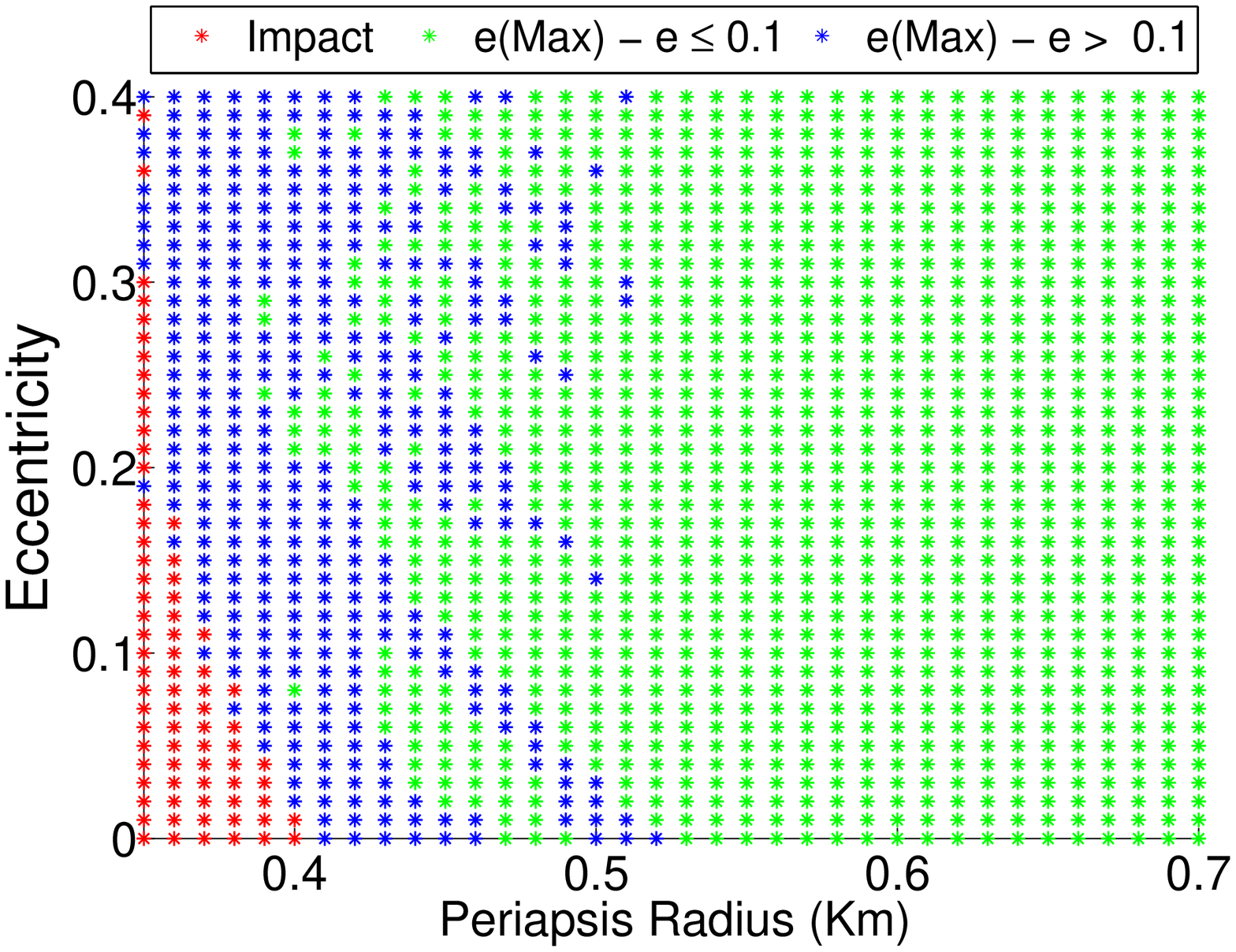}
 \textcolor{white}{.}\\ \hspace{0.5cm} $180^{\circ}$ \hspace{8cm} $270^{\circ}$\\
 \includegraphics[width=0.48\linewidth]{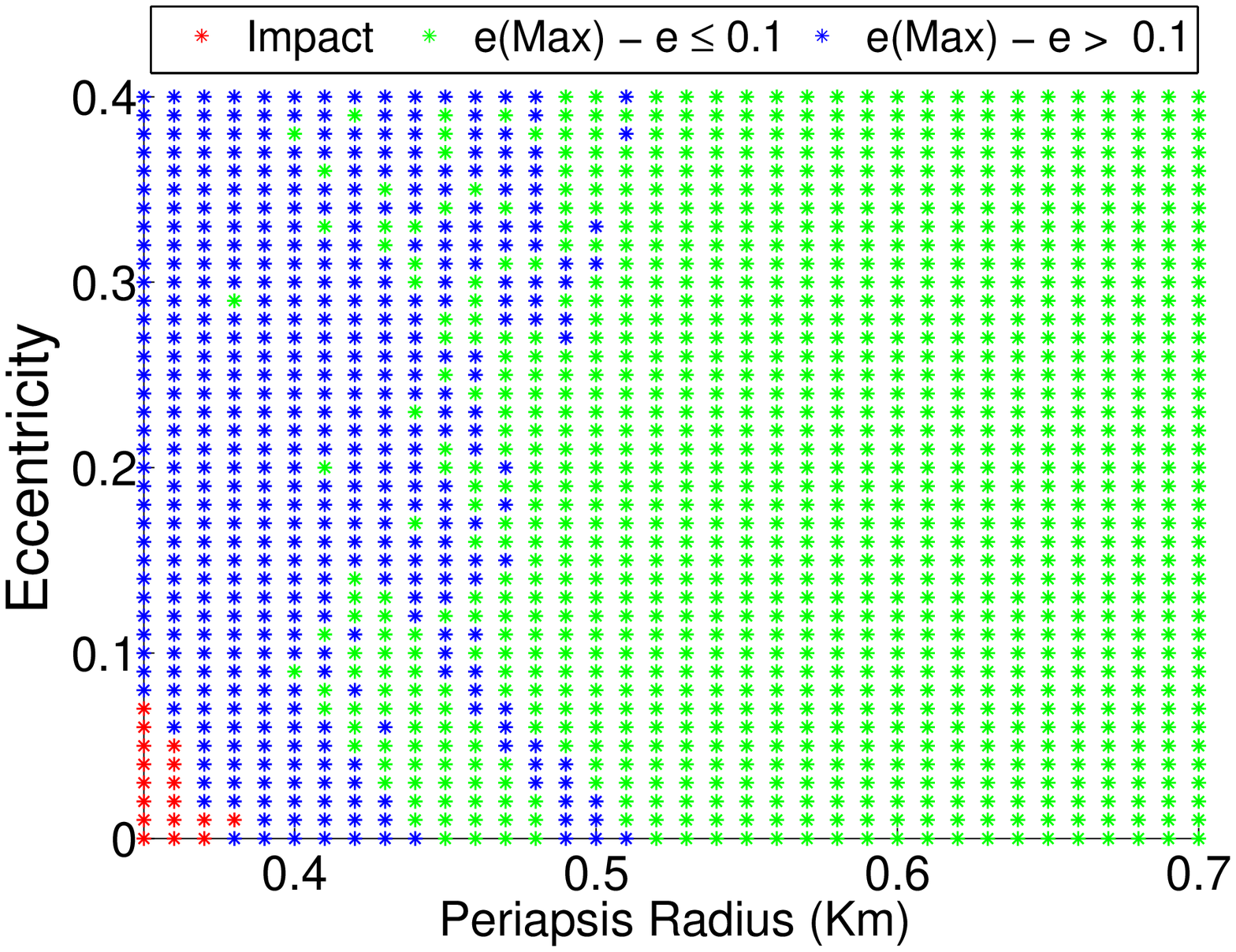}
 \includegraphics[width=0.48\linewidth]{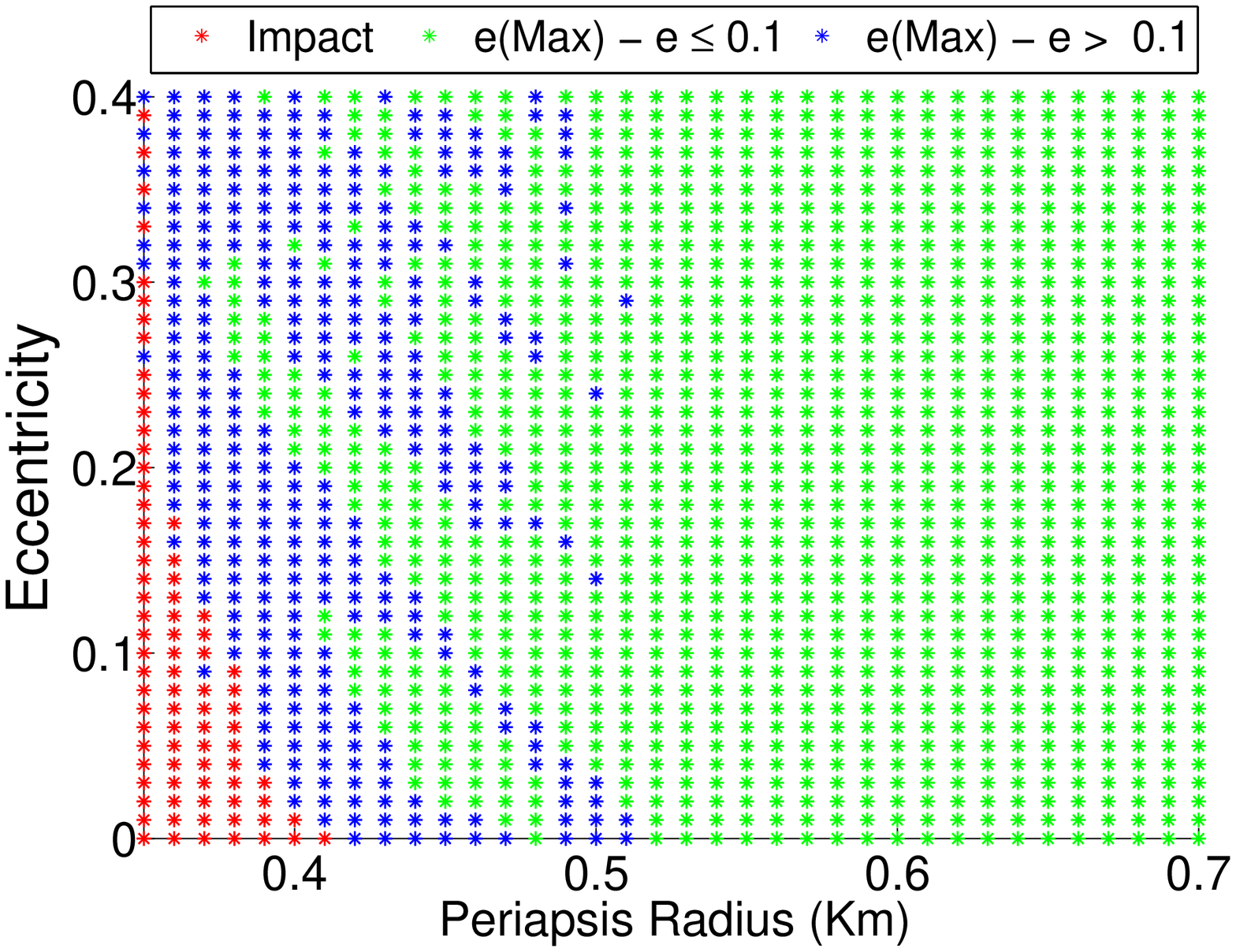}\\
     \caption{Stability maps of the equatorial orbits relative to (101955) Bennu with initial longitudes $\lambda = 0 $, $\lambda = 90^o $, $\lambda = 180^o $ and $\lambda = 270^o $ after $60$ days. The initial eccentricity $e$ goes from 0 up to 0.4 at their periapsis radius and is compared with the maximal eccentricity of the orbit. The initial periapsis radii are from 0.35 to 0.7 $Km$. The solar radiation pressure is not accounted for.
}
         \label{Fig7} 
\end{figure*} 
\subsection[]{Results of the numerical simulations}
It is reasonable to consider that the gravitational perturbation due to Bennu shape should be weaker than the analogous perturbations for asteroids like (216) Kleopatra or (433) Eros. As we neglected other effects beyond the gravitational and SRP, the solar radiation pressure becomes a relevant perturbation.  Orbits that increase less than 0.01 in their eccentricities will be considered as stable and marked in green in the figures. Furthermore, the trajectories that collide with the asteroid will be marked in red. In an attempt to model a close encounter of the Osiris-Rex probe with the asteroid (101955) Bennu, we followed the trajectories in the 3-D space to verify their ultimate fate. The trajectories are launched from four initial longitudes in the equatorial plane, where, in principle, the highest and lowest values of the energy potential occur, as shown in Table 4. Taking into account the whole trajectories, we generated Fig. 7 after an integration of $60$ days, or more than 330 (101955) Bennu rotation periods. Here, we choose eccentricities compatible with an observation mission ($e = 0$ up to $e = 0.4$). The solar radiation pressure is not accounted for. Orbits with initial $\lambda = 0 $ appear to be less subject to collision, which may keep low eccentric orbits near the body for more time. On the contrary, orbits with initial $\lambda = 90^o $ and $\lambda = 270^o $ are more likely to collide when the solar radiation pressure is not accounted for. When compared with the asteroid (216) Kleopatra or (433) Eros \citep{Chanut_2014,Chanut_2015b}, we need to add a factor $\pi$. That is probably due to the fact that asteroid (101955) Bennu is undergoing retrograde rotation. Moreover, below 0.34 km, the orbits are unstable, and we checked that the limit radius for direct, initially equatorial circular orbits that cannot impact with (101955) Bennu surface is, indeed, 0.42 km. For the initial conditions that we choose, it clearly appears that most of the orbits are stable beyond 0.5 km. However, below this distance, orbits with initial $\lambda = 90^o $ and $\lambda = 270^o $ seem to find more stable conditions. The results are shown in Fig. 7.

\begin{figure*}
   \centering
 SRP ($g=3.037665 \times 10^{-11}km\cdot s^{-2}$,  $R=1.3559AU$) \\
{\small \textcolor{white}{.} \hspace{0.75cm} $0^{\circ}$ \hspace{3.75cm} $90^{\circ}$ \hspace{3.75cm}  $180^{\circ}$ \hspace{3.75cm} $270^{\circ}$}\\
1) \includegraphics[width=0.24\linewidth]{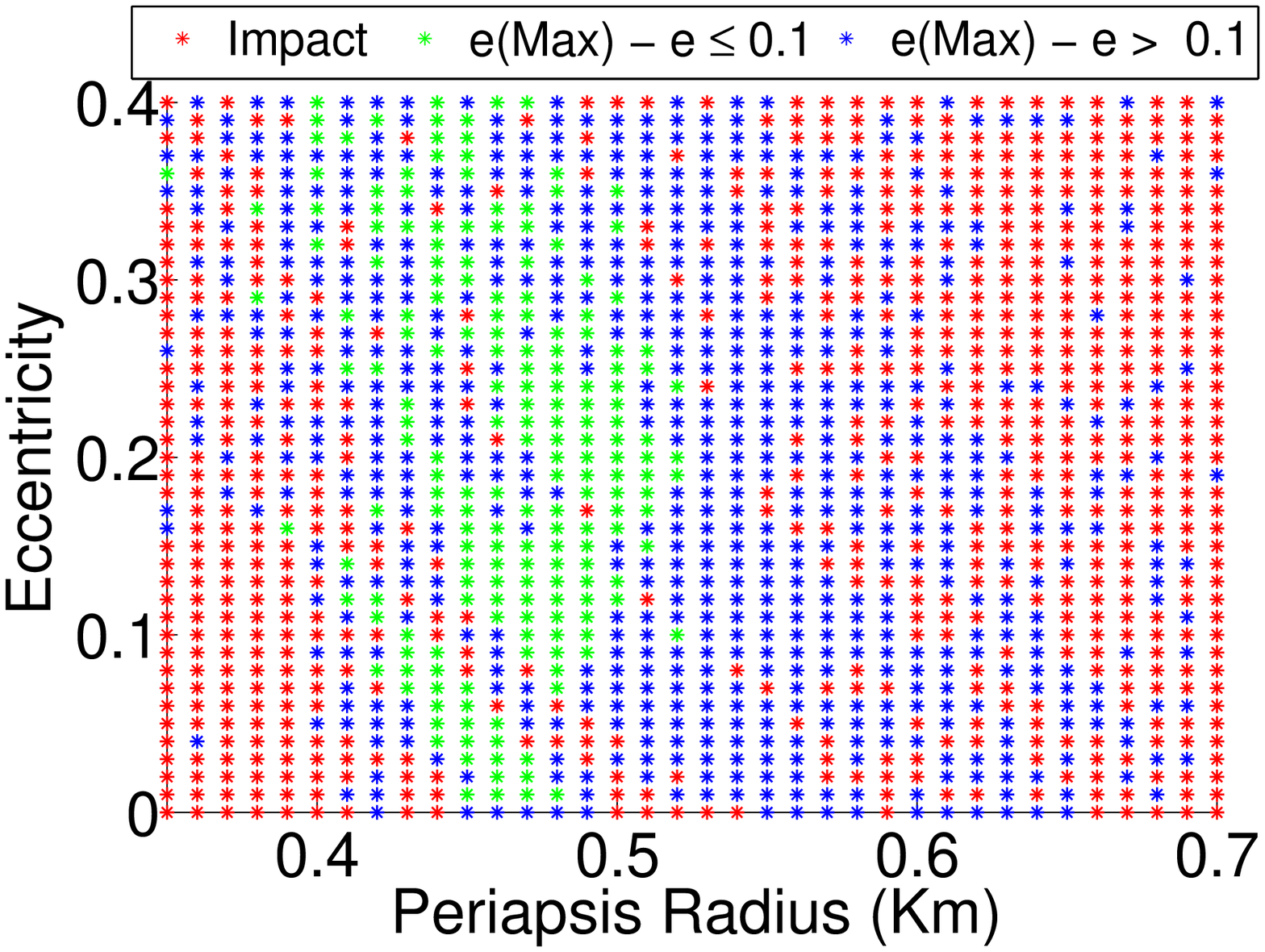}
 \includegraphics[width=0.24\linewidth]{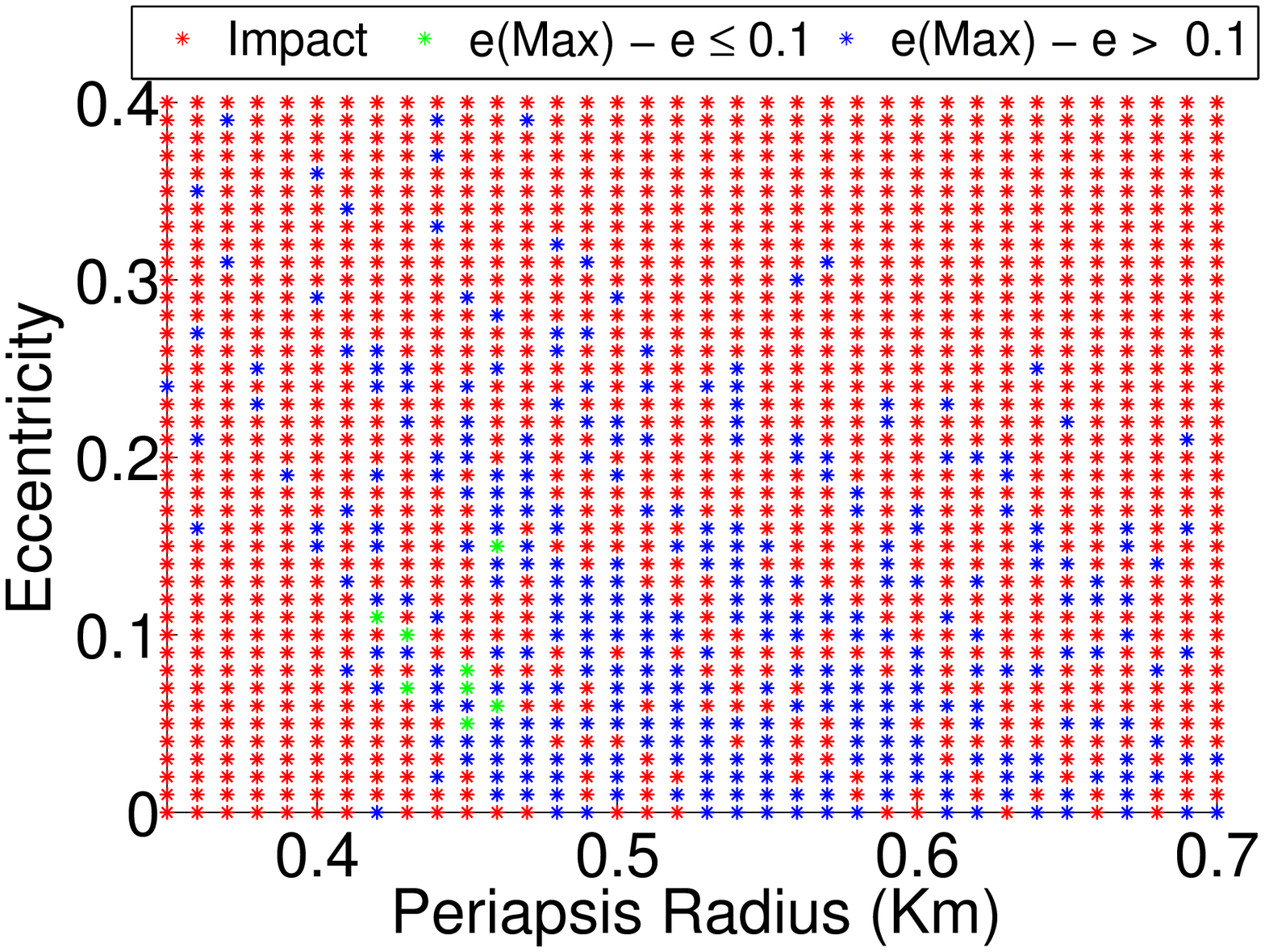}
 \includegraphics[width=0.24\linewidth]{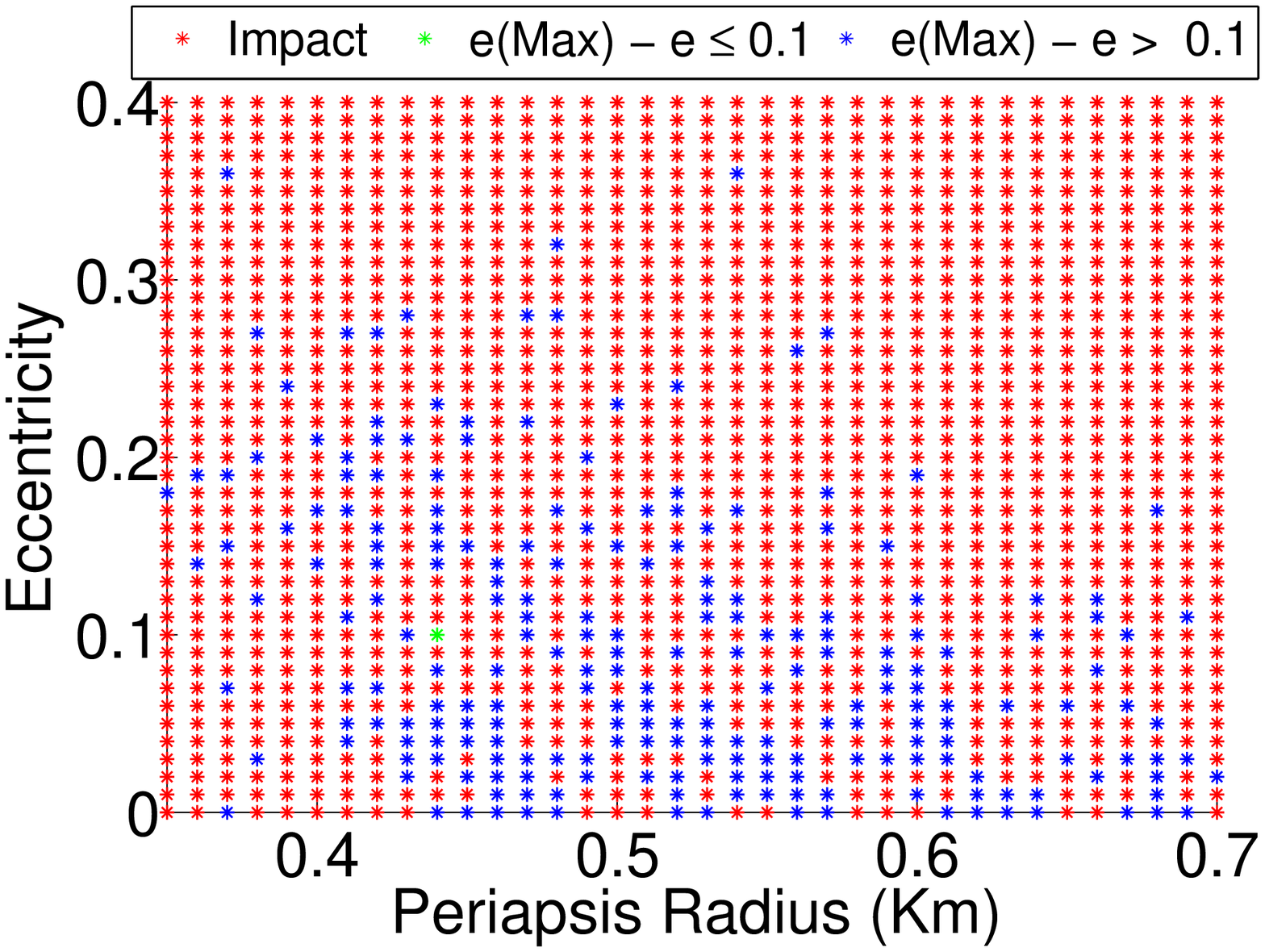}
 \includegraphics[width=0.24\linewidth]{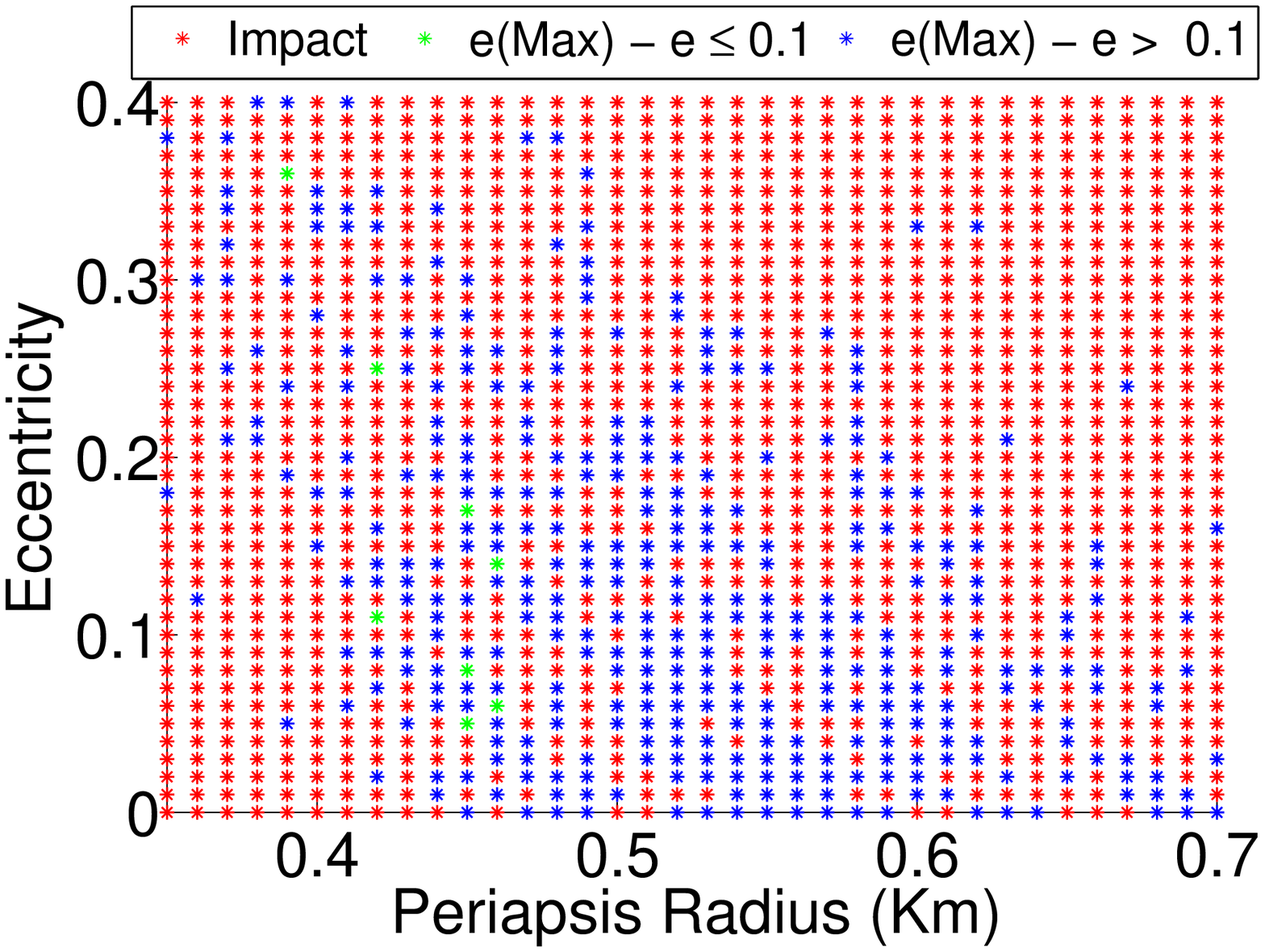}\\
 {\small \textcolor{white}{.} \hspace{0.75cm} $0^{\circ}$ \hspace{3.75cm} $90^{\circ}$ \hspace{3.75cm}  $180^{\circ}$ \hspace{3.75cm} $270^{\circ}$}\\
 2) \includegraphics[width=0.24\linewidth]{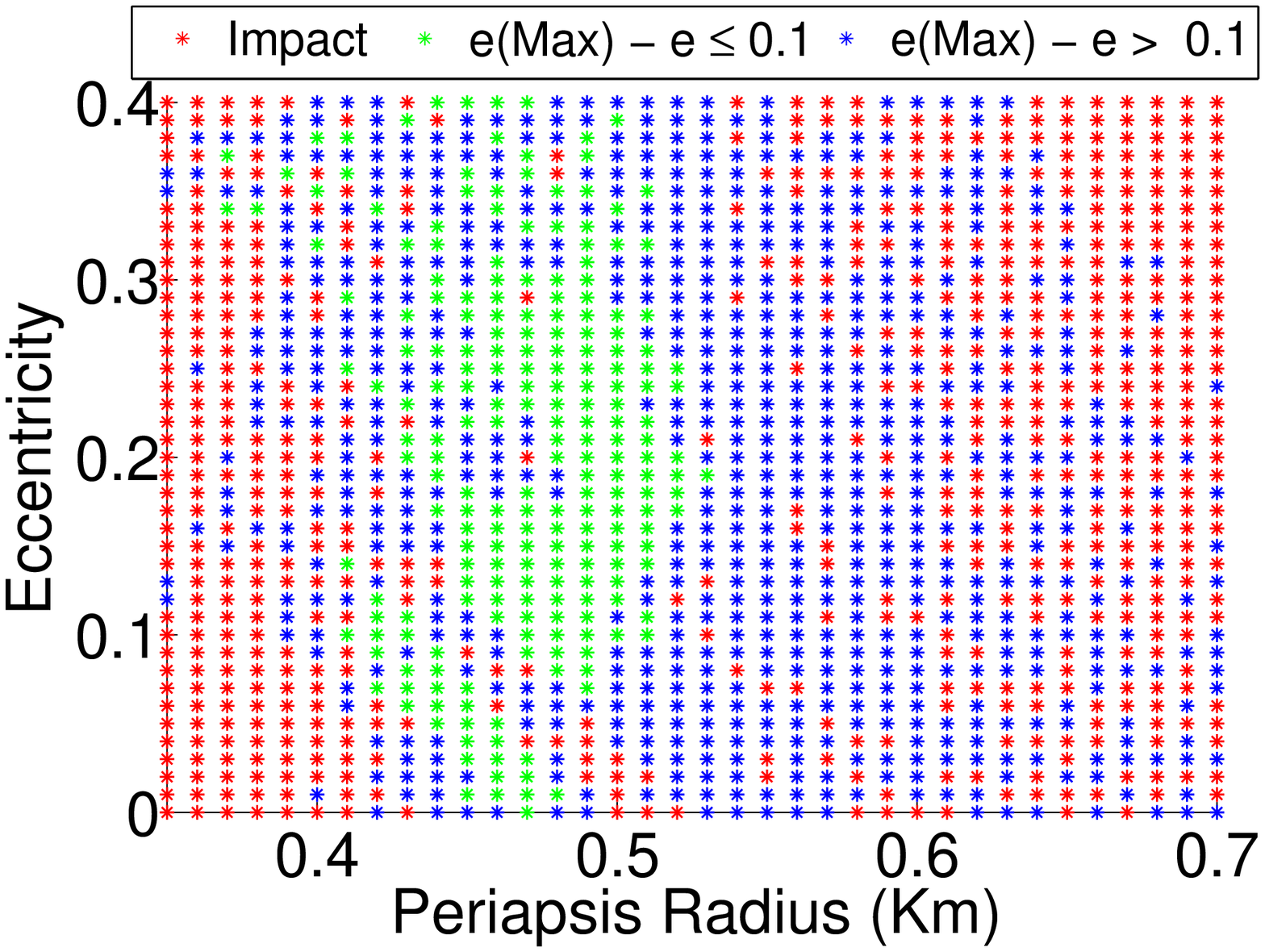}
 \includegraphics[width=0.24\linewidth]{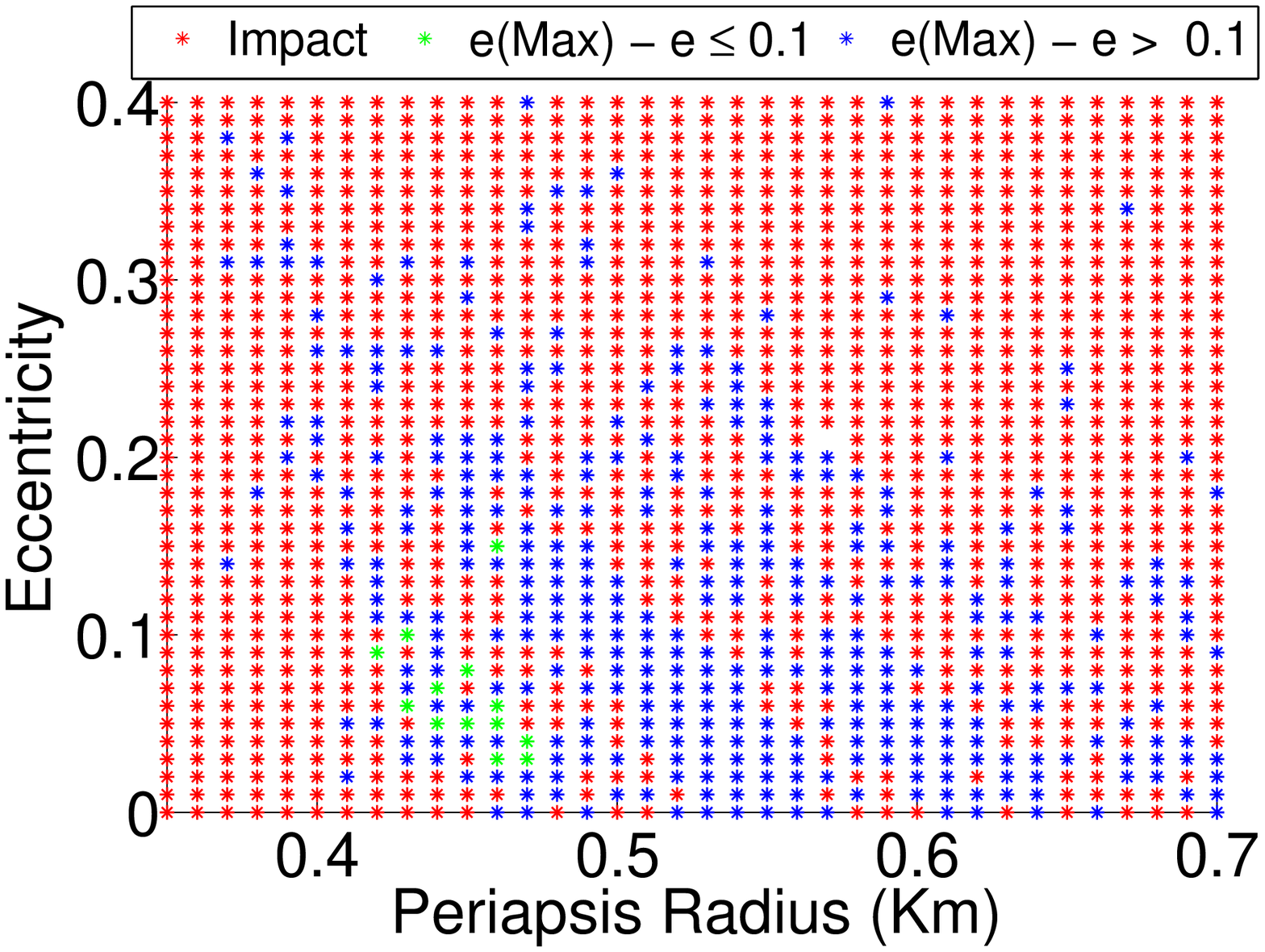}
 \includegraphics[width=0.24\linewidth]{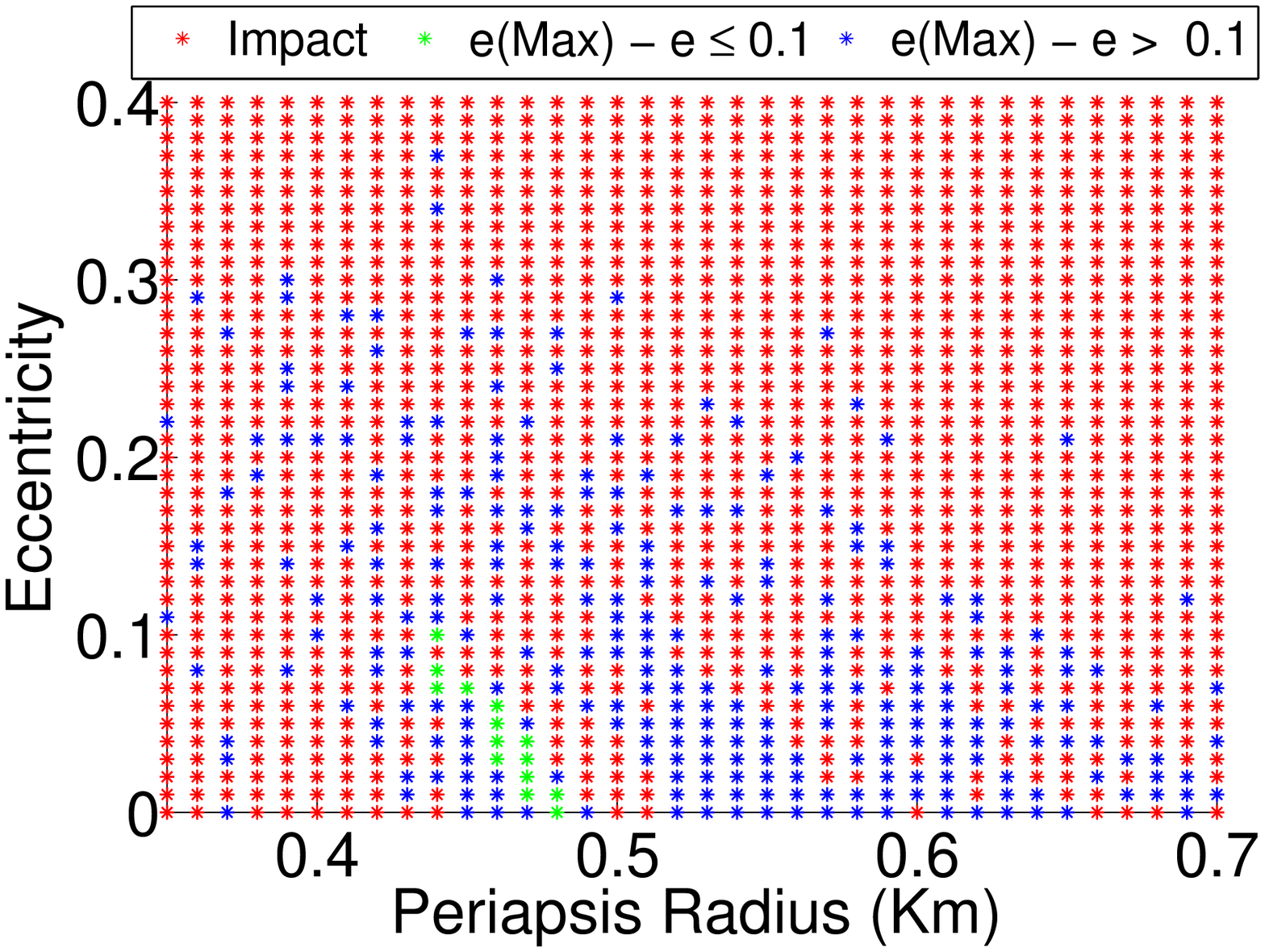}
 \includegraphics[width=0.24\linewidth]{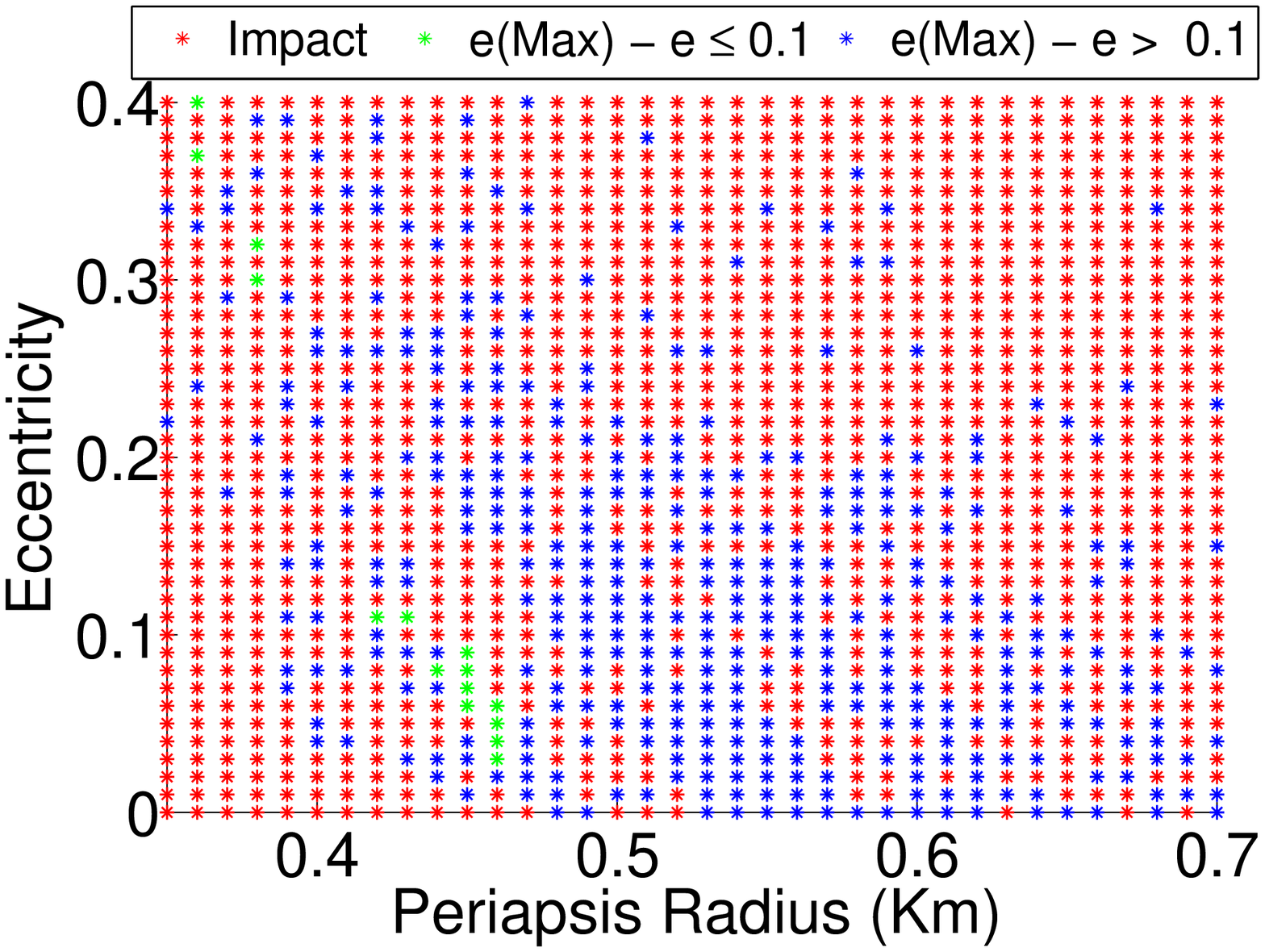}\\
 SRP ($g=4.405370 \times 10^{-11}km\cdot s^{-2}$,  $R=1.1264AU$) \\
{\small \textcolor{white}{.} \hspace{0.75cm} $0^{\circ}$ \hspace{3.75cm} $90^{\circ}$ \hspace{3.75cm}  $180^{\circ}$ \hspace{3.75cm} $270^{\circ}$}\\
1) \includegraphics[width=0.24\linewidth]{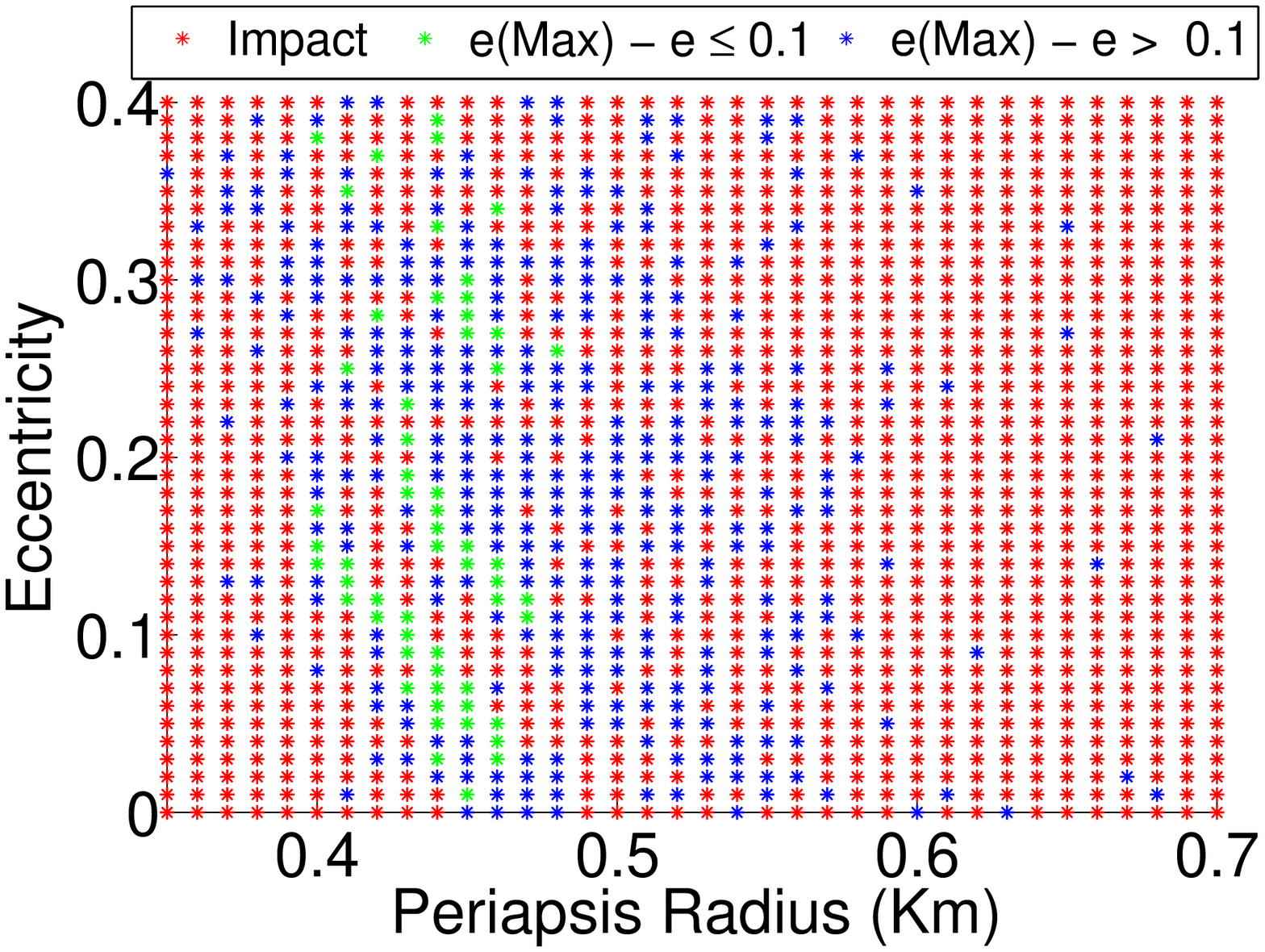}
 \includegraphics[width=0.24\linewidth]{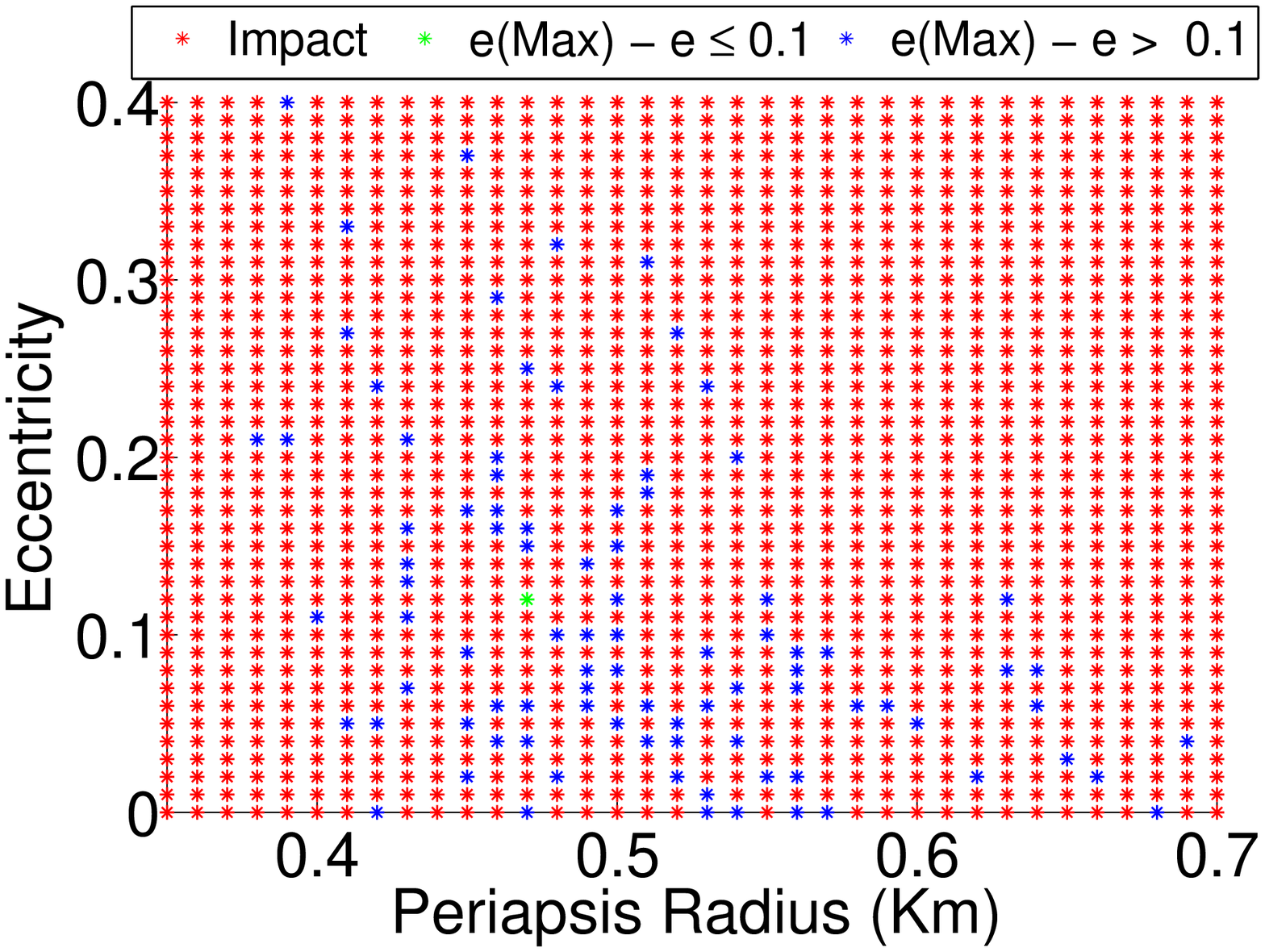}
 \includegraphics[width=0.24\linewidth]{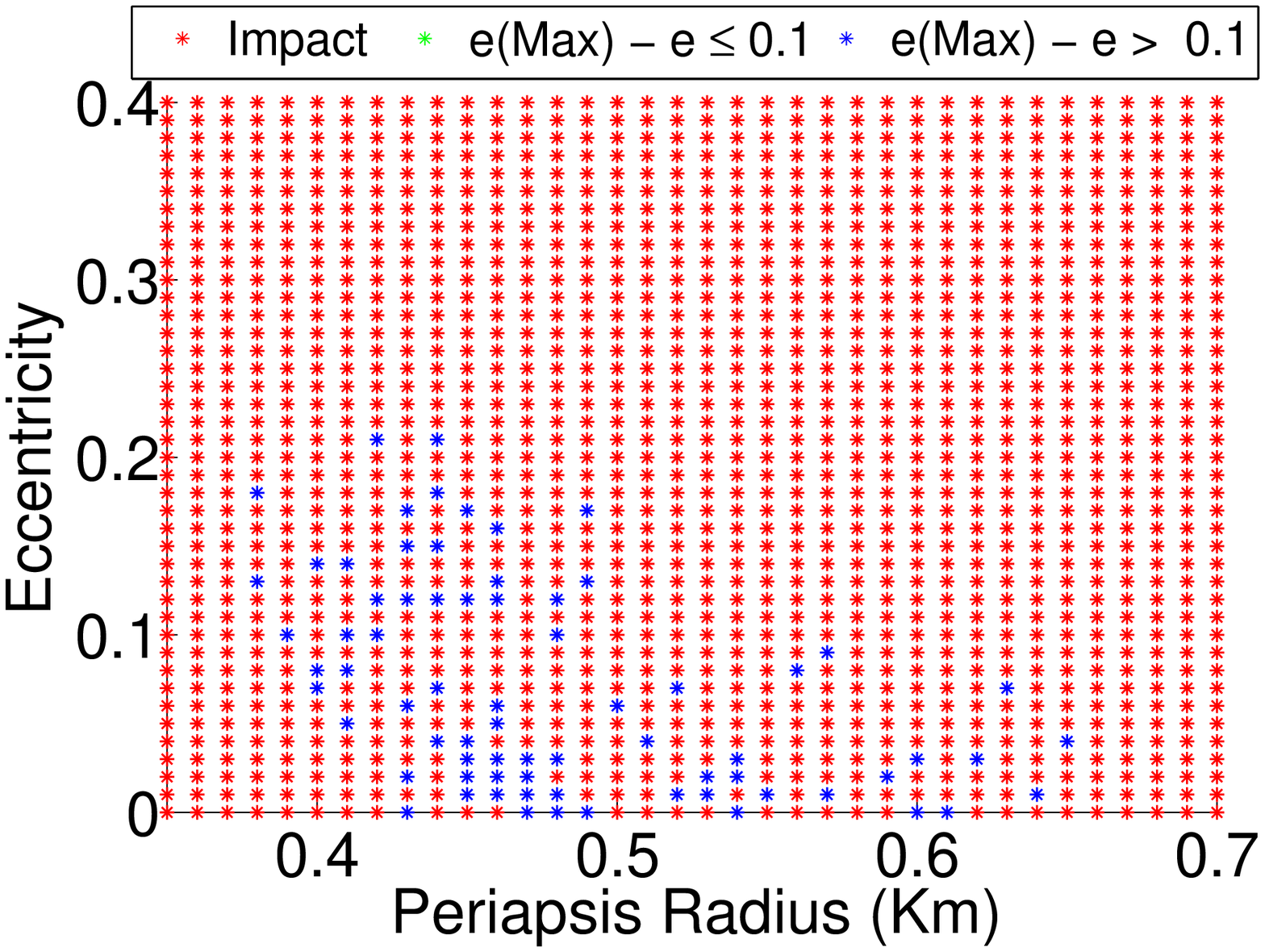}
 \includegraphics[width=0.24\linewidth]{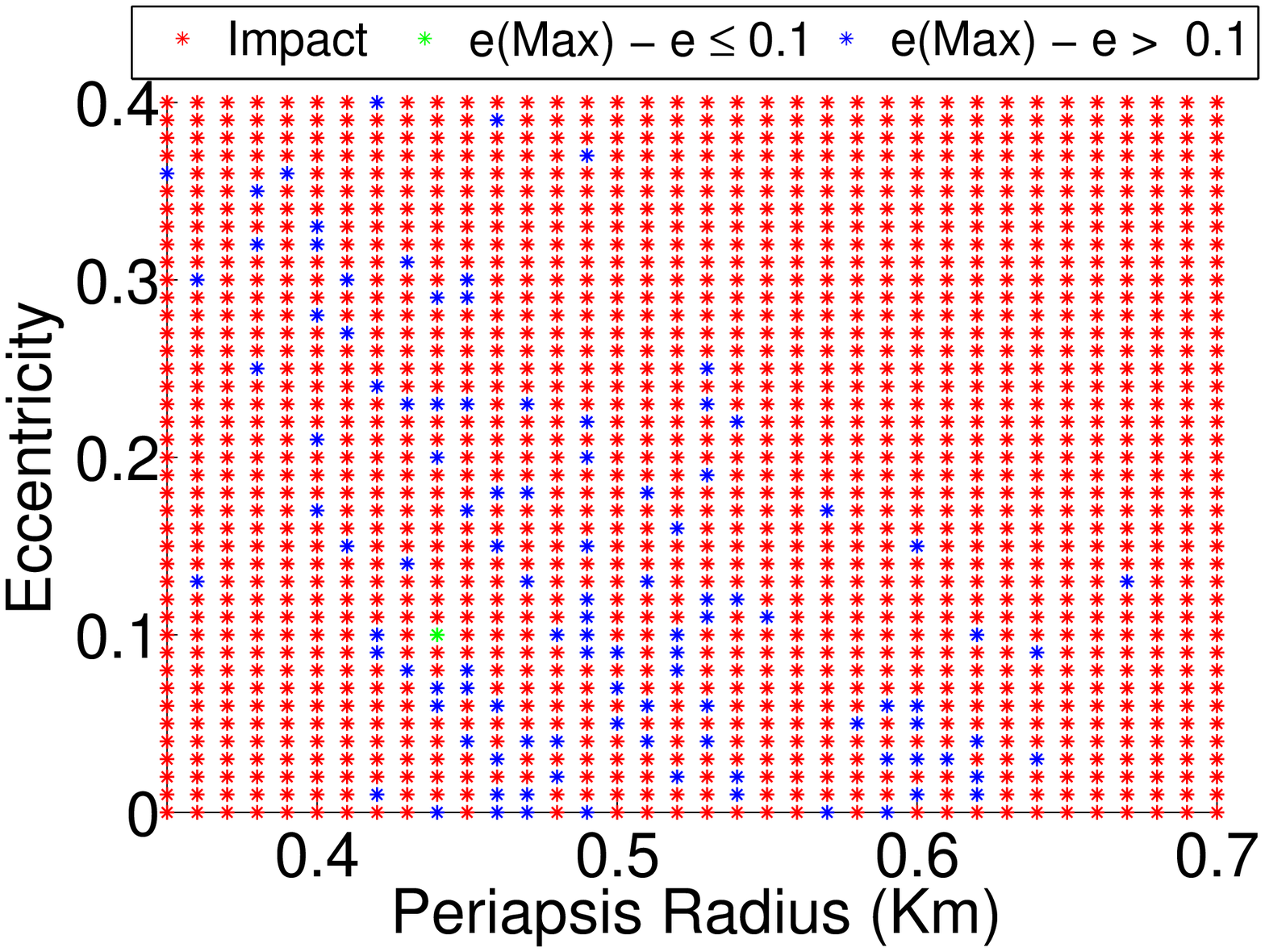}\\
 {\small \textcolor{white}{.} \hspace{0.75cm} $0^{\circ}$ \hspace{3.75cm} $90^{\circ}$ \hspace{3.75cm}  $180^{\circ}$ \hspace{3.75cm} $270^{\circ}$}\\
2) \includegraphics[width=0.24\linewidth]{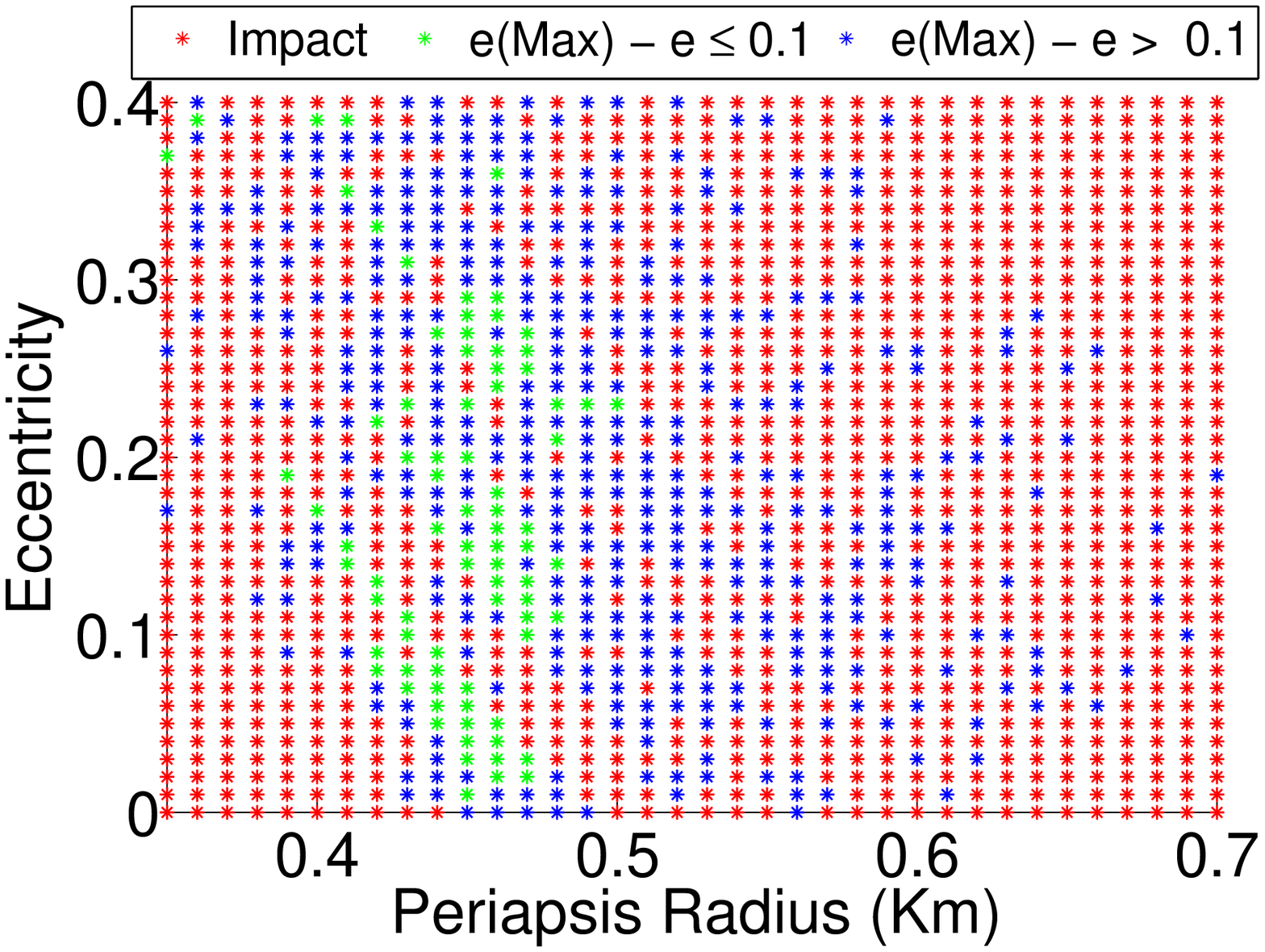}
 \includegraphics[width=0.24\linewidth]{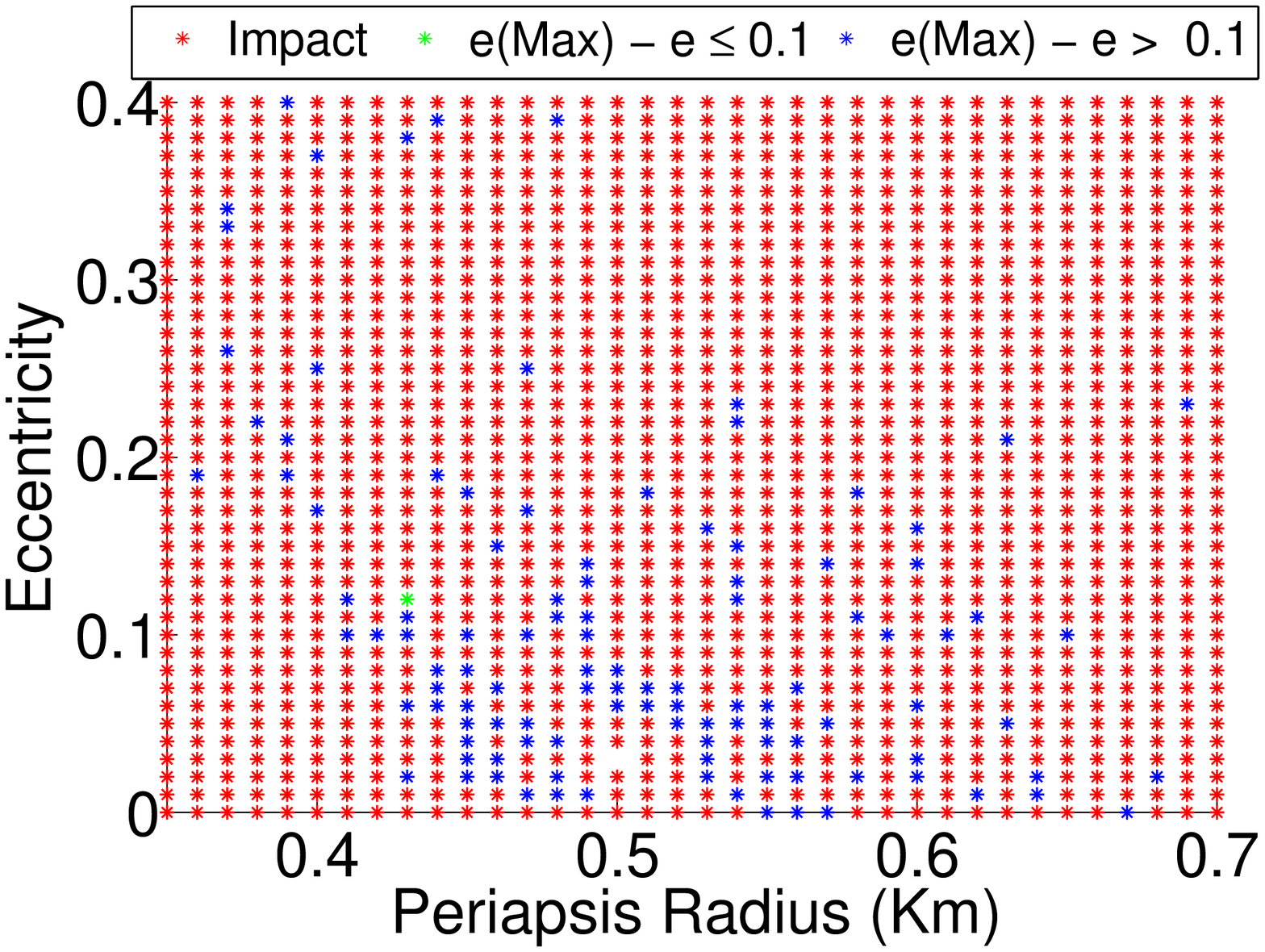}
 \includegraphics[width=0.24\linewidth]{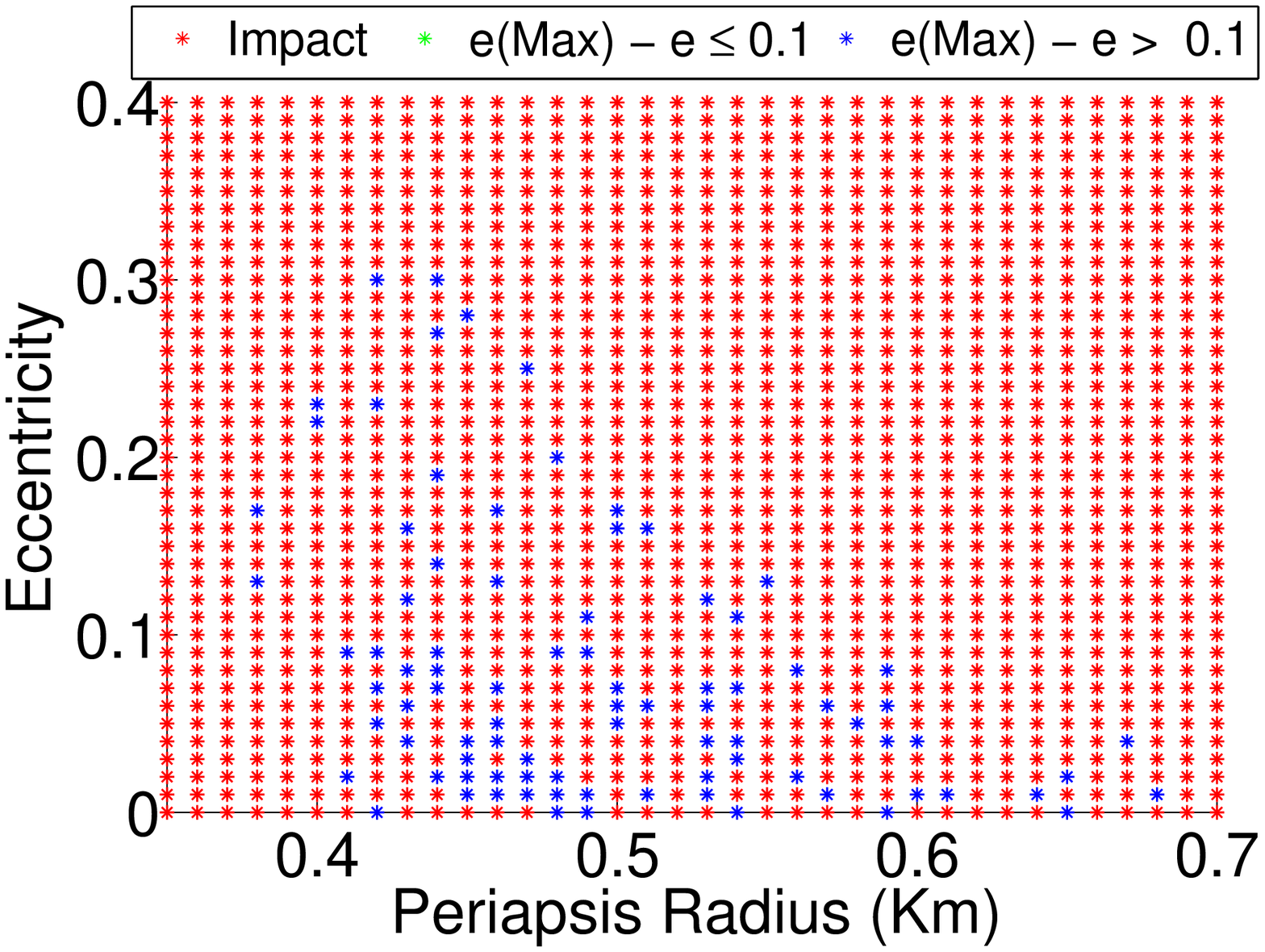}
 \includegraphics[width=0.24\linewidth]{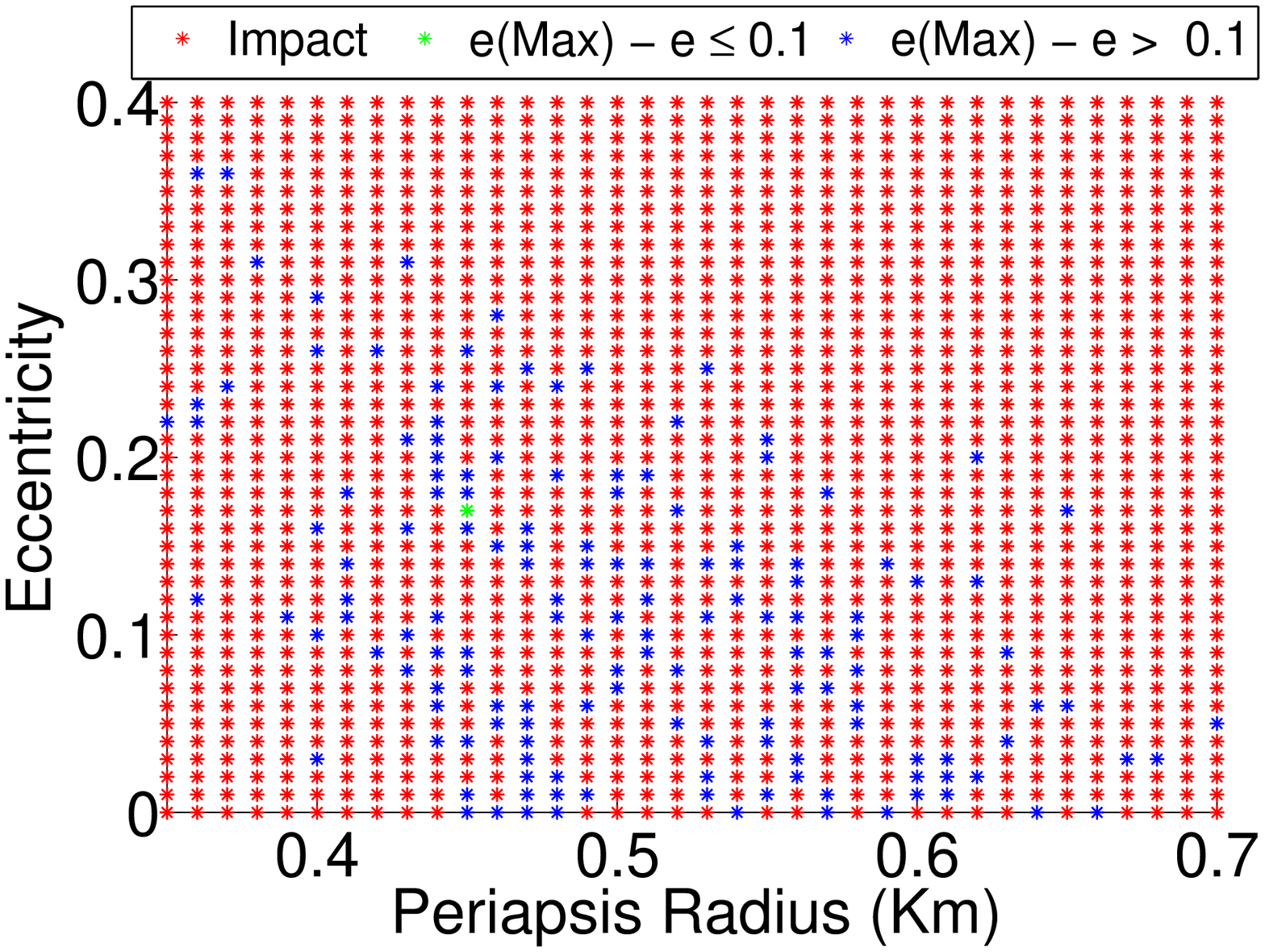}\\
 SRP ($g=6.941835 \times 10^{-11}km\cdot s^{-2}$, $R=0.8969AU$) \\
{\small \textcolor{white}{.} \hspace{0.75cm} $0^{\circ}$ \hspace{3.75cm} $90^{\circ}$ \hspace{3.75cm}  $180^{\circ}$ \hspace{3.75cm} $270^{\circ}$}\\
1) \includegraphics[width=0.24\linewidth]{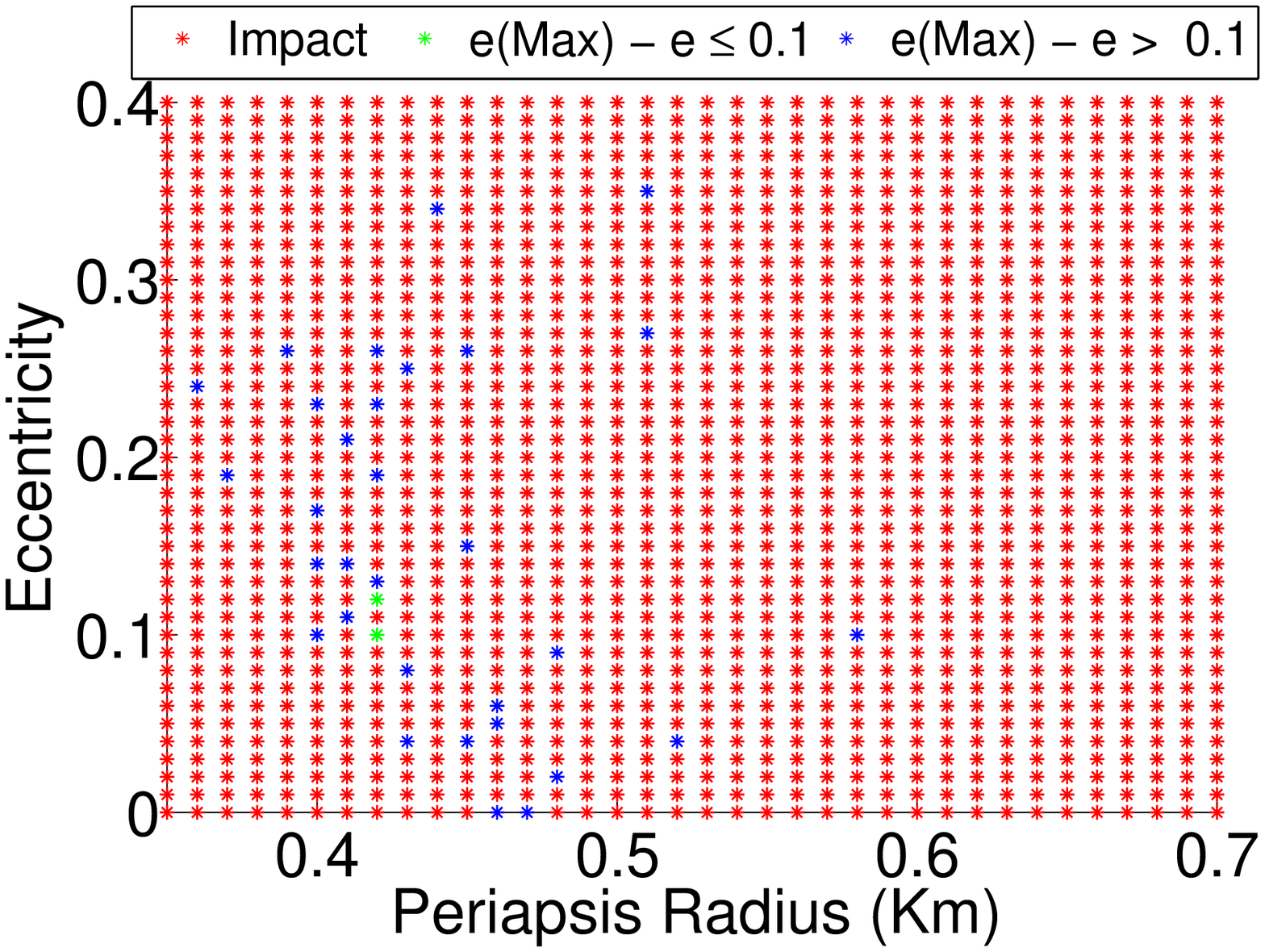}
 \includegraphics[width=0.24\linewidth]{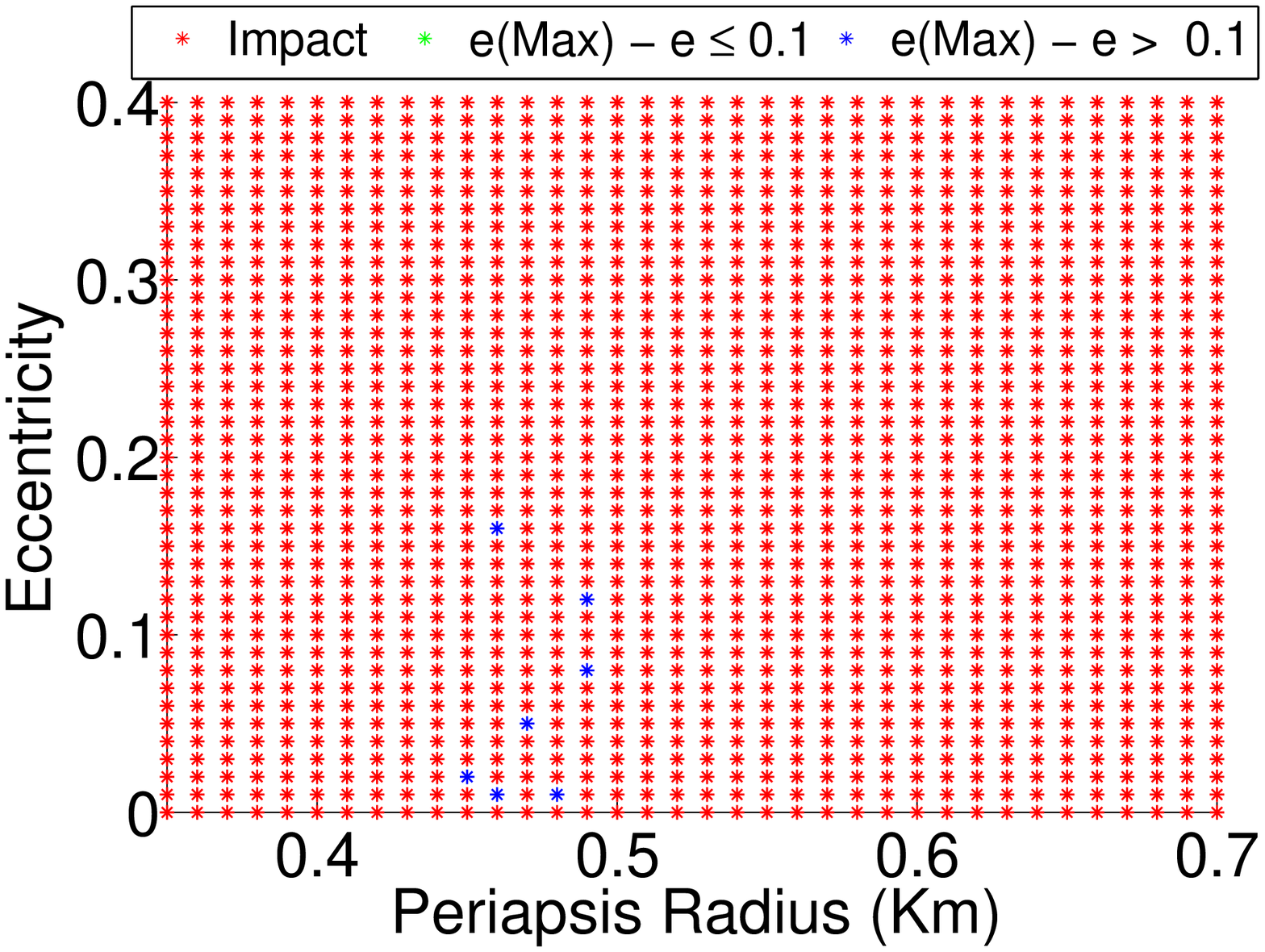}
 \includegraphics[width=0.24\linewidth]{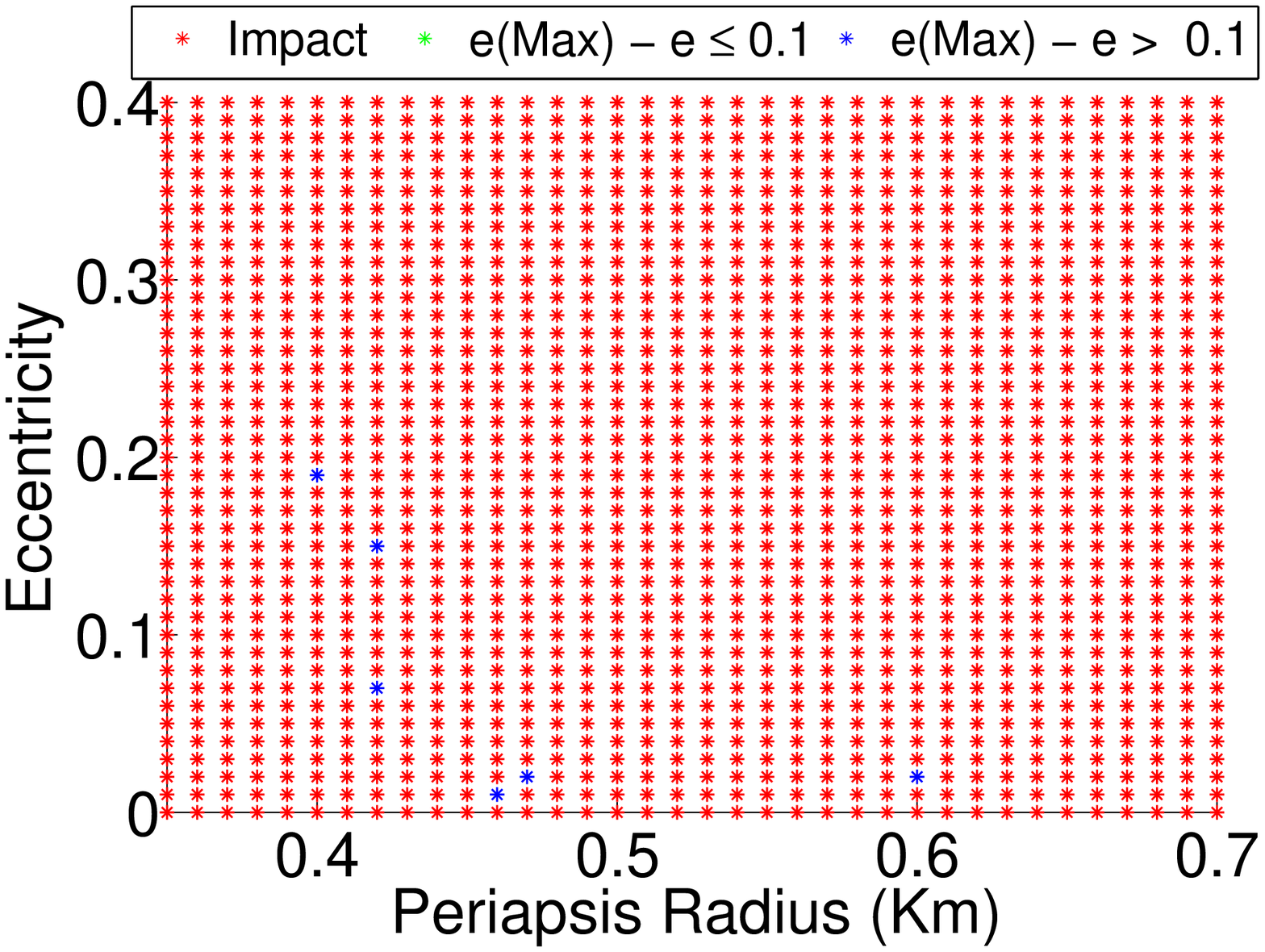}
 \includegraphics[width=0.24\linewidth]{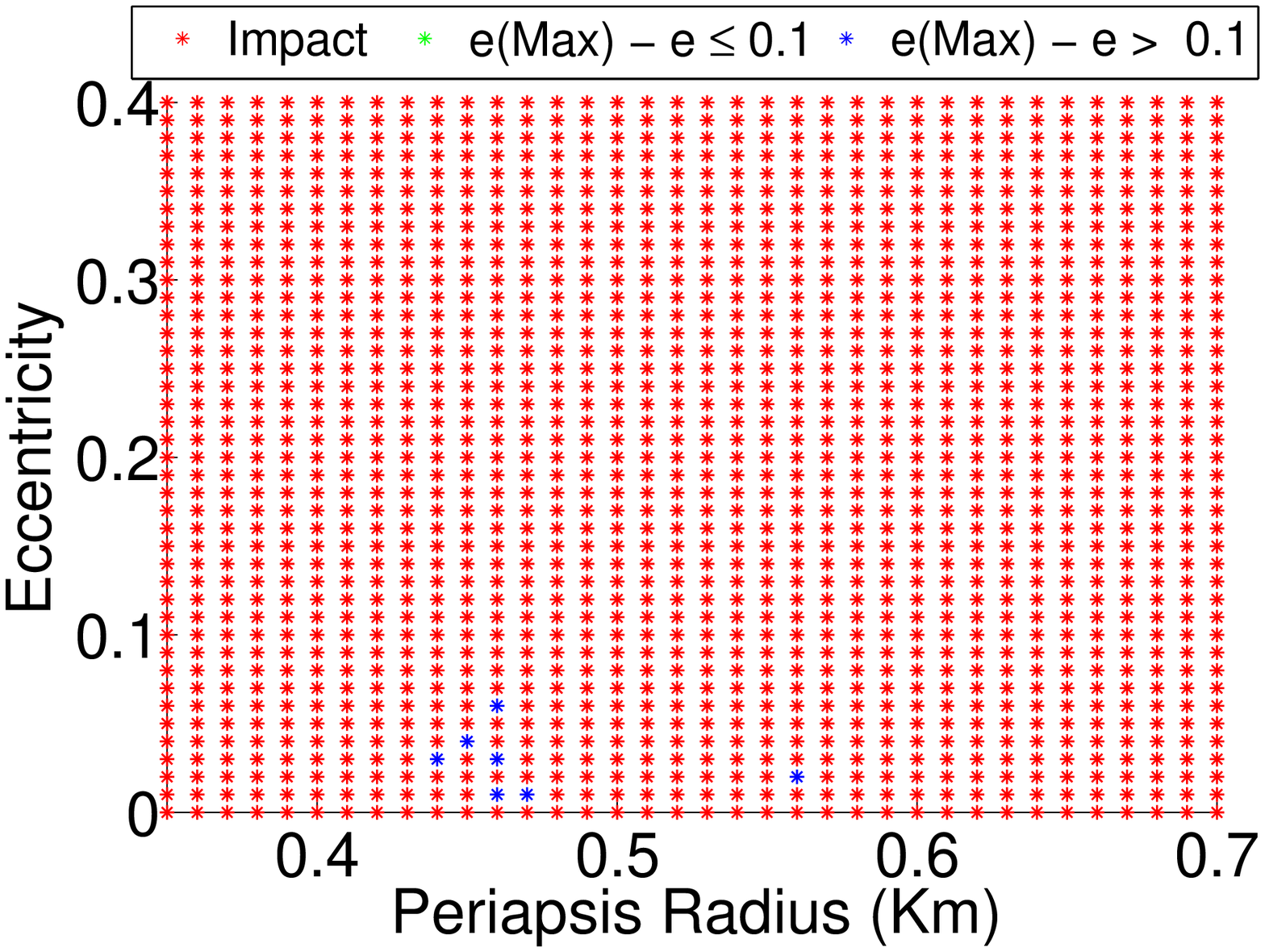}\\ 
{\small \textcolor{white}{.} \hspace{0.75cm} $0^{\circ}$ \hspace{3.75cm} $90^{\circ}$ \hspace{3.75cm}  $180^{\circ}$ \hspace{3.75cm} $270^{\circ}$}\\
 2) \includegraphics[width=0.24\linewidth]{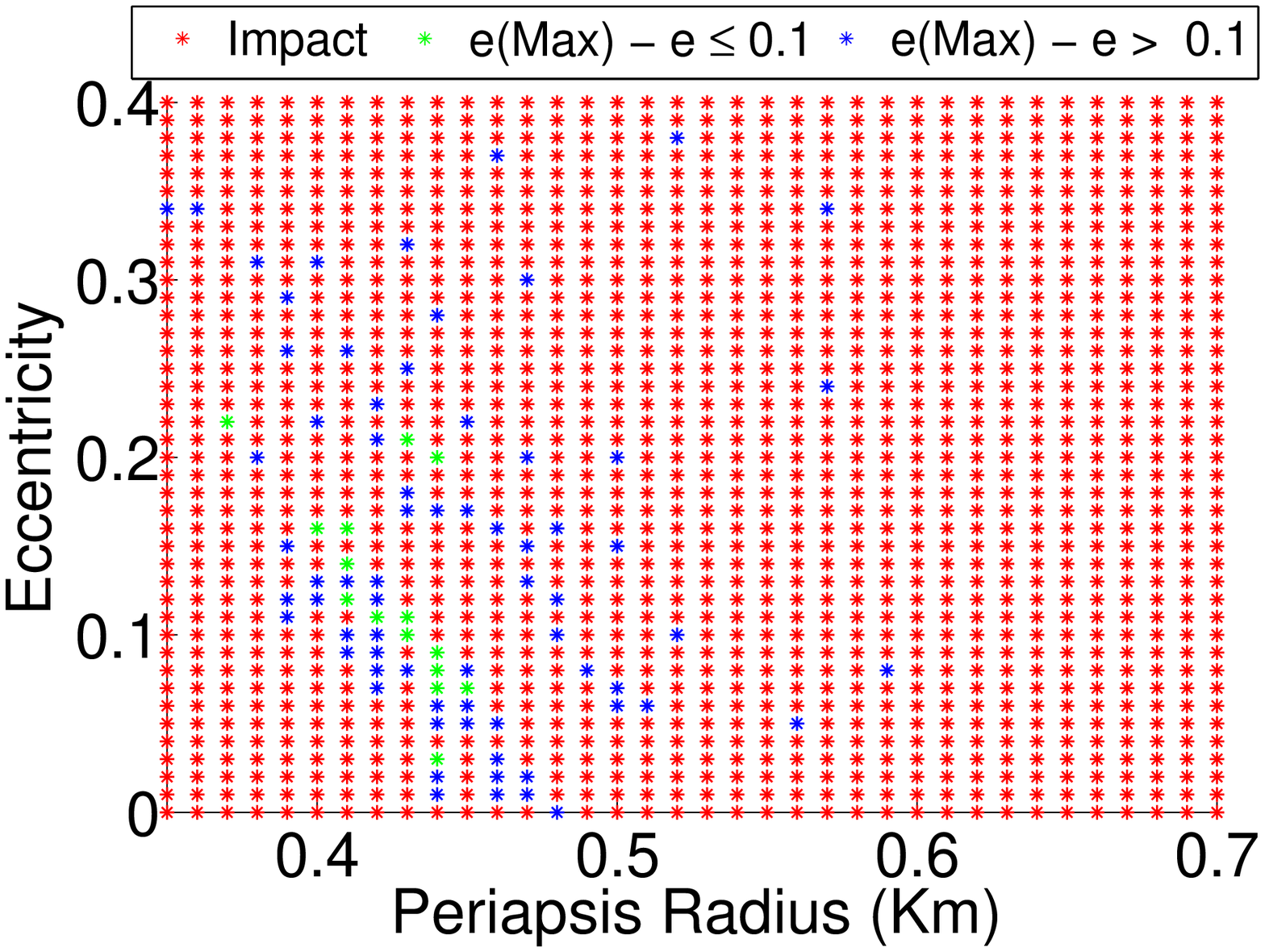}
 \includegraphics[width=0.24\linewidth]{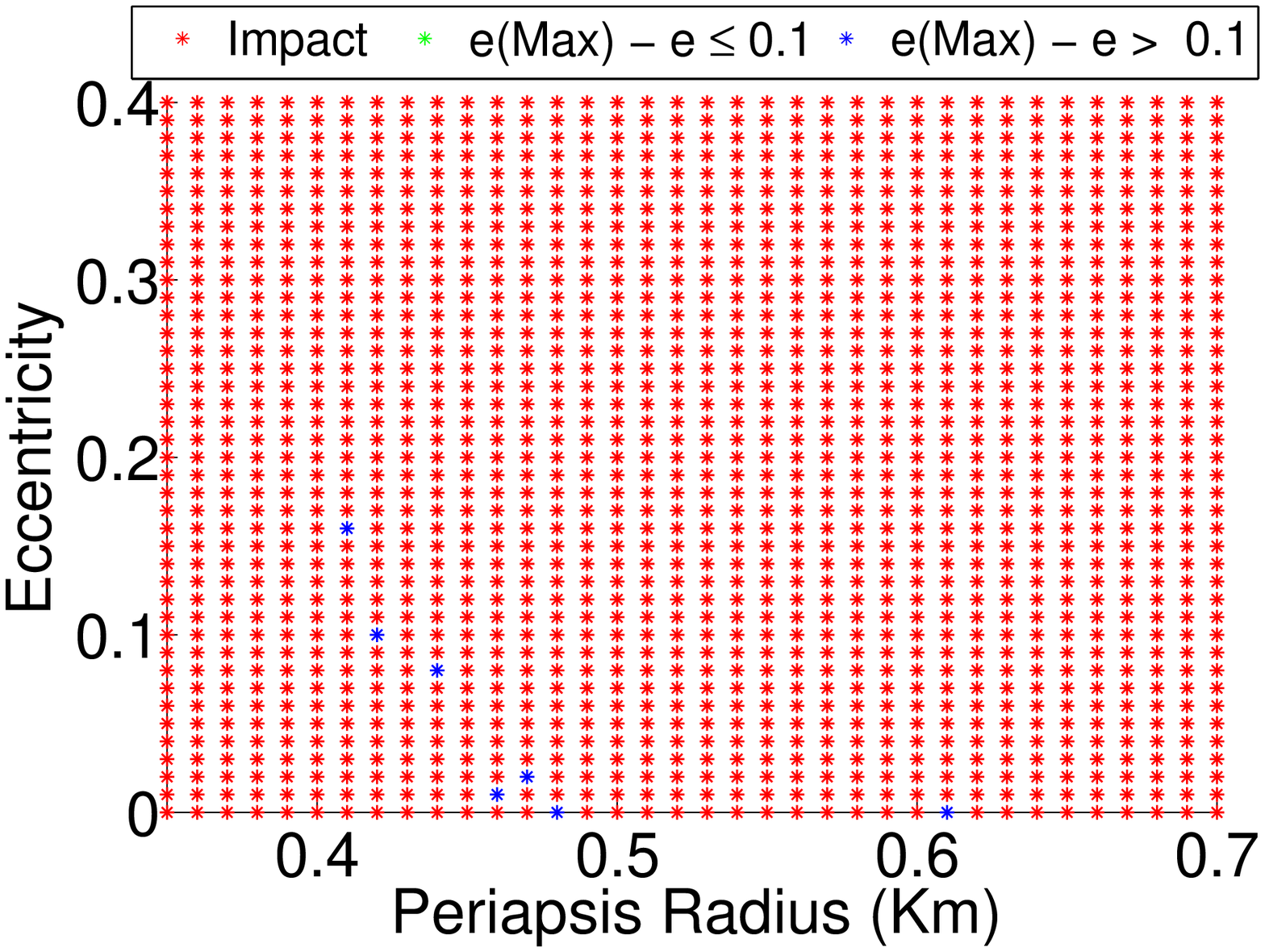}
 \includegraphics[width=0.24\linewidth]{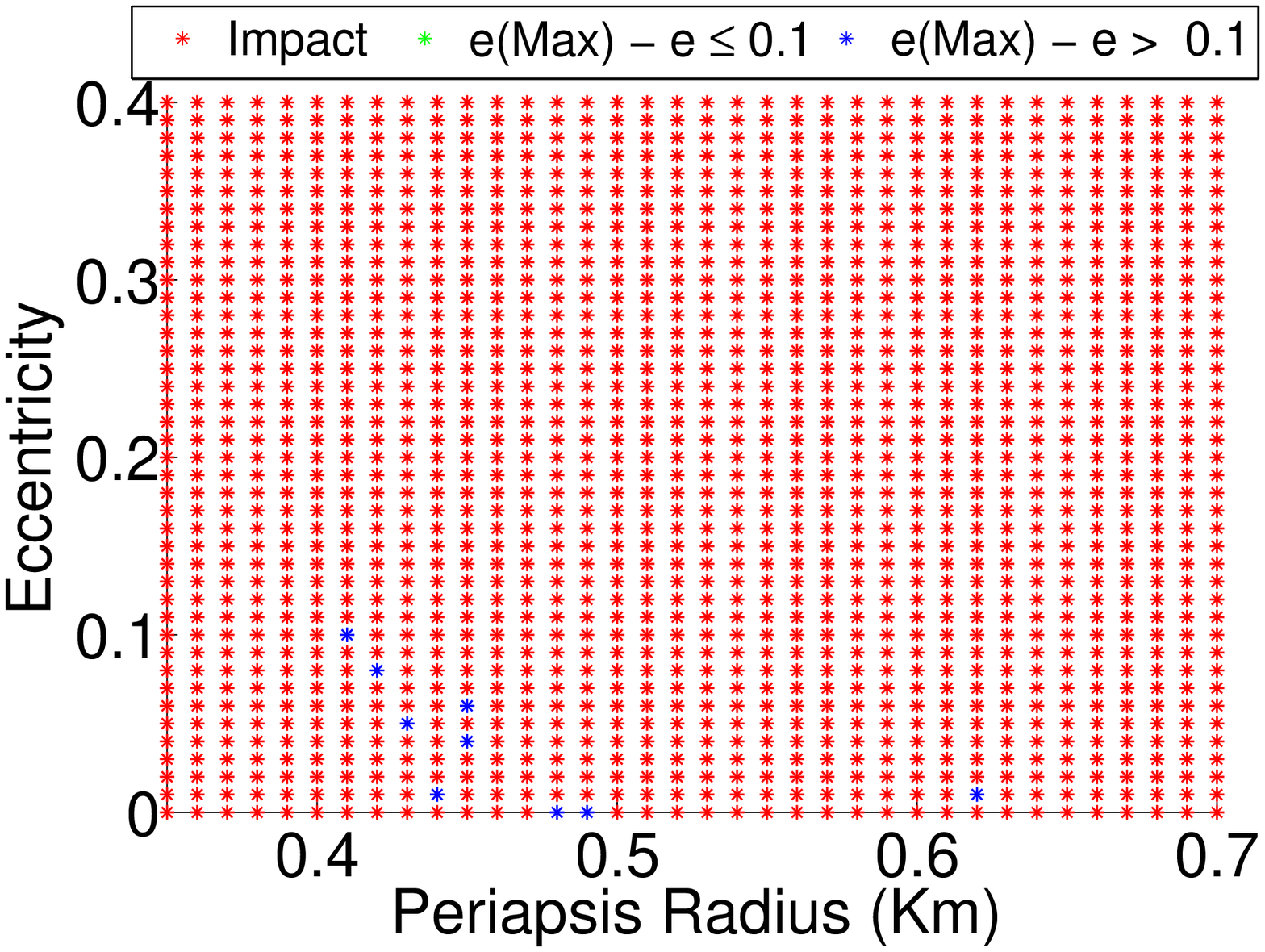}
 \includegraphics[width=0.24\linewidth]{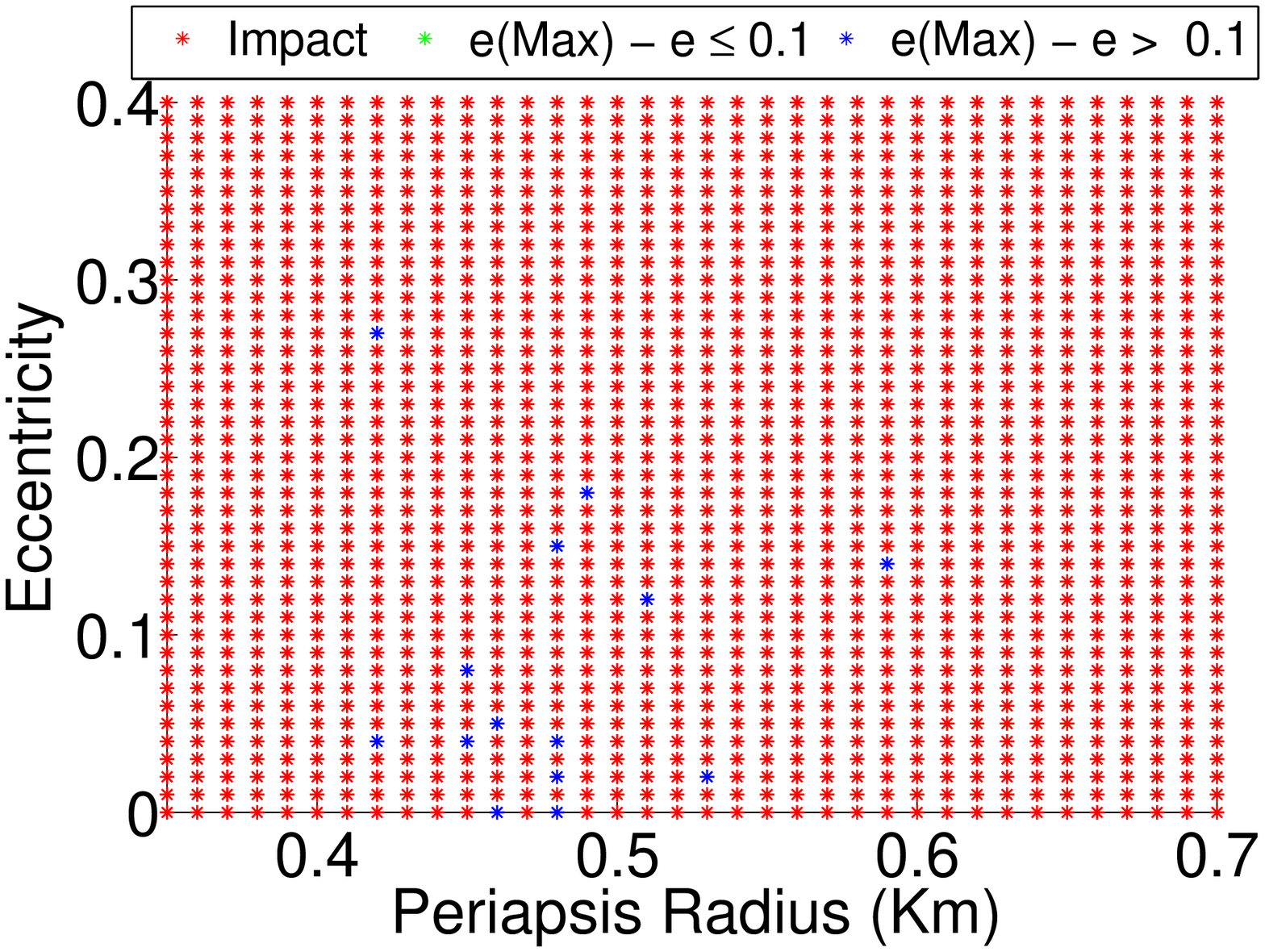}\\
      
 \caption{Stability maps of the equatorial orbits relative to (101955) Bennu with the same initial conditions of Fig.7. The solar radiation pressure is accounted for with the Sun initial longitude $\psi_0 = -180^o$. The (101955) Bennu distance from the Sun $R$ is noted on the top of the related grafics and the eclipse is not taken into account in 1) and accounted for in 2).  }
  \label{Fig8} 
\end{figure*}

\begin{figure*}
   \centering
 SRP ($g=3.037665 \times 10^{-11}km\cdot s^{-2}$,  $R=1.3559AU$)\\
{\small \textcolor{white}{.} \hspace{0.75cm} $0^{\circ}$ \hspace{3.75cm} $90^{\circ}$ \hspace{3.75cm}  $180^{\circ}$ \hspace{3.75cm} $270^{\circ}$}\\
1) \includegraphics[width=0.24\linewidth]{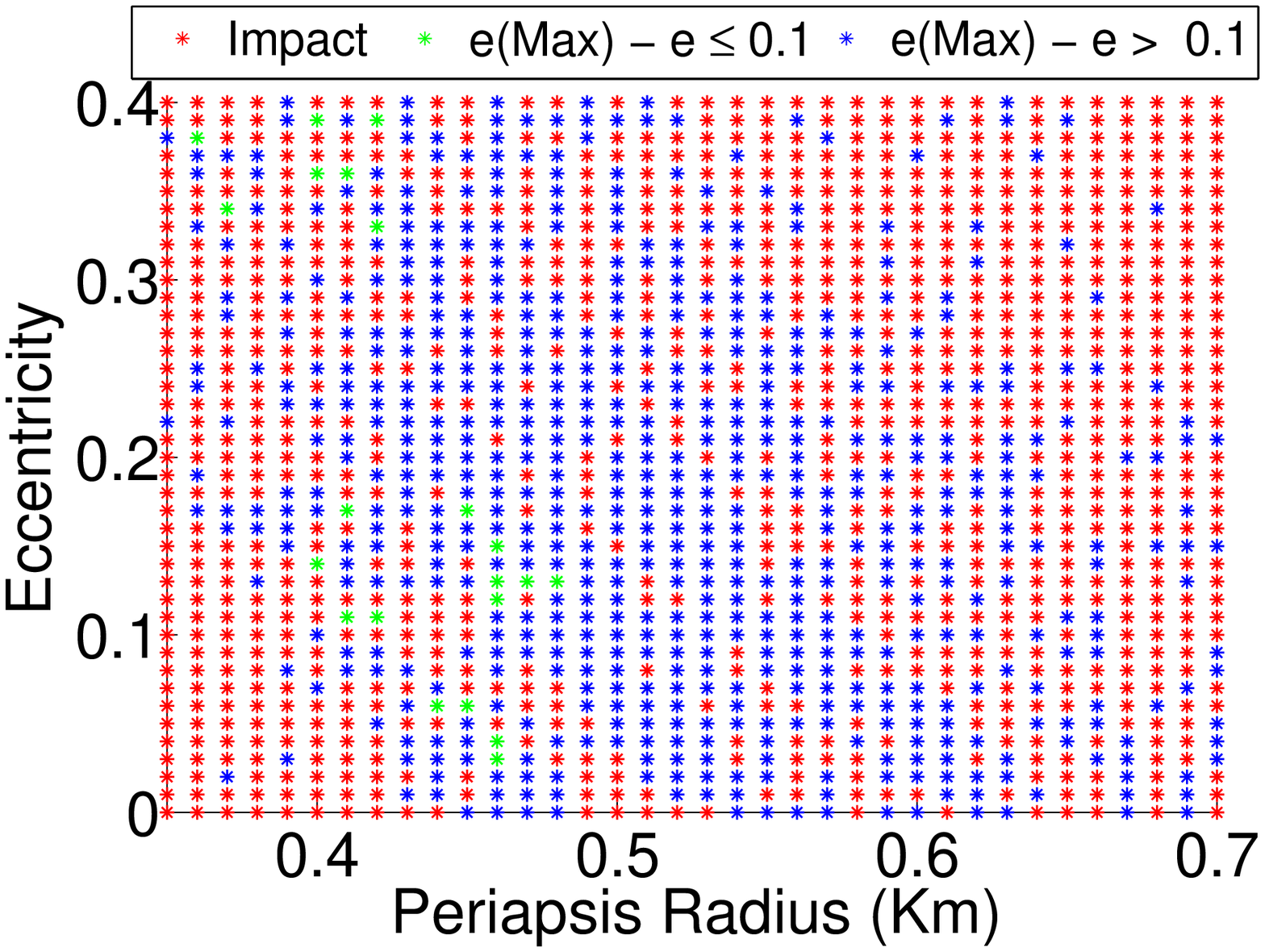}
 \includegraphics[width=0.24\linewidth]{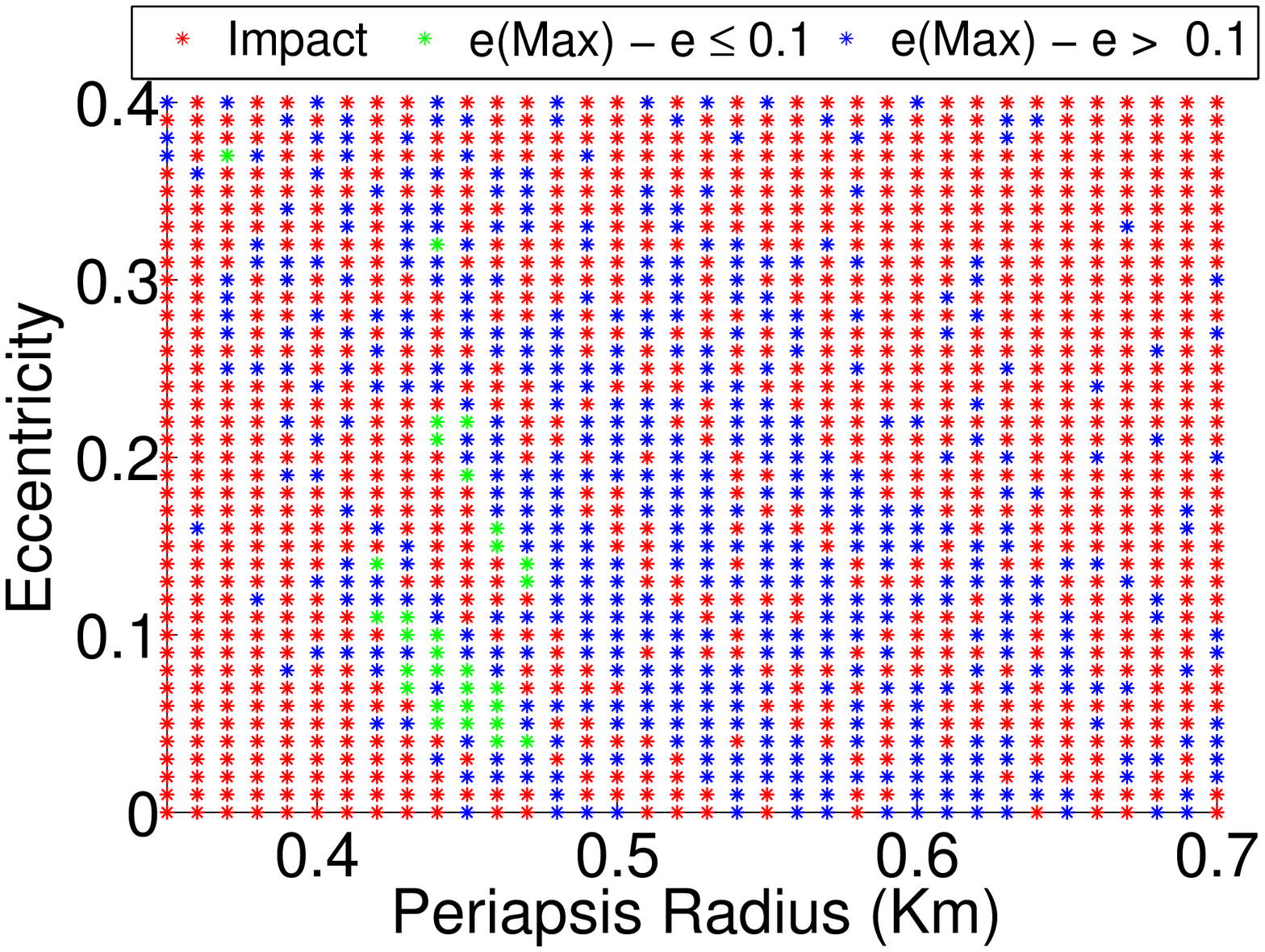}
 \includegraphics[width=0.24\linewidth]{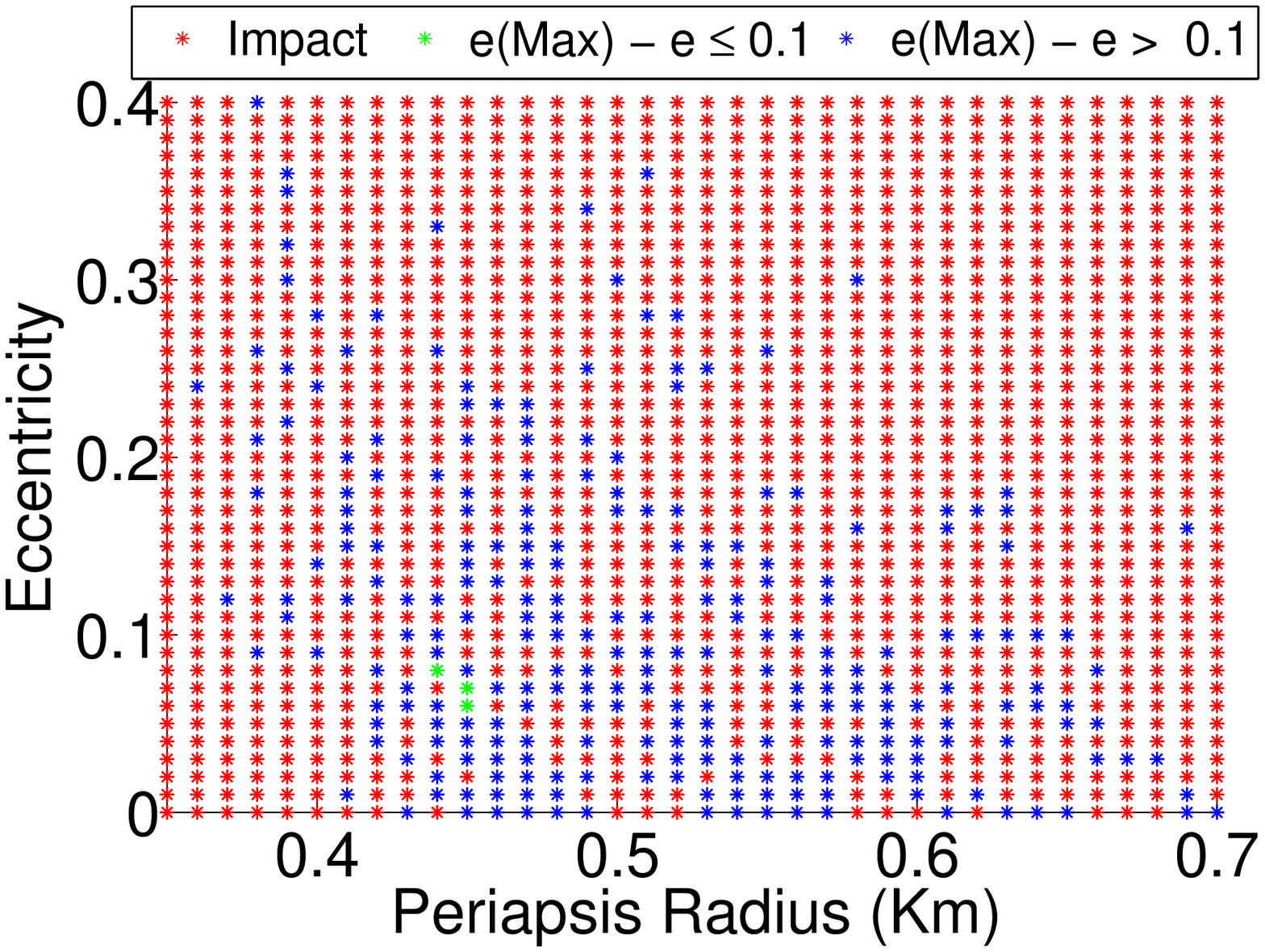}
 \includegraphics[width=0.24\linewidth]{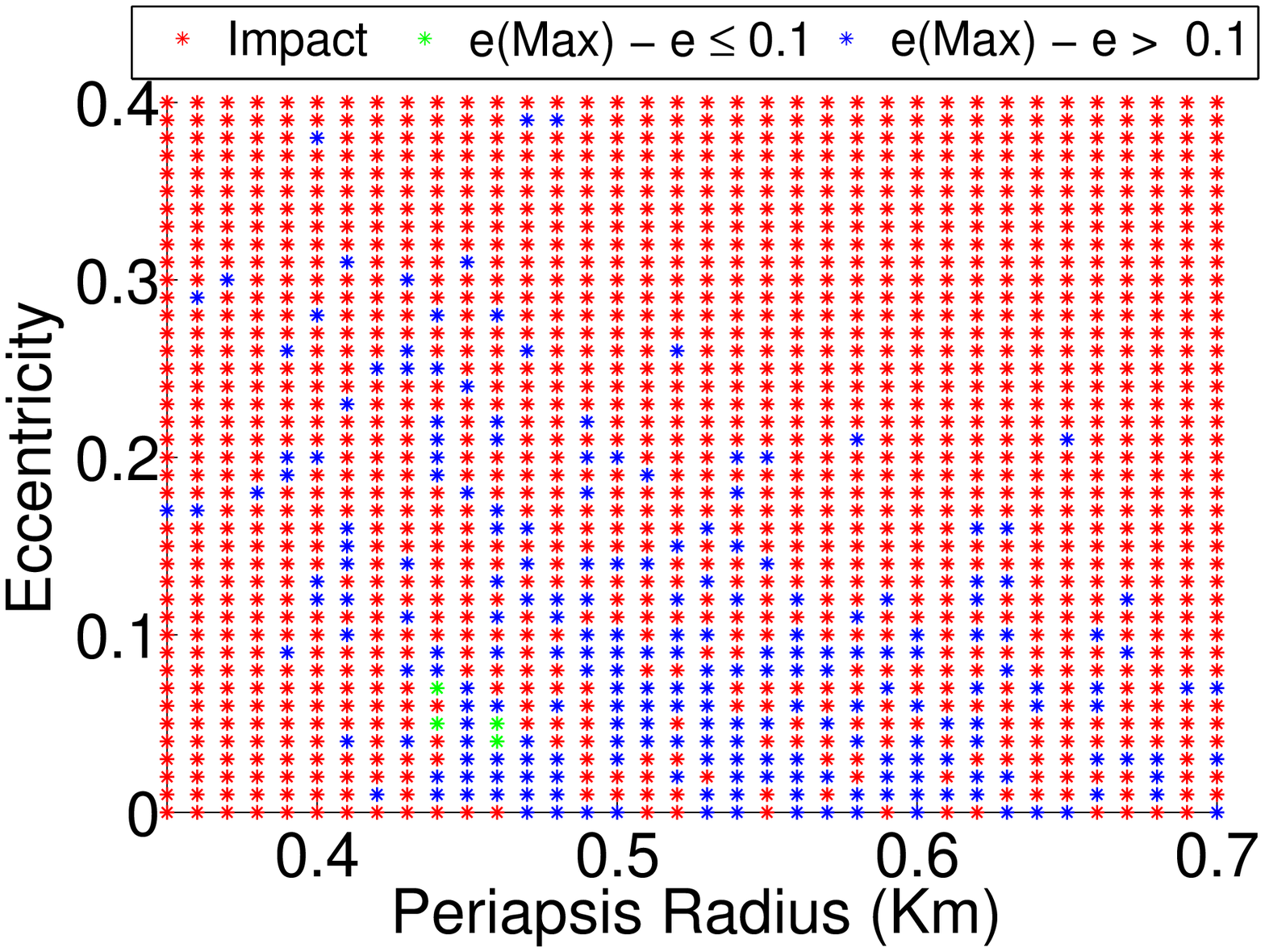}\\
 {\small \textcolor{white}{.} \hspace{0.75cm} $0^{\circ}$ \hspace{3.75cm} $90^{\circ}$ \hspace{3.75cm}  $180^{\circ}$ \hspace{3.75cm} $270^{\circ}$}\\
2)  \includegraphics[width=0.24\linewidth]{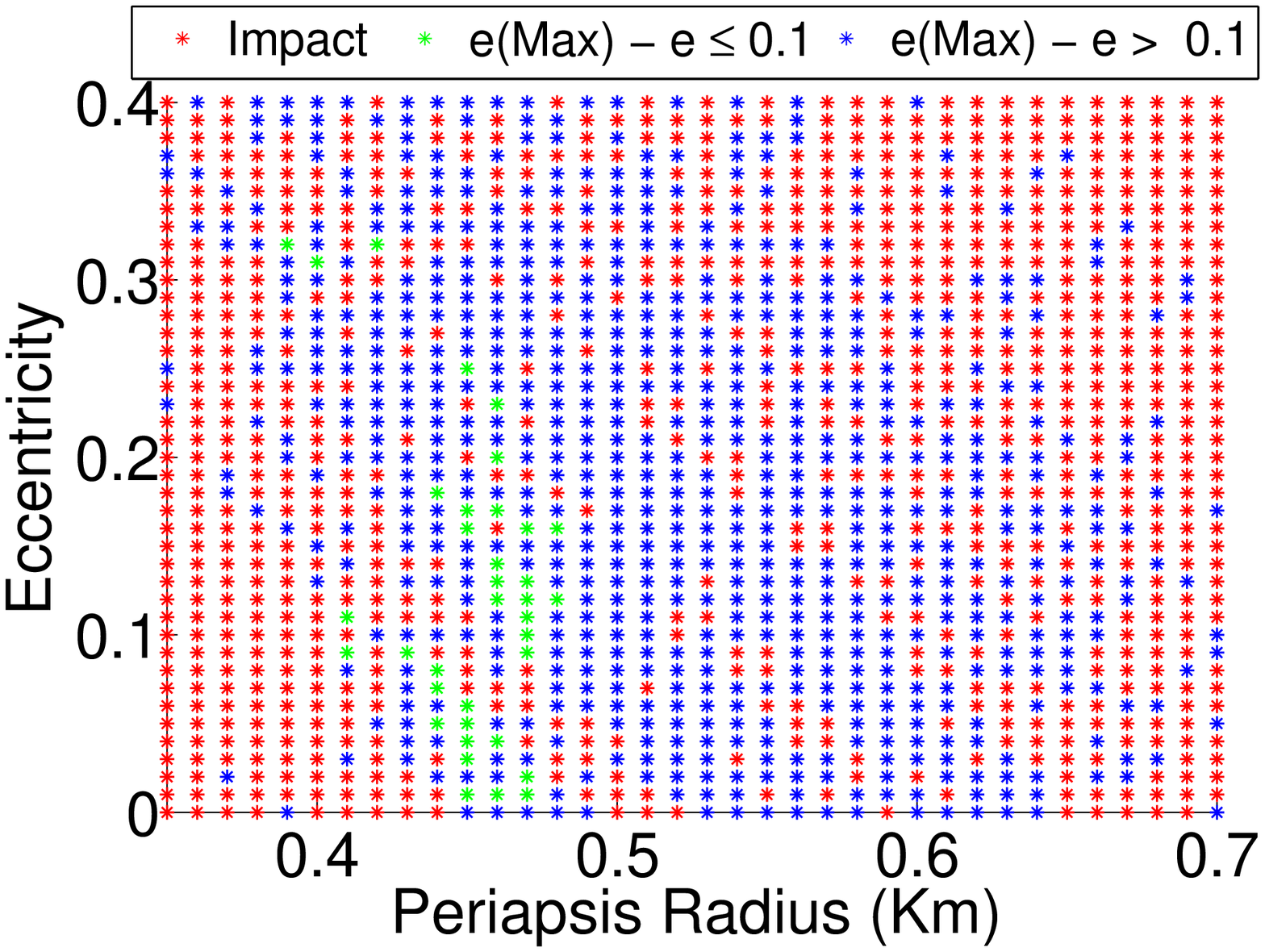}
 \includegraphics[width=0.24\linewidth]{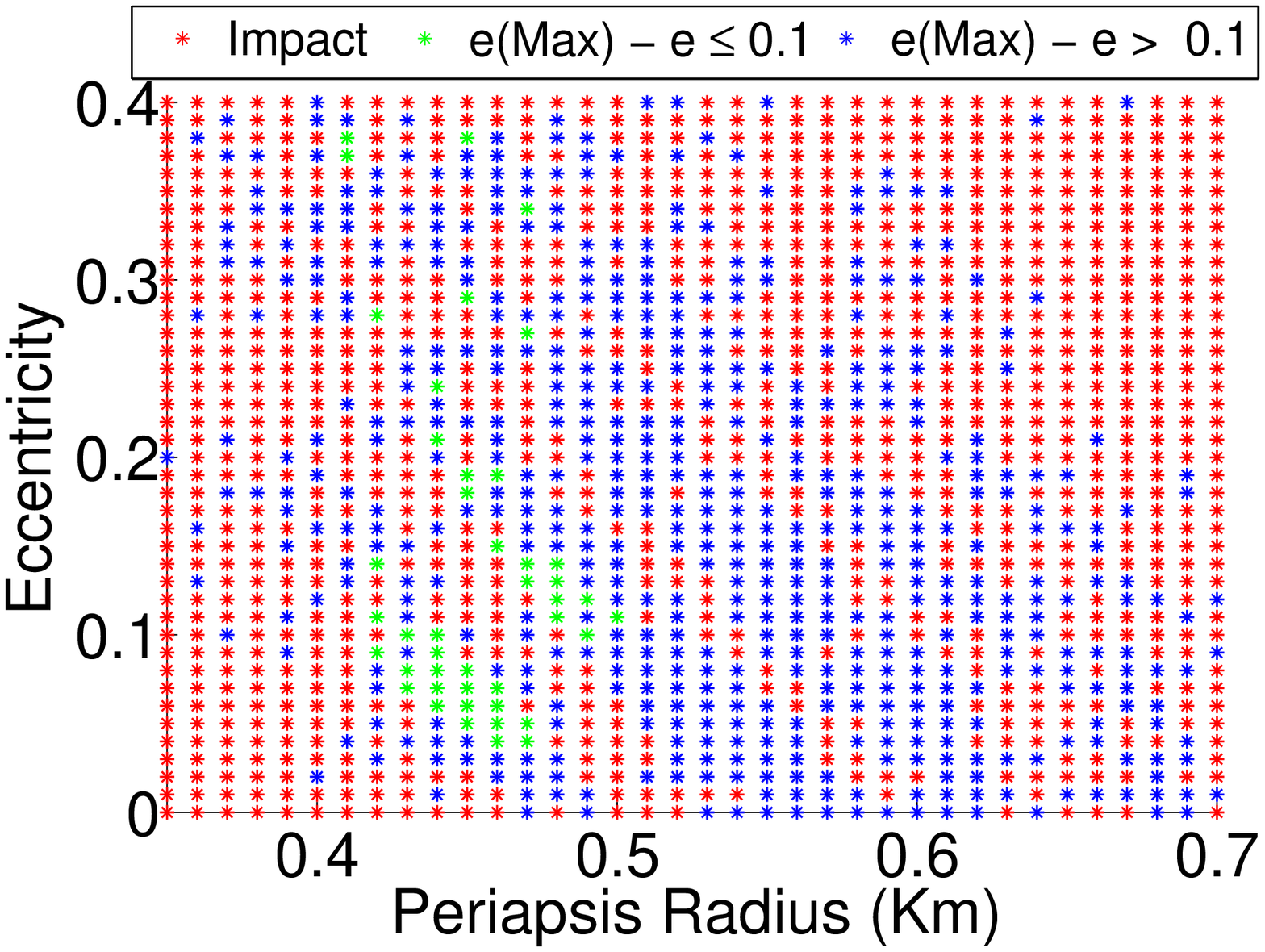}
 \includegraphics[width=0.24\linewidth]{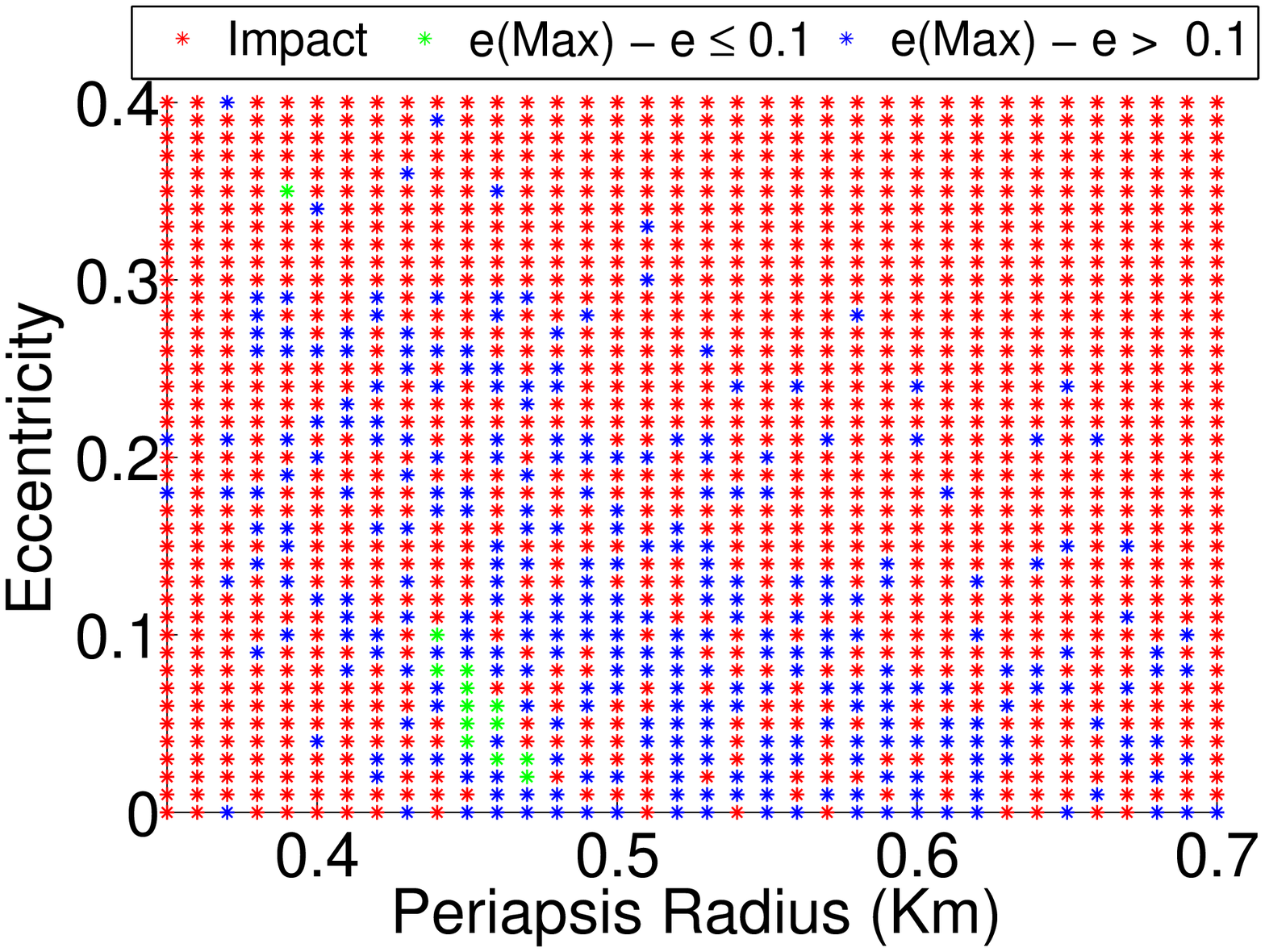}
 \includegraphics[width=0.24\linewidth]{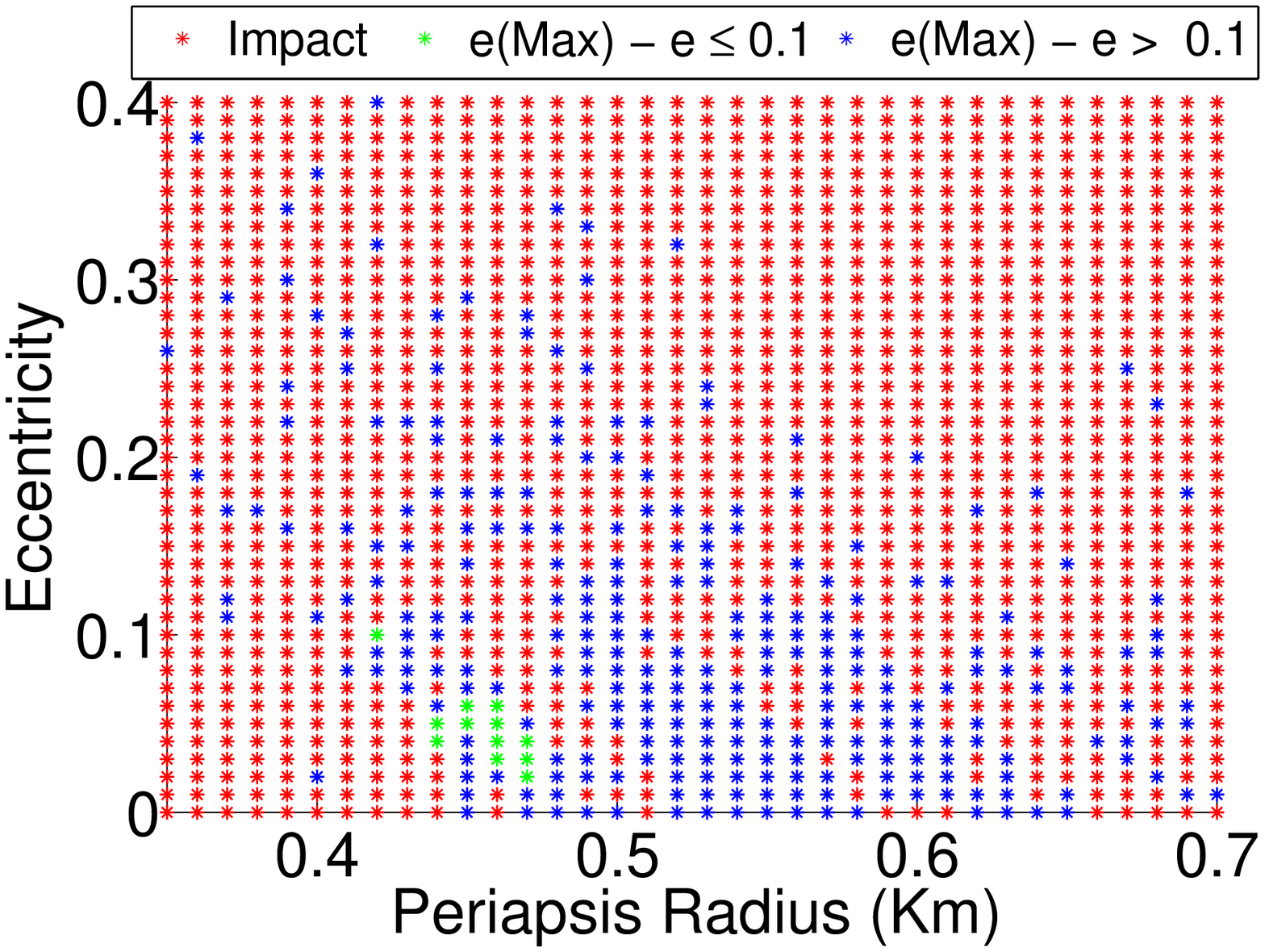}\\
 SRP ($g=4.405370 \times 10^{-11}km\cdot s^{-2}$,  $R=1.1264AU$) \\
{\small \textcolor{white}{.} \hspace{0.75cm} $0^{\circ}$ \hspace{3.75cm} $90^{\circ}$ \hspace{3.75cm}  $180^{\circ}$ \hspace{3.75cm} $270^{\circ}$}\\
 1)\includegraphics[width=0.24\linewidth]{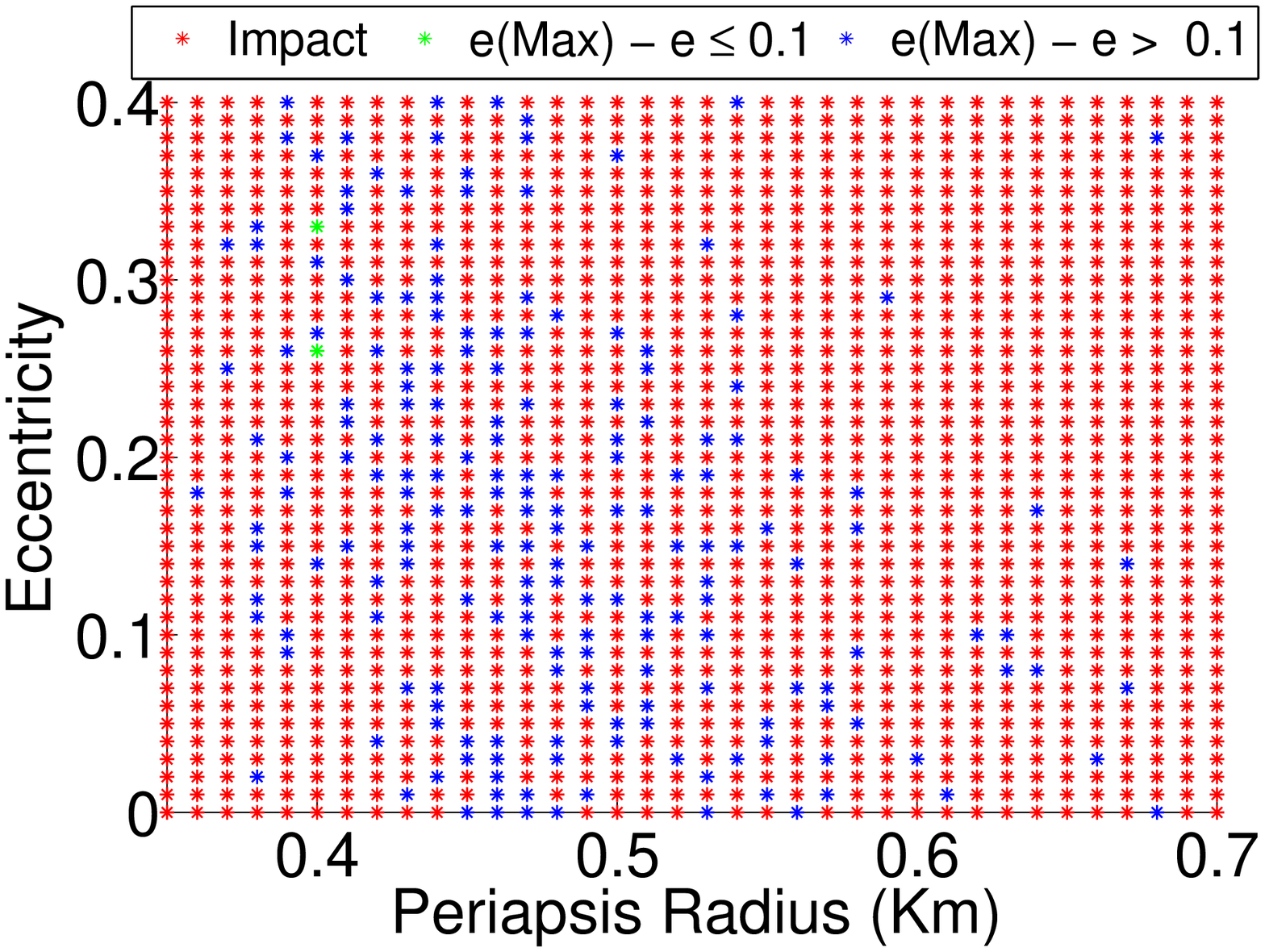}
 \includegraphics[width=0.24\linewidth]{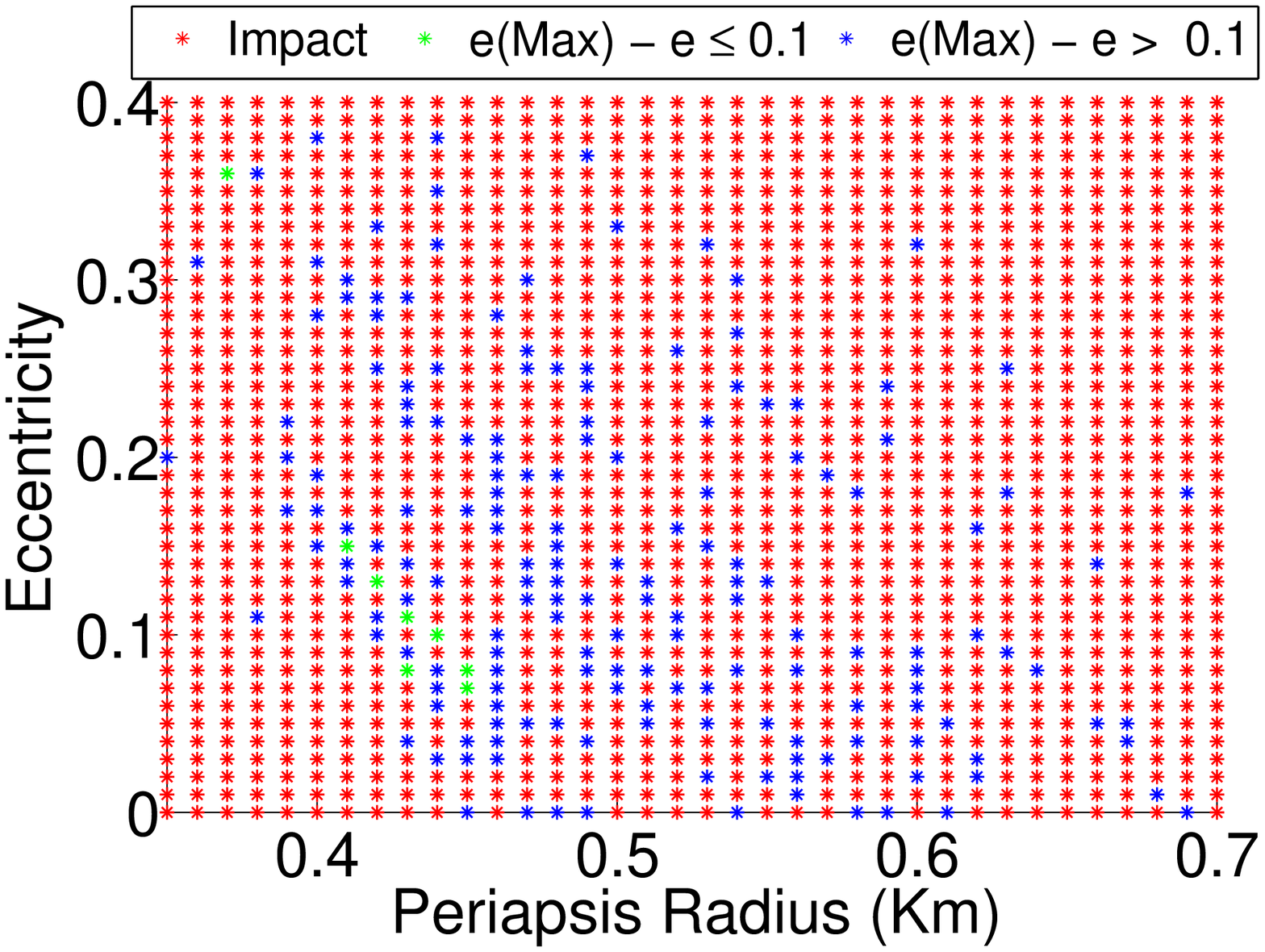}
 \includegraphics[width=0.24\linewidth]{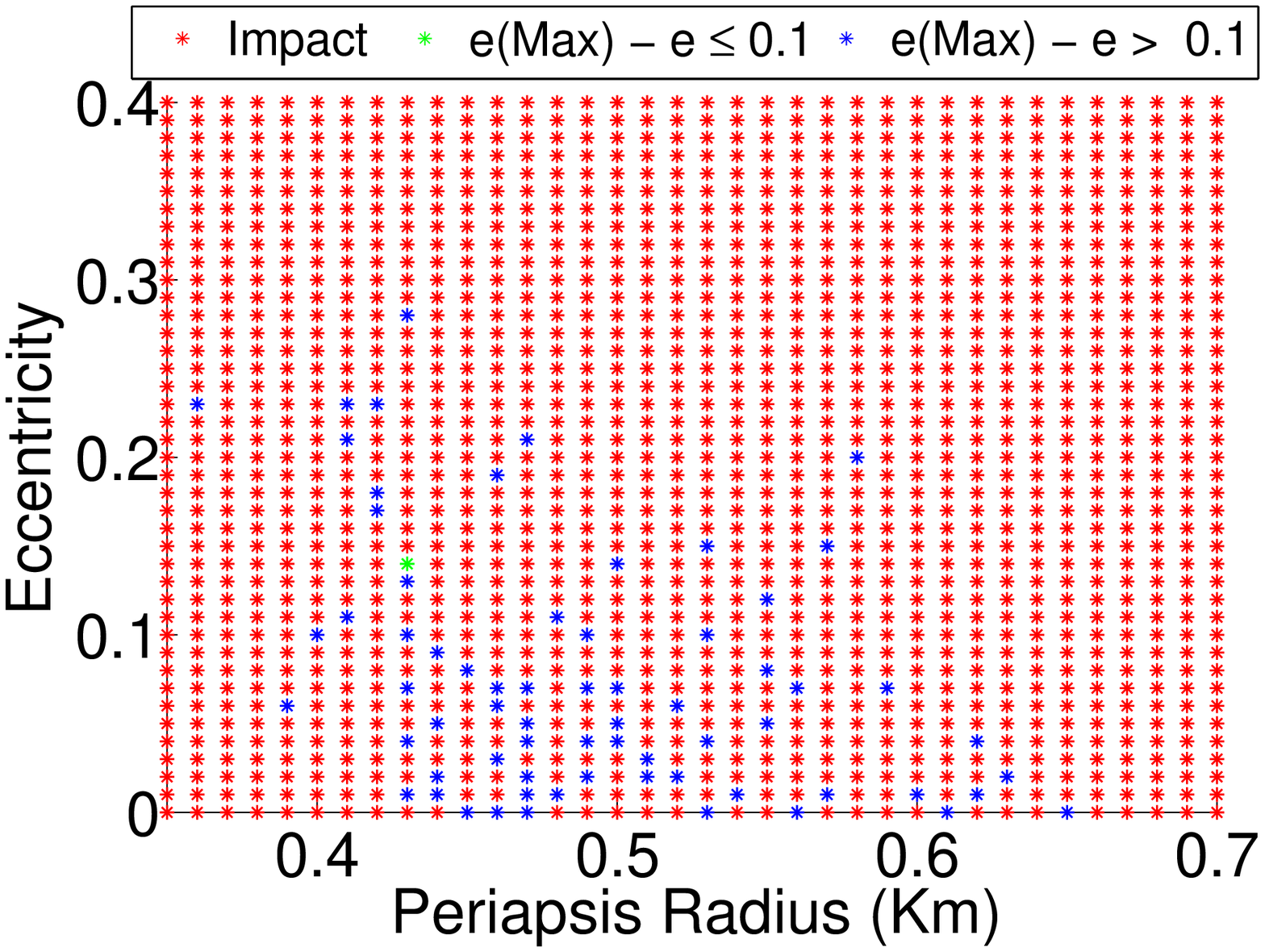}
 \includegraphics[width=0.24\linewidth]{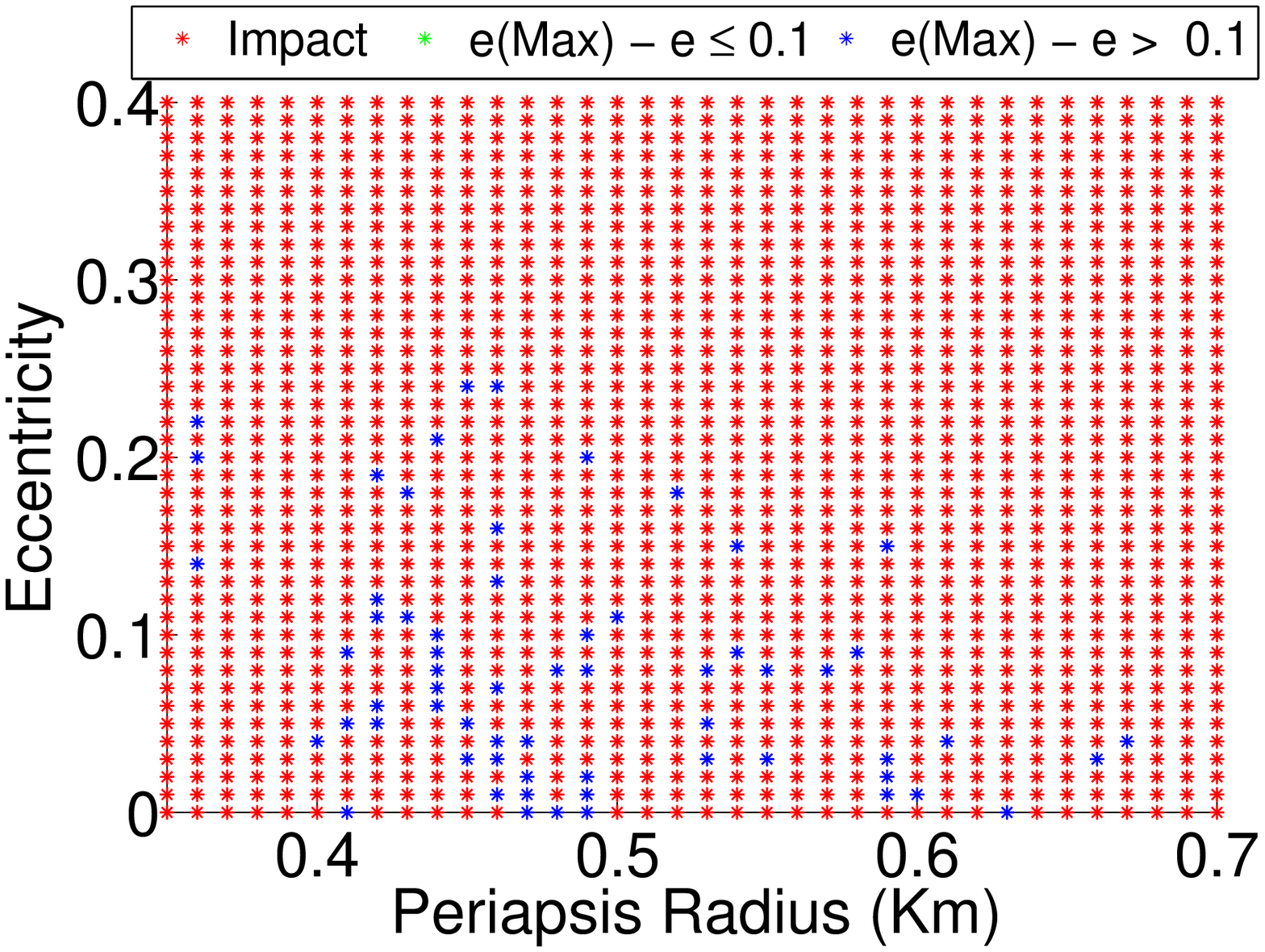}\\
{\small \textcolor{white}{.} \hspace{0.75cm} $0^{\circ}$ \hspace{3.75cm} $90^{\circ}$ \hspace{3.75cm}  $180^{\circ}$ \hspace{3.75cm} $270^{\circ}$}\\
 2) \includegraphics[width=0.24\linewidth]{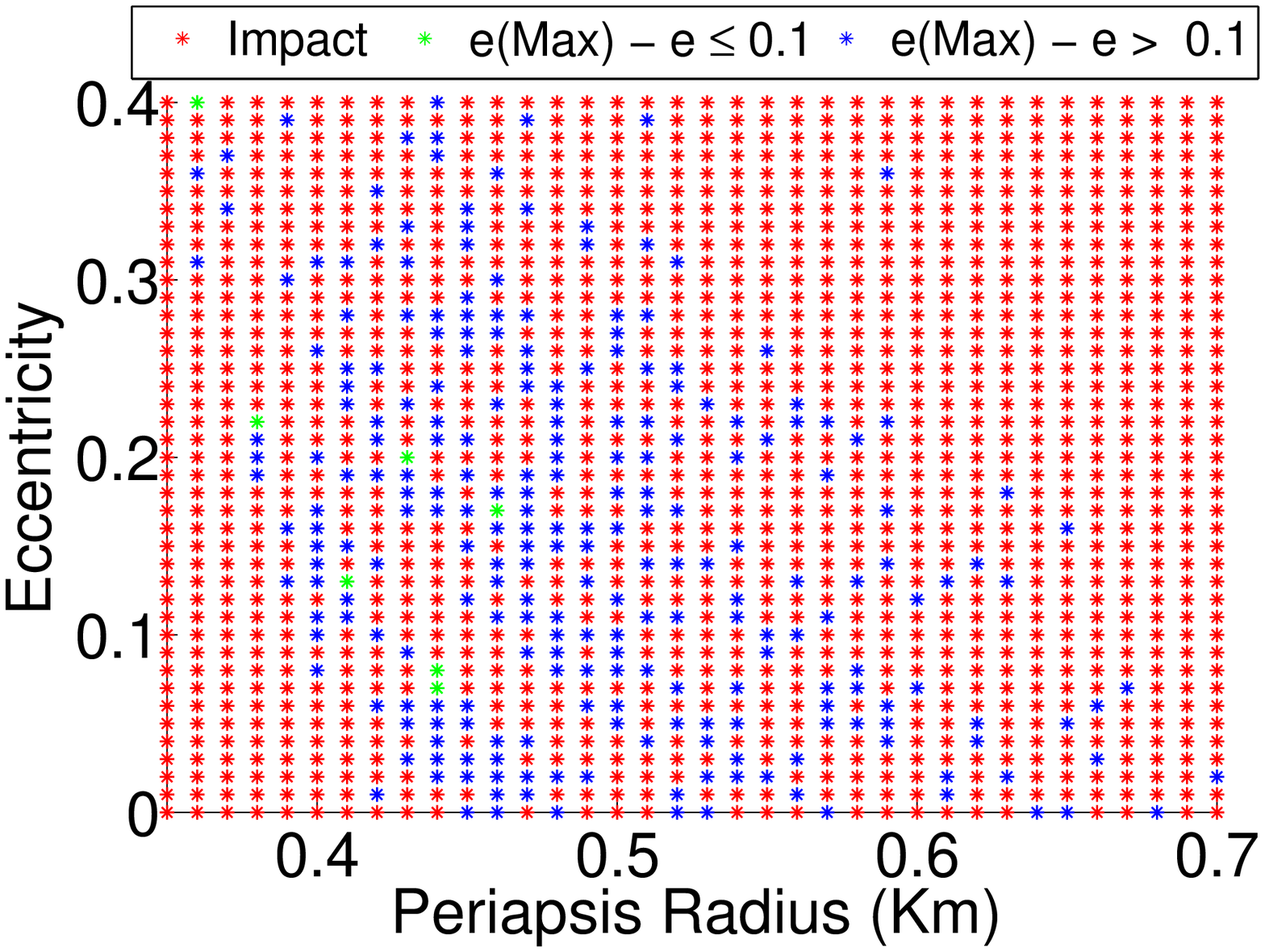}
 \includegraphics[width=0.24\linewidth]{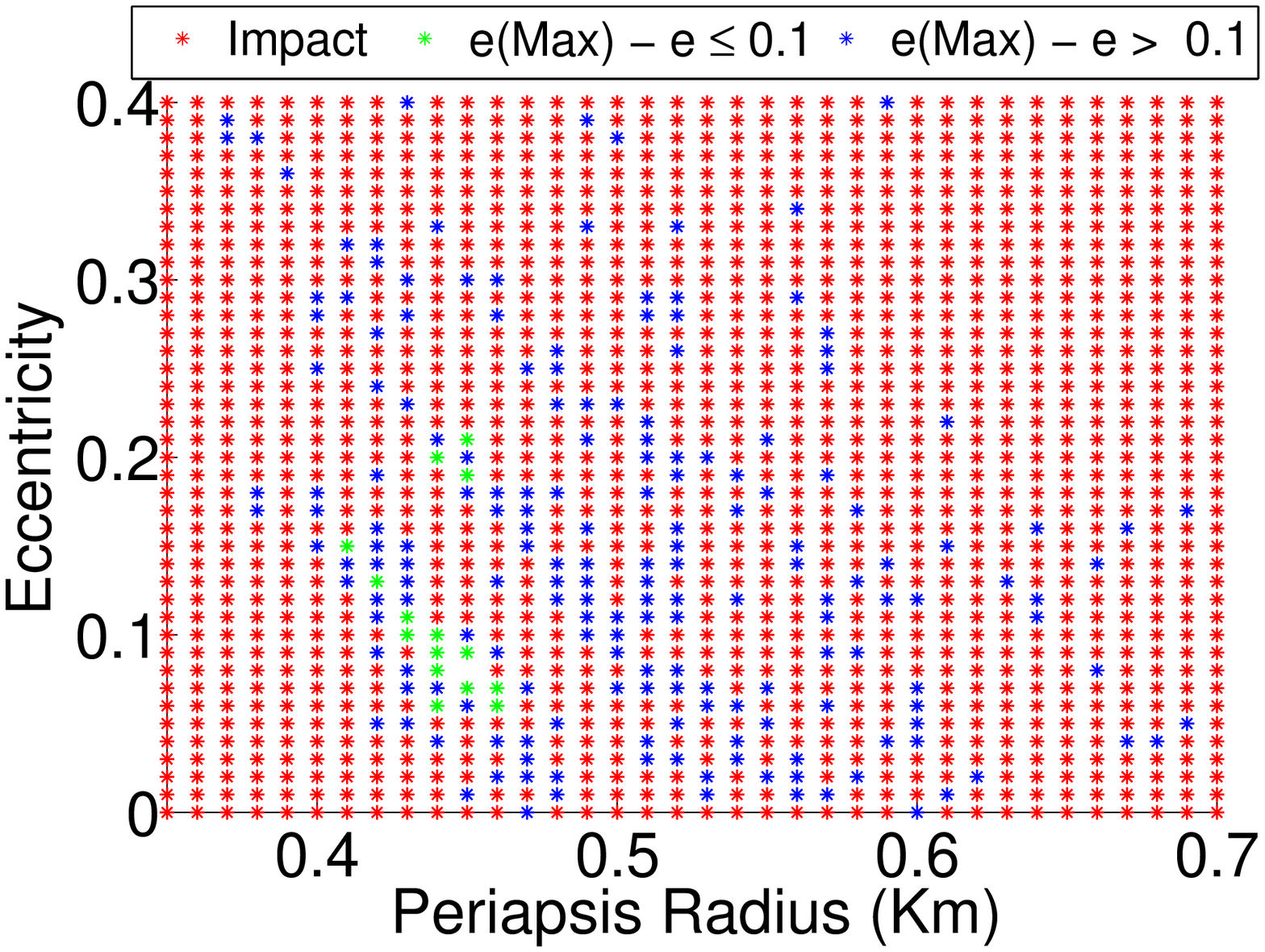}
 \includegraphics[width=0.24\linewidth]{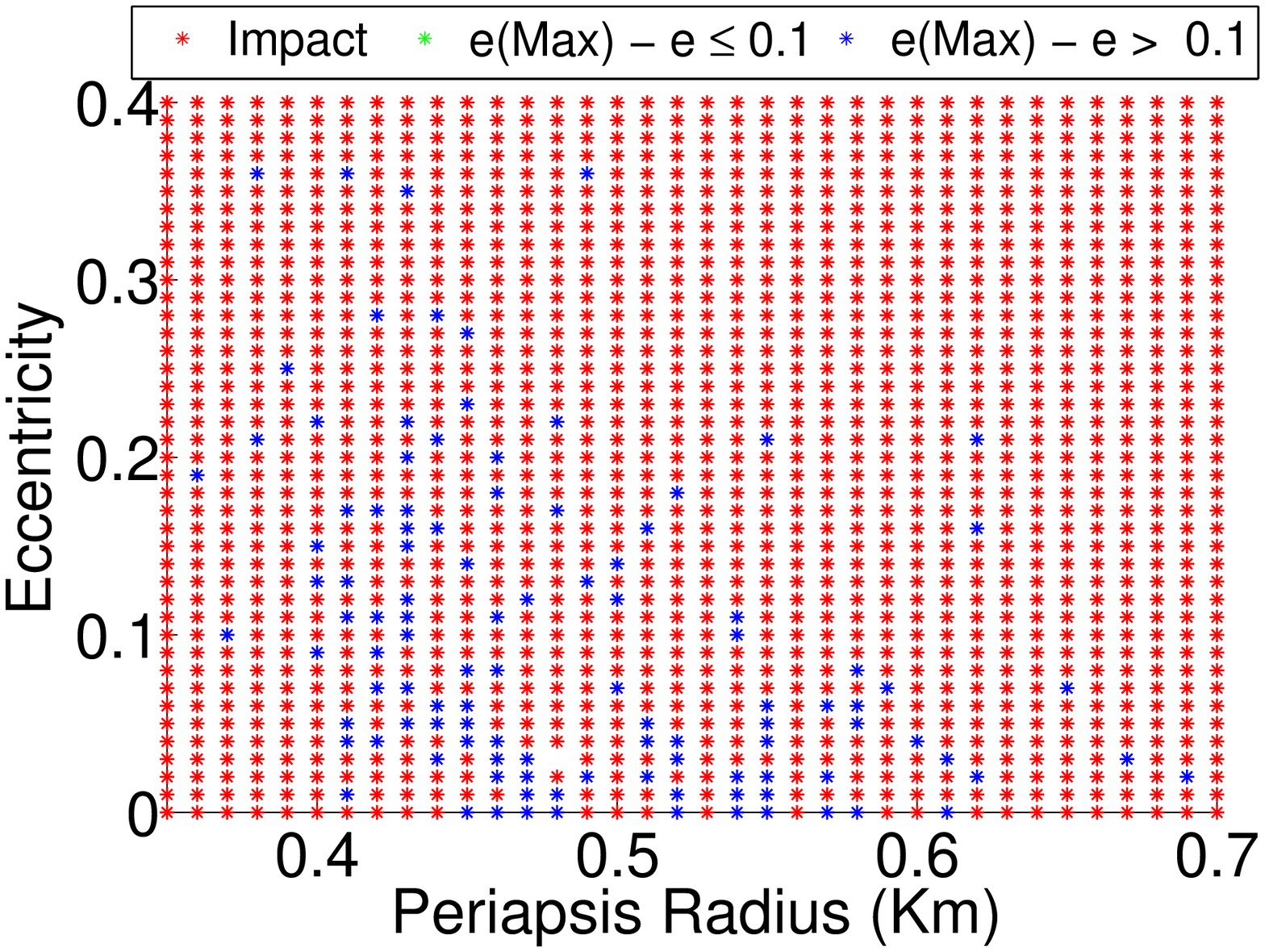}
 \includegraphics[width=0.24\linewidth]{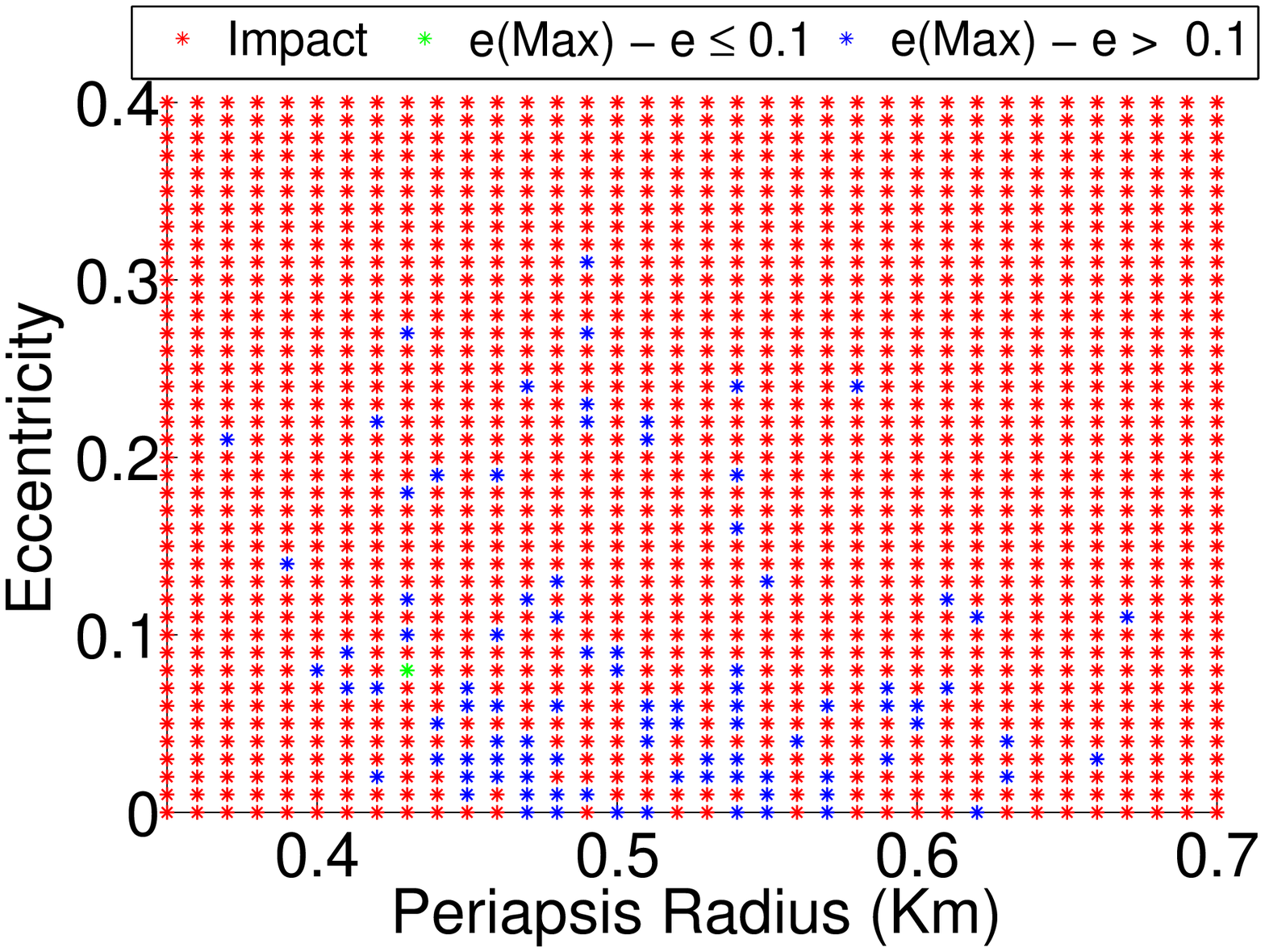}\\
       
 \caption{Stability maps of the equatorial orbits relative to (101955) Bennu where the solar radiation pressure is accounted for with the Sun initial longitude $\psi_0 = -135^o$. As in Fig.8, the (101955) Bennu's distance from the Sun $R$ is noted on the top of the related grafics and the eclipse is not taken into account in 1) and accounted for in 2). }
  \label{Fig9} 
\end{figure*}
We wish to show the effect of solar radiation pressure (SRP) on the close approach of the spacecraft at different distances of (101955) Bennu's orbit relative to the Sun. We set the solar radiation pressure in six configurations: at the points of maximal, medium and minimal perturbation and for the characteristics of Osiris-Rex Spacecraft when the eclipse is accounted for or not. As seen previously, we choose two initial longitudes for the Sun.
The results for the Sun initial longitude $\psi_0 = -180^o$ are shown in Fig. 8 and for $\psi_0 = -135^o$ in Fig. 9.
In a general way, we can see in Fig. 8 that the solar radiation pressure disturbs the orbits and a great number of collisions occurs where initially stable orbits existed. Furthermore, stable orbits that started at longitude $0^o$ occur in a great number at periapsis radius closer to (101955) Bennu, between 0.4 and 0.5 km. We also find somewhat more stability for the orbits launched from the longitude $270^o$. 
This fact was observed under the conditions that can be encountered by the Osiris-Rex probe close at the aphelion distance of the asteroid. However, we can highlight that the radiation pressure destabilises the spacecraft's orbits further, causing them to collide when the asteroid approaches its perihelion. The orbits that start at the longitude $\lambda = 0^o $ appear to be more stable even when the solar radiation pressure reaches its maximum. When the passage of the spacecraft in the shadow is accounted for, the effects of solar radiation pressure are softer and we find a larger portion of stable orbits. We encounter the same behaviour at the longitude $\lambda = 90^o $ when the Sun initial longitude is $\psi_0 = -90^o$ (Fig A2 in appendix A). This shows that orbits that began in the projected shadow seem to be more stable.
We encounter some stability still below 0.5 km and with a relatively low initial eccentricity. It happens for the other initial longitudes when the shadow effect is accounted for and the solar radiation pressure is at its lowest value. When we change the initial longitude of the Sun to $\psi_0 = -135^o$, Fig 9 shows that the orbits with initial longitude $\lambda = 90^0 $ appear to be more stable than those of Fig. 8. It also occurs when the orbit begins in $\lambda = 90^0 $ with the initial longitude of the Sun $\psi_0 = -45^o$. So, we notice a certain symmetry regarding the effects of solar radiation pressure on the behaviour of the spacecraft around asteroid (101955) Bennu. It is certainly due to its shape, with a small ellipticity, as shown in Fig 1.
Furthermore, the orbits that start at the longitude $\lambda = 0^o $ are less stable for $\psi_0 = -135^o$. Thus, we show the results in appendix A for all the initial longitudes when the asteroid reaches its minimum distance from the Sun (perihelion) and in this case, we find almost none stability for all the initial longitudes when the asteroid reaches its minimum distance from the Sun (perihelion). 
\begin{figure*}
   \centering
      SRP ($g=4.405370 \times 10^{-11}km\cdot s^{-2}$,  $R=1.1264AU$) \\
      \small \textcolor{white}{.}\\
   1) \includegraphics[width=0.25\linewidth]{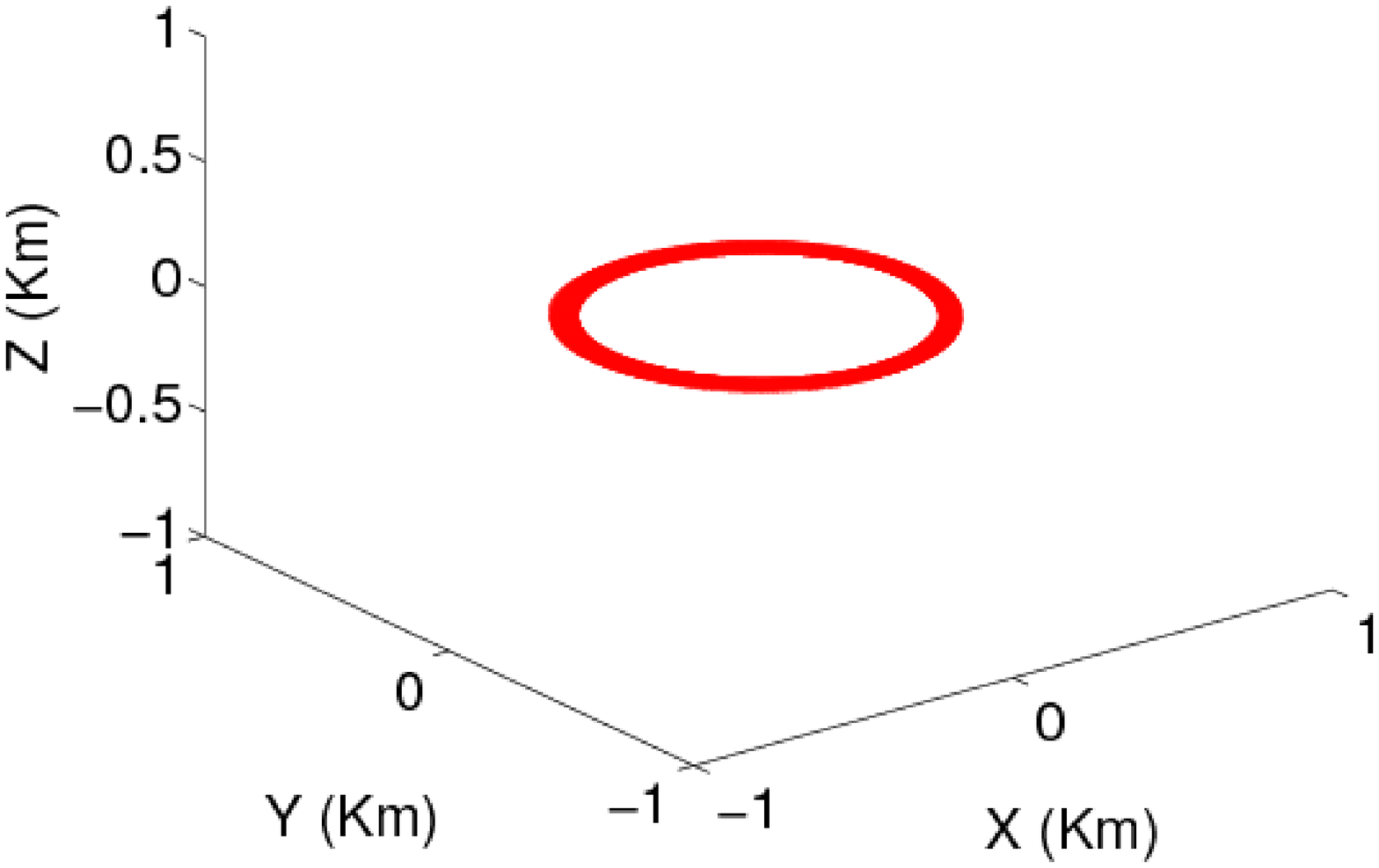}
     \includegraphics[width=0.25\linewidth]{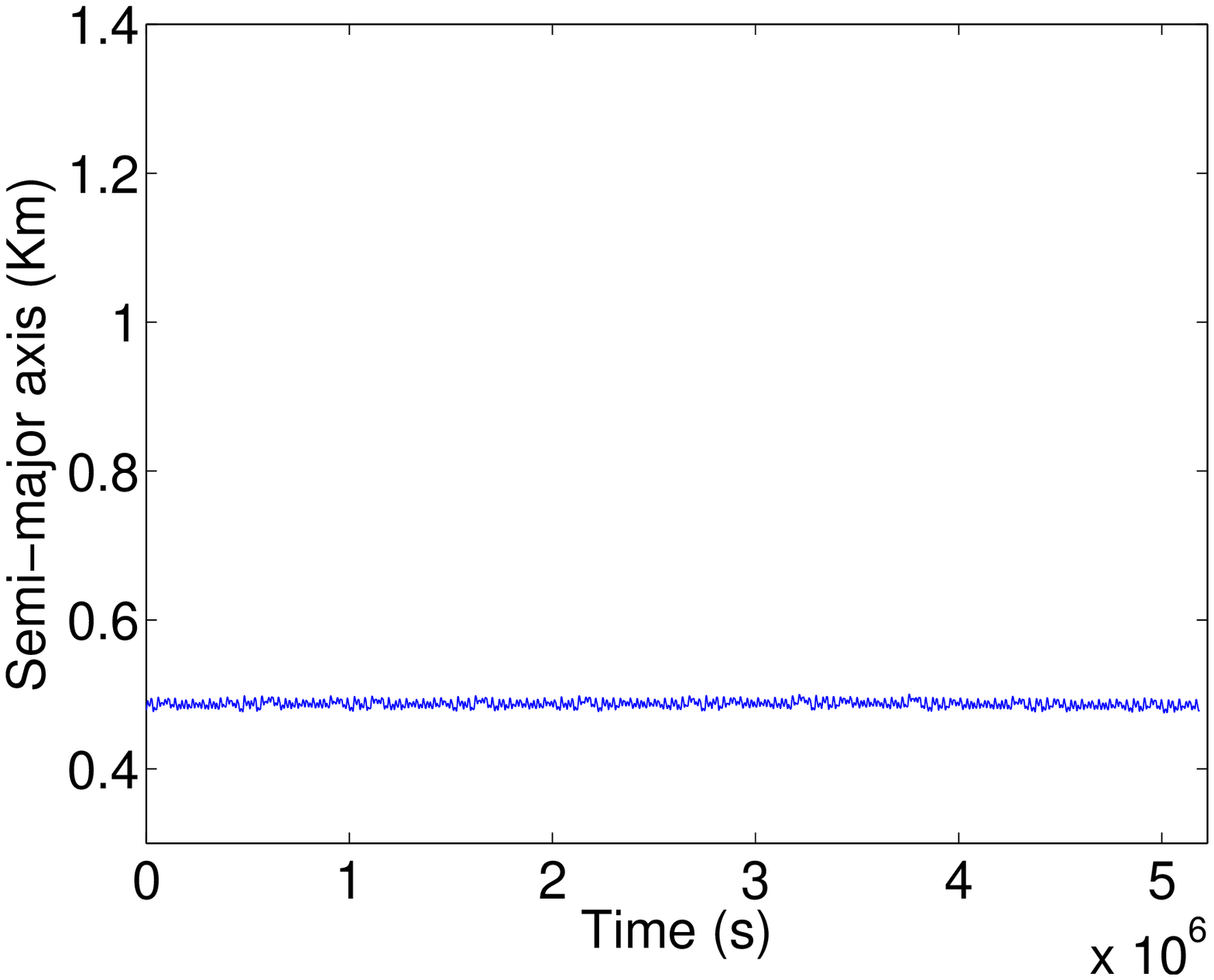}
     \includegraphics[width=0.25\linewidth]{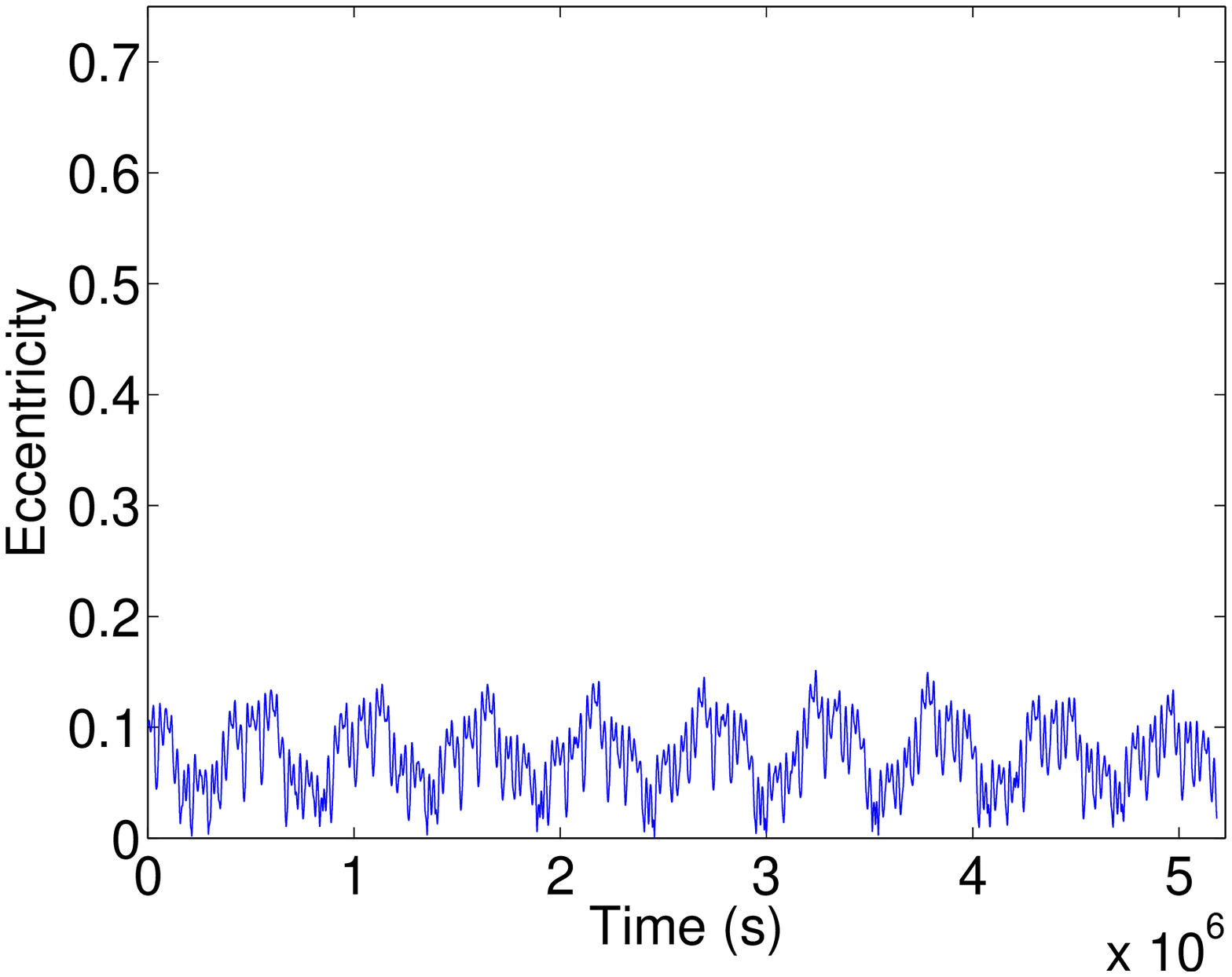}\\
  2) \includegraphics[width=0.25\linewidth]{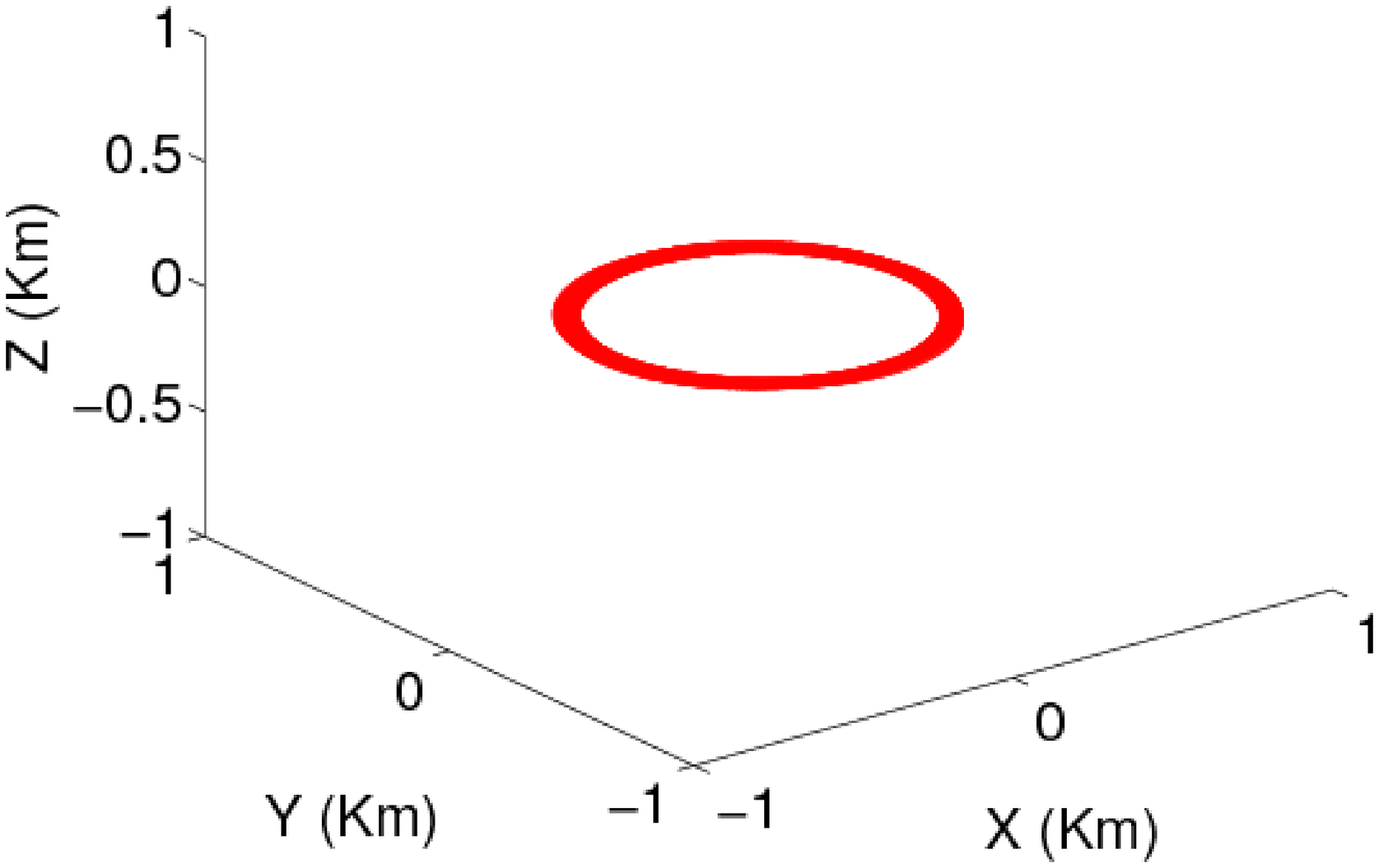}
     \includegraphics[width=0.25\linewidth]{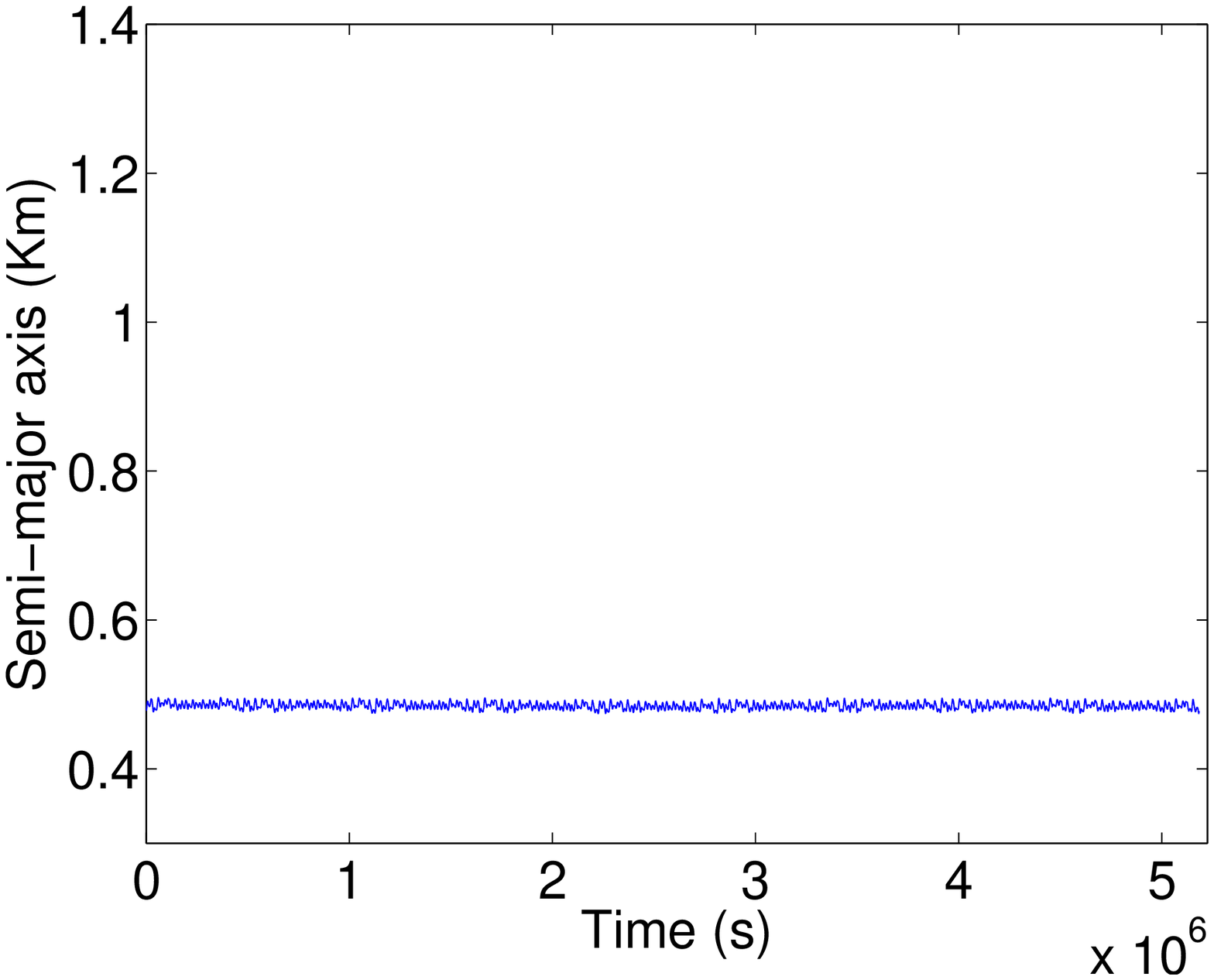}
     \includegraphics[width=0.25\linewidth]{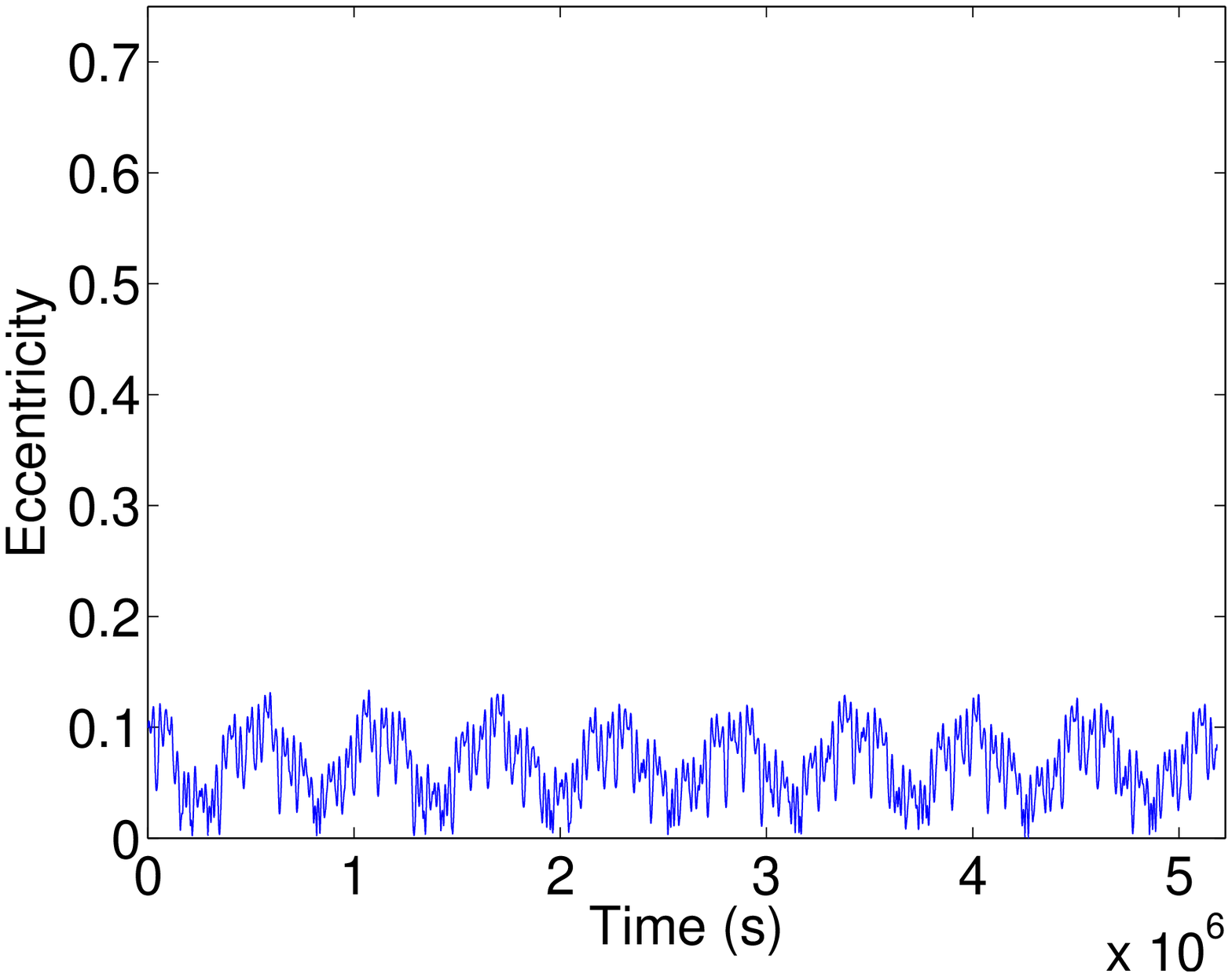}\\
     \small \textcolor{white}{.}\\
     SRP ($g=6.941835 \times 10^{-11}km\cdot s^{-2}$, $R=0.8969AU$) \\
     \small \textcolor{white}{.}\\
    1) \includegraphics[width=0.25\linewidth]{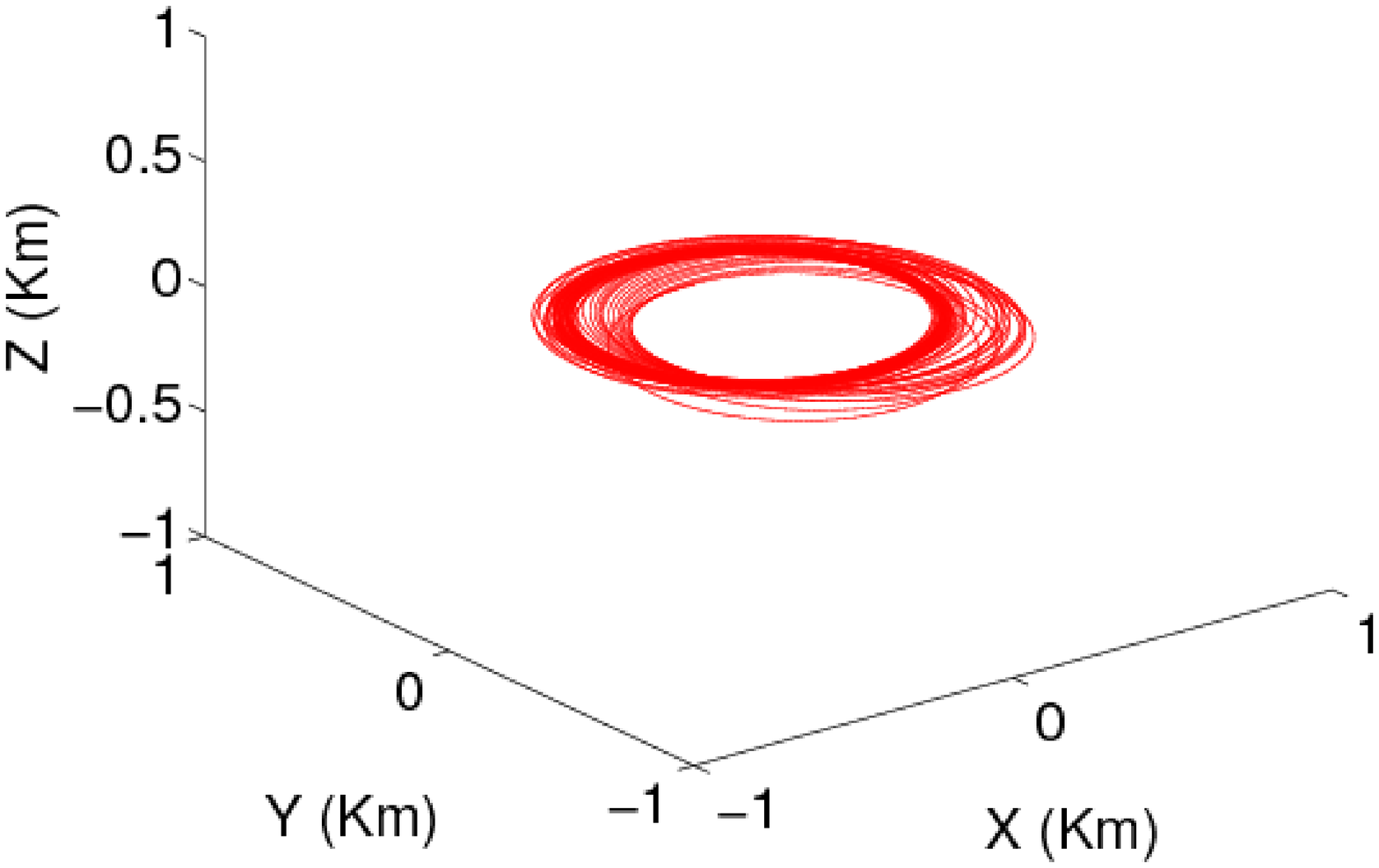}
     \includegraphics[width=0.25\linewidth]{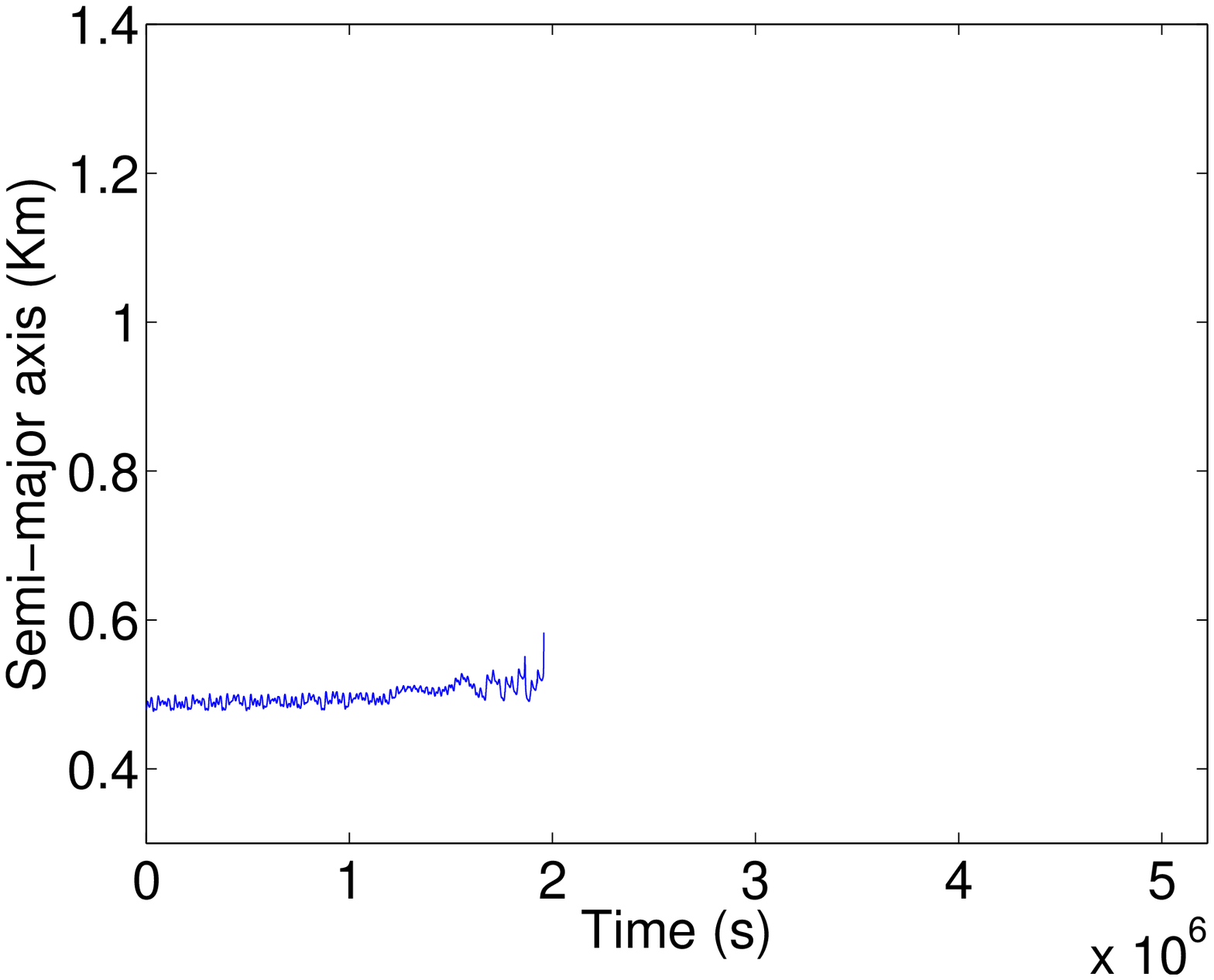}
     \includegraphics[width=0.25\linewidth]{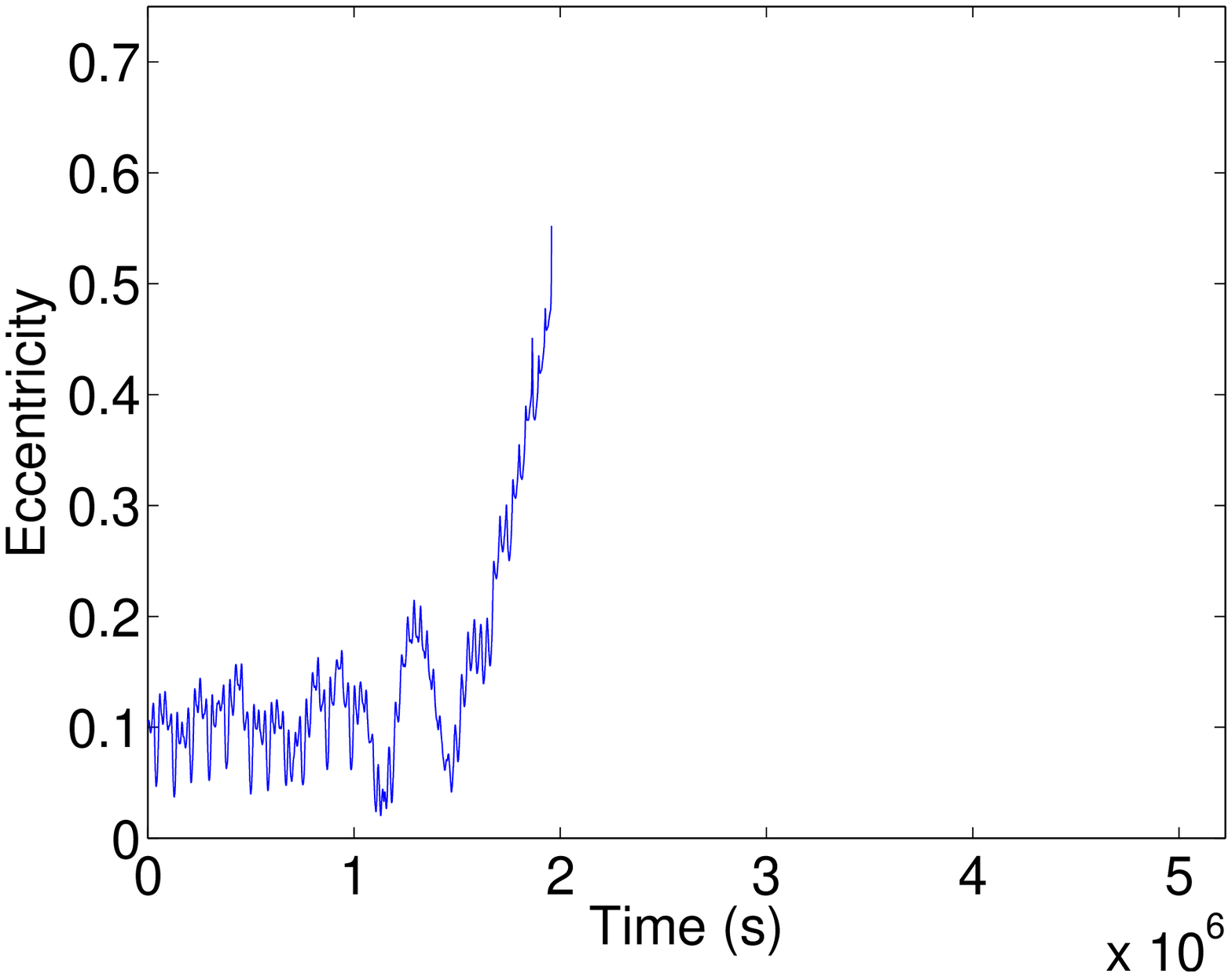}\\  
     2)\includegraphics[width=0.25\linewidth]{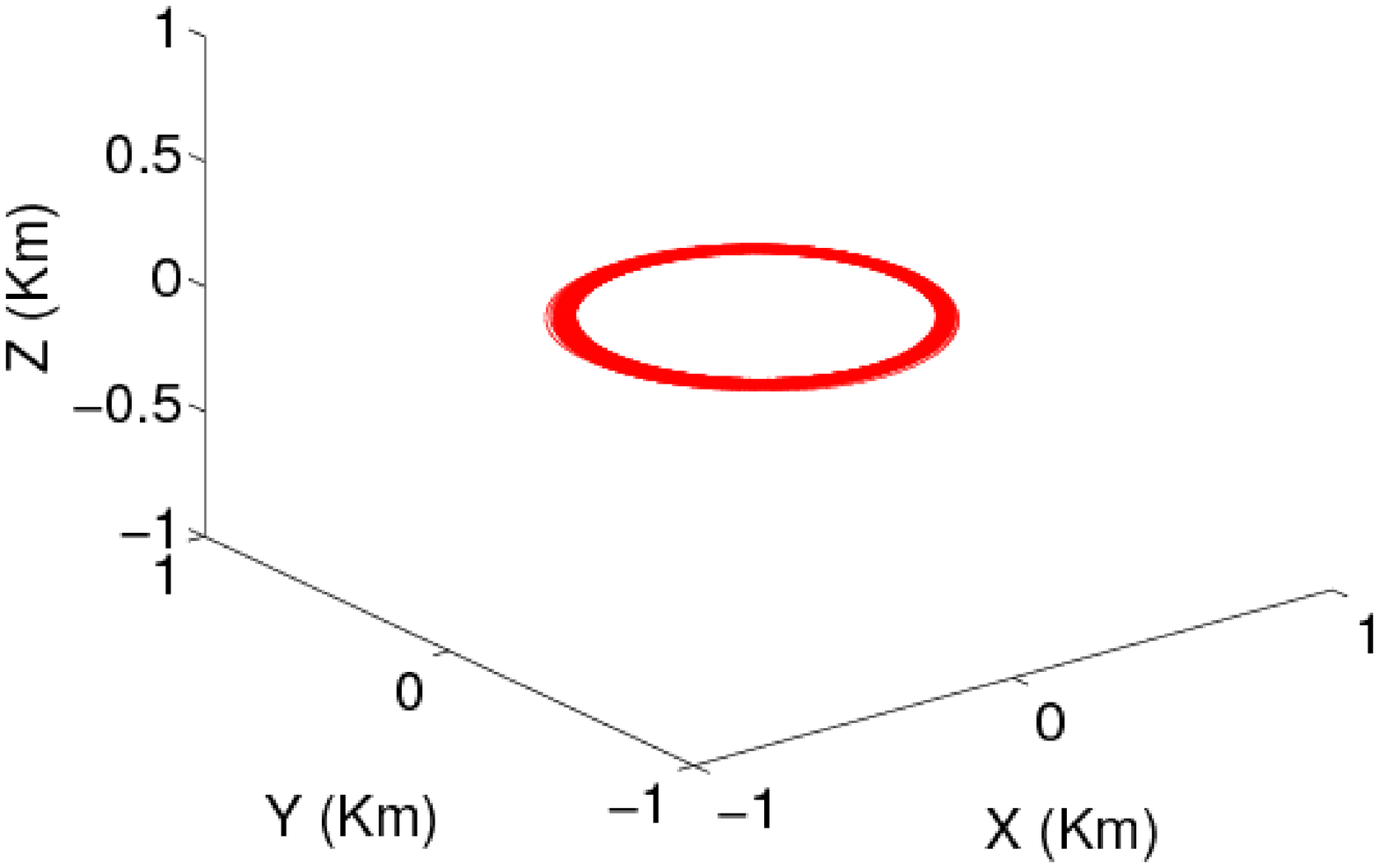}
     \includegraphics[width=0.25\linewidth]{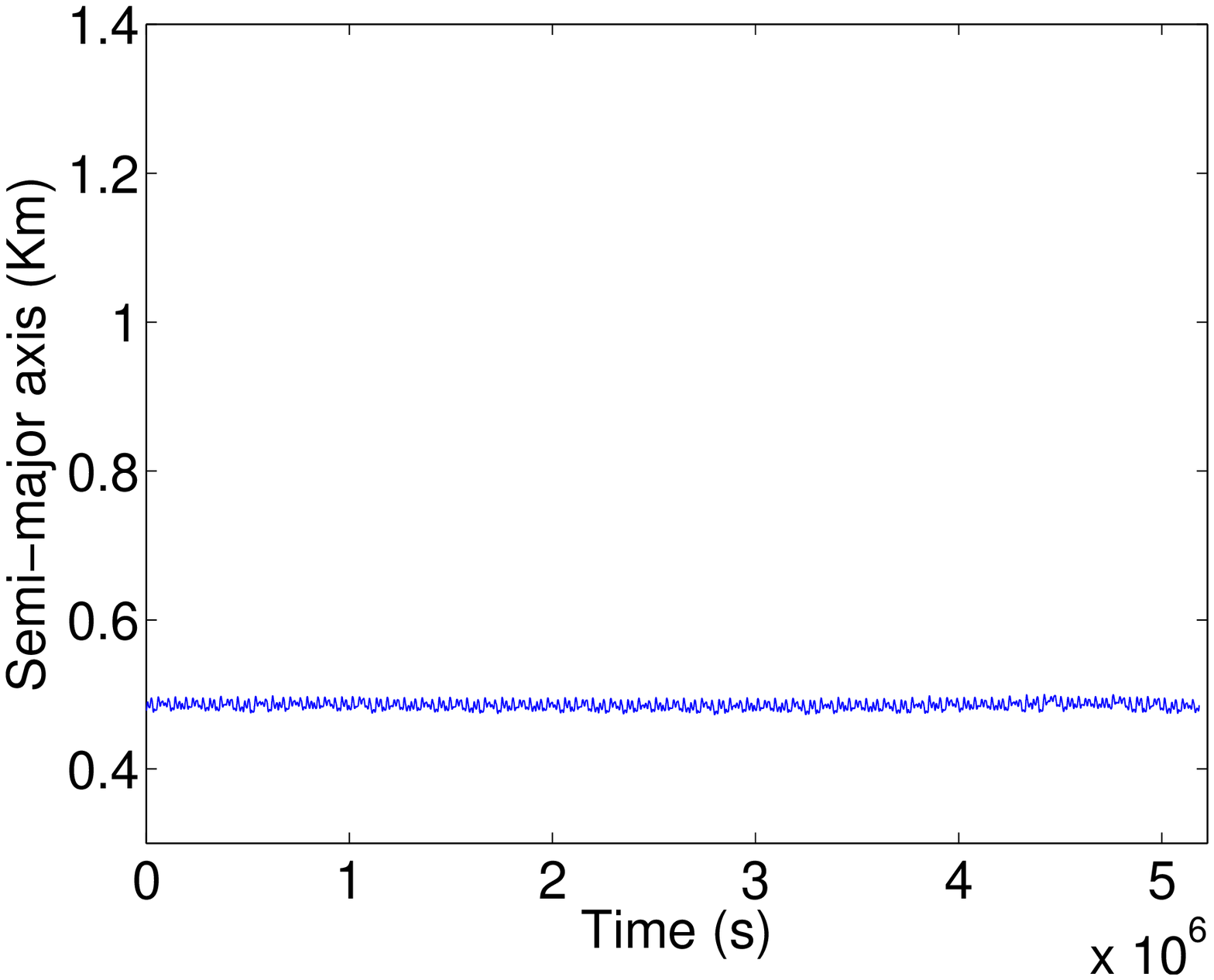}
     \includegraphics[width=0.25\linewidth]{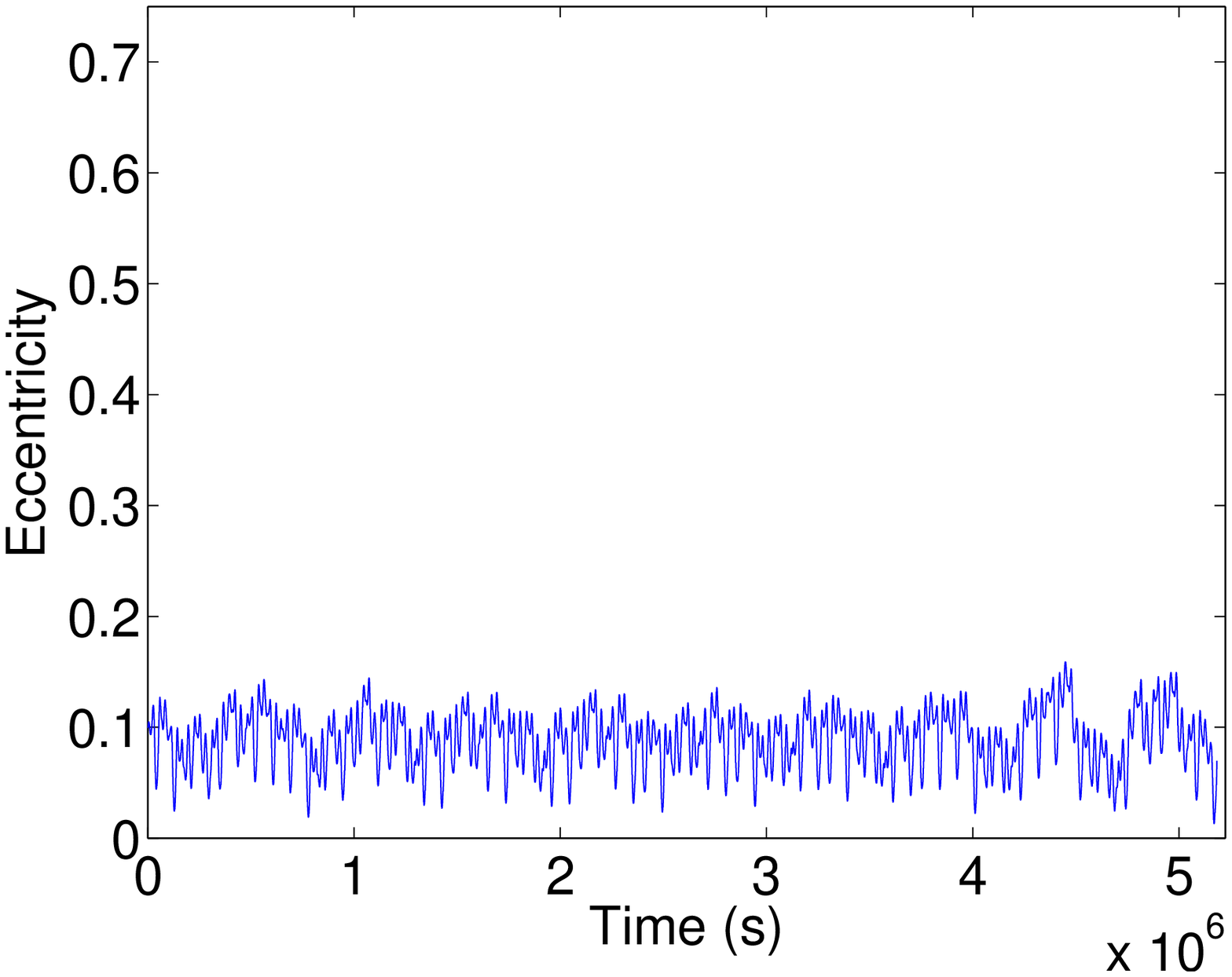}\\
     \caption{3D equatorial orbits around (101955) Bennu after $60$ days. These orbits are launched from the longitude $\lambda =0^o$. The initial eccentricity is 0.1 and its initial periapsis radius is 0.43 $km$. The solar radiation pressure  is accounted for without the shadow in 1) and with the passage in the shadow in 2). The behaviour of the semi-major axis is represented in the middle column and the eccentricity in the right-hand column.
}
         \label{Fig10} 
\end{figure*}

\begin{figure*}
   \centering
      SRP ($g=4.405370 \times 10^{-11}km\cdot s^{-2}$,  $R=1.1264AU$) \\
      \small \textcolor{white}{.}\\
   1) \includegraphics[width=0.25\linewidth]{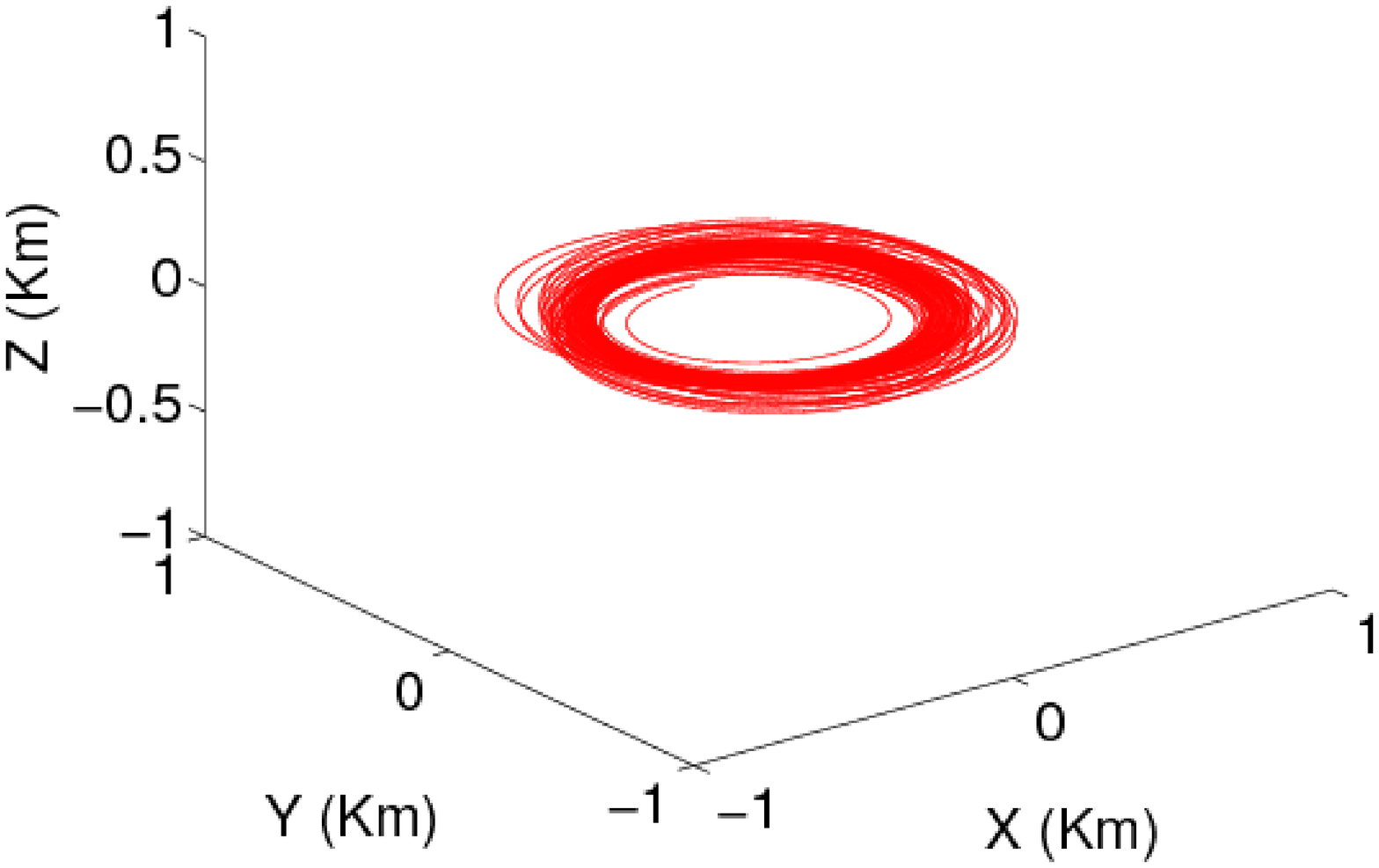}
     \includegraphics[width=0.25\linewidth]{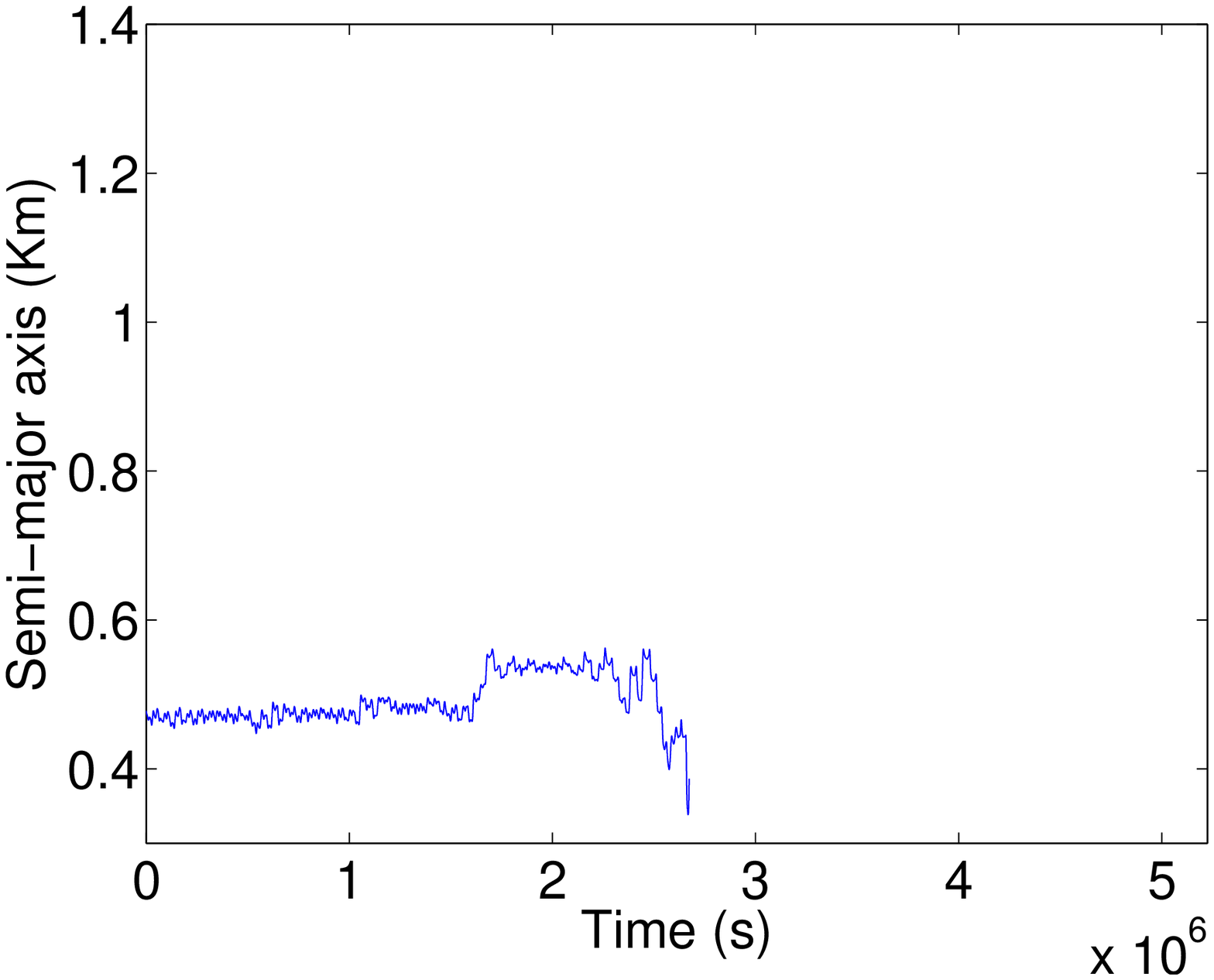}
     \includegraphics[width=0.25\linewidth]{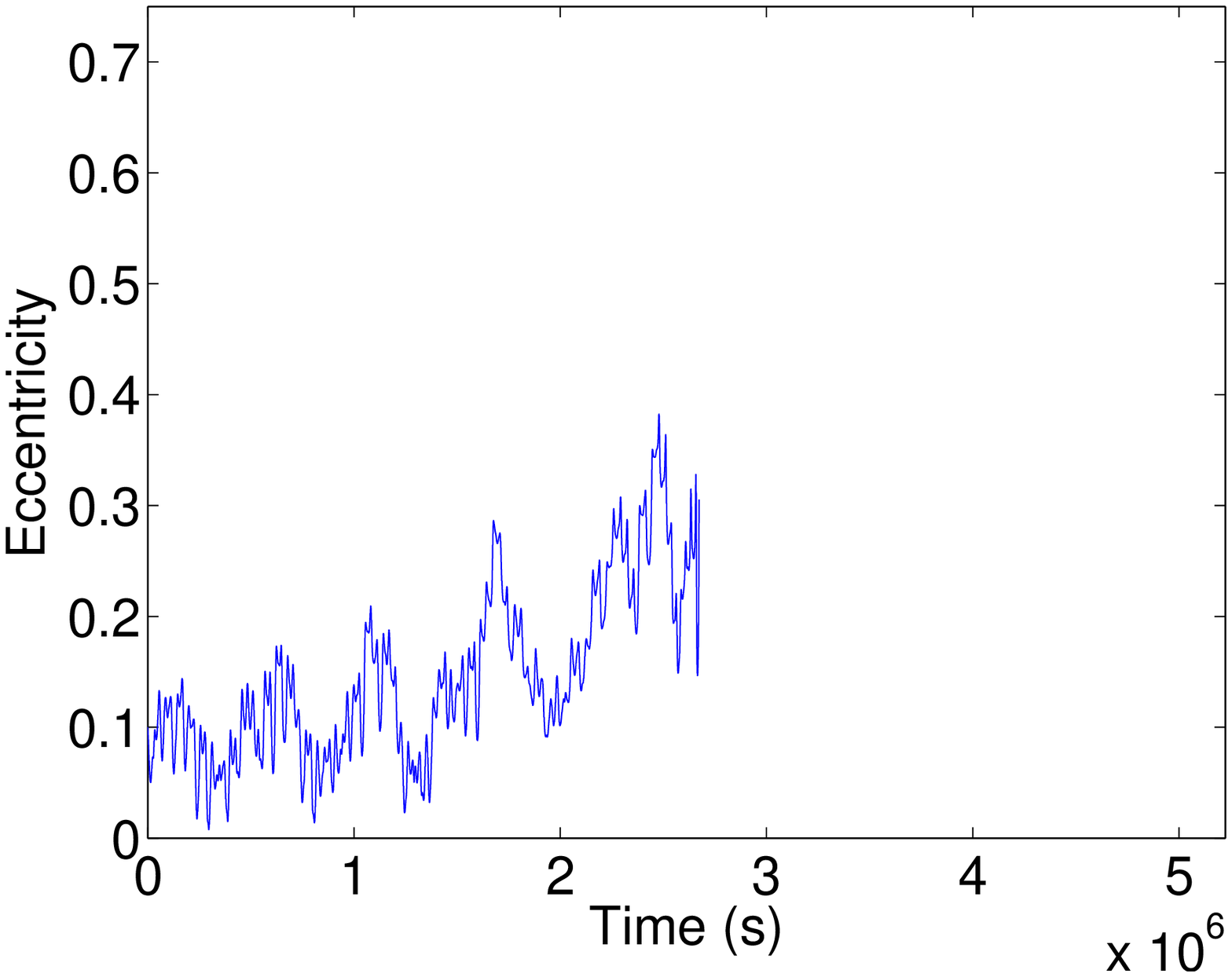}\\
    2) \includegraphics[width=0.25\linewidth]{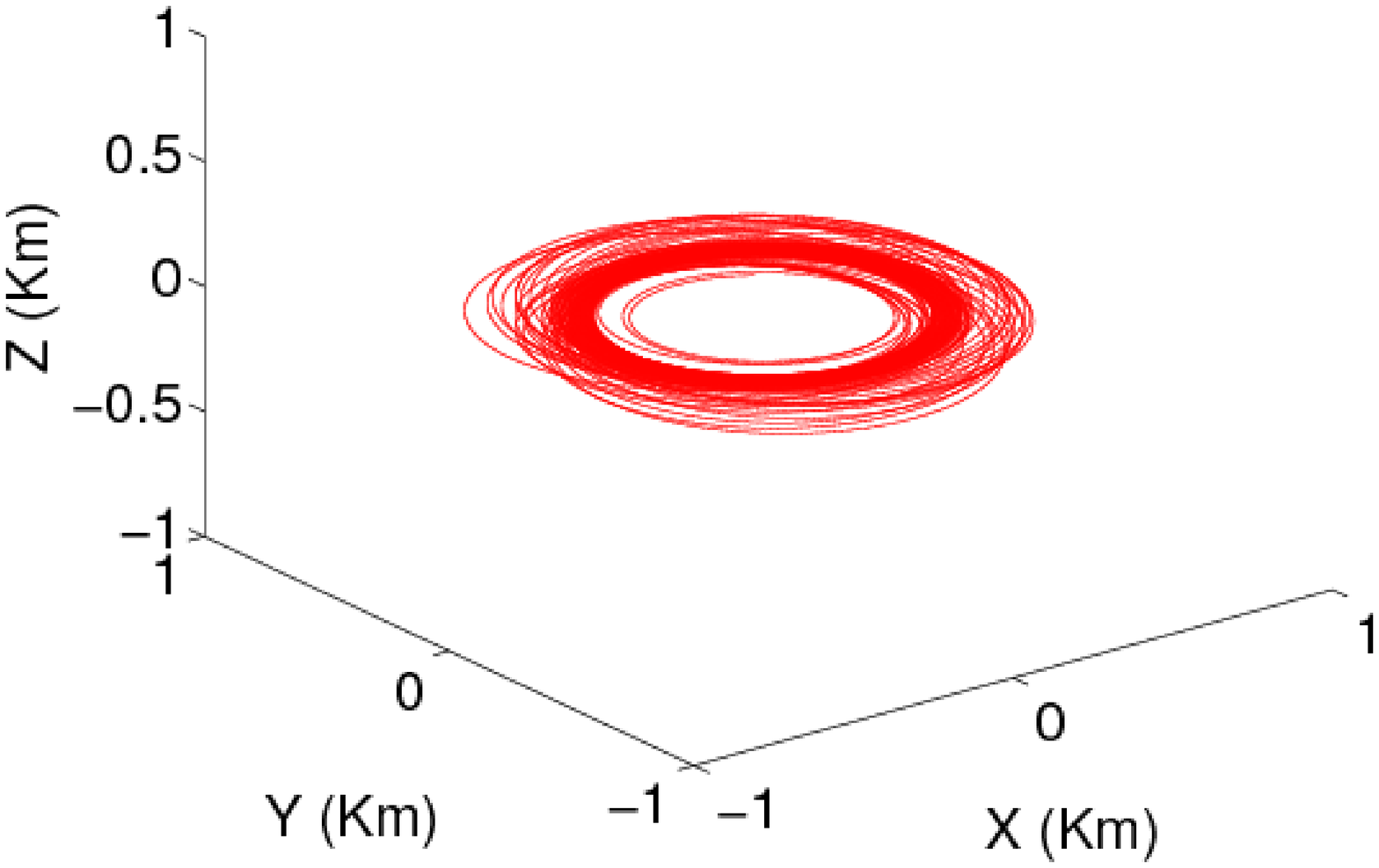}
     \includegraphics[width=0.25\linewidth]{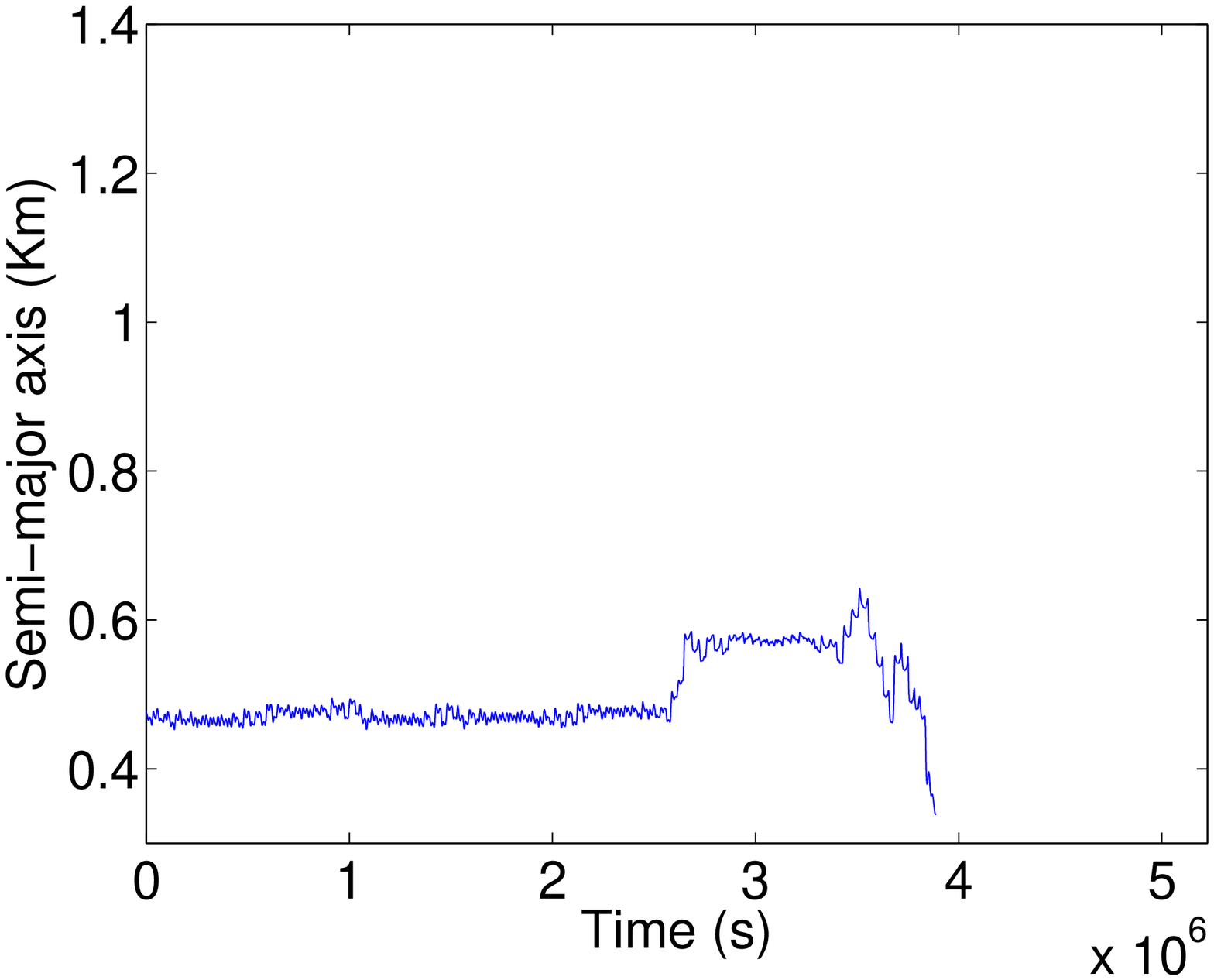}
     \includegraphics[width=0.25\linewidth]{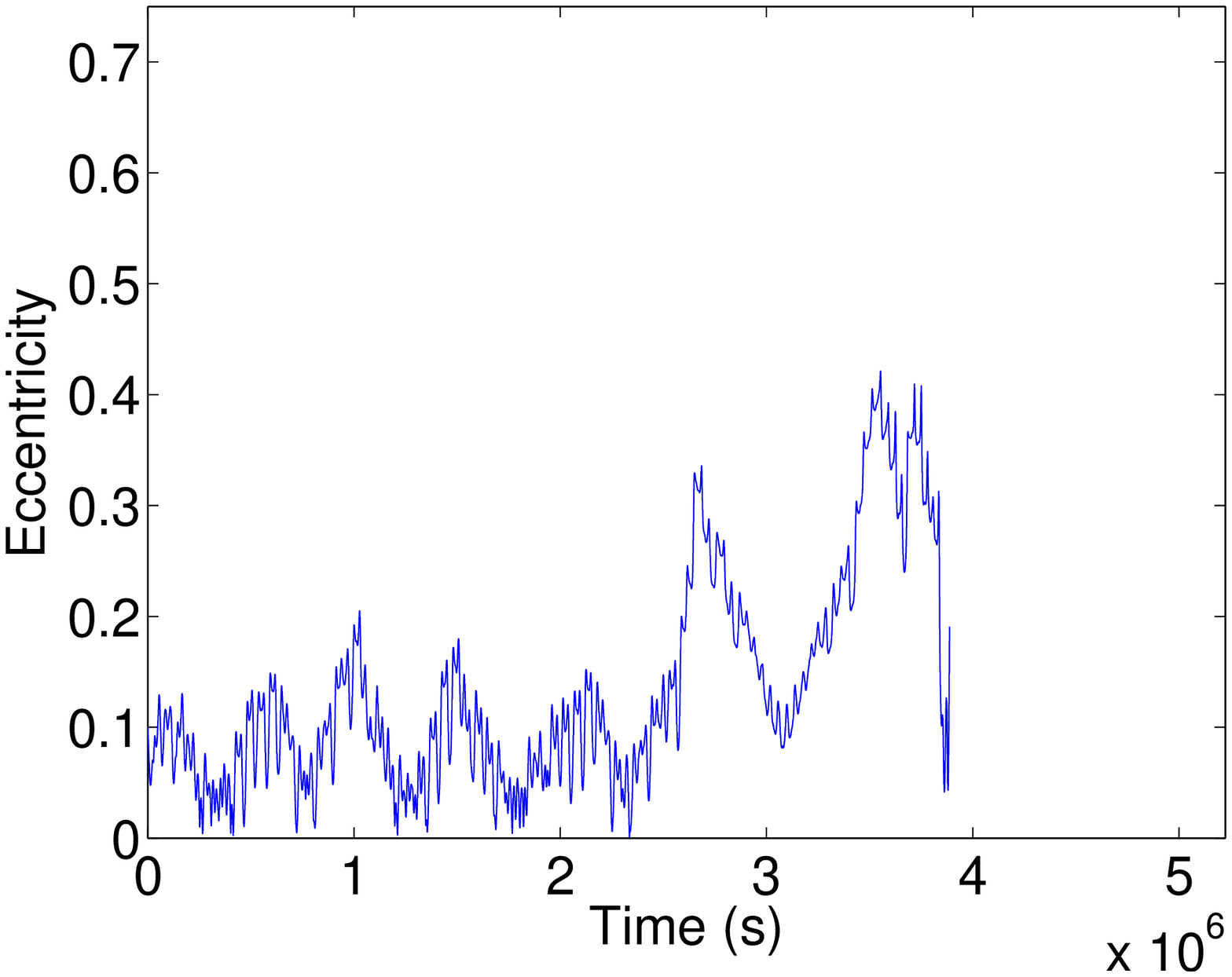}\\
     \small \textcolor{white}{.}\\
     SRP ($g=6.941835 \times 10^{-11}km\cdot s^{-2}$, $R=0.8969AU$)\\
     \small \textcolor{white}{.}\\
    1) \includegraphics[width=0.25\linewidth]{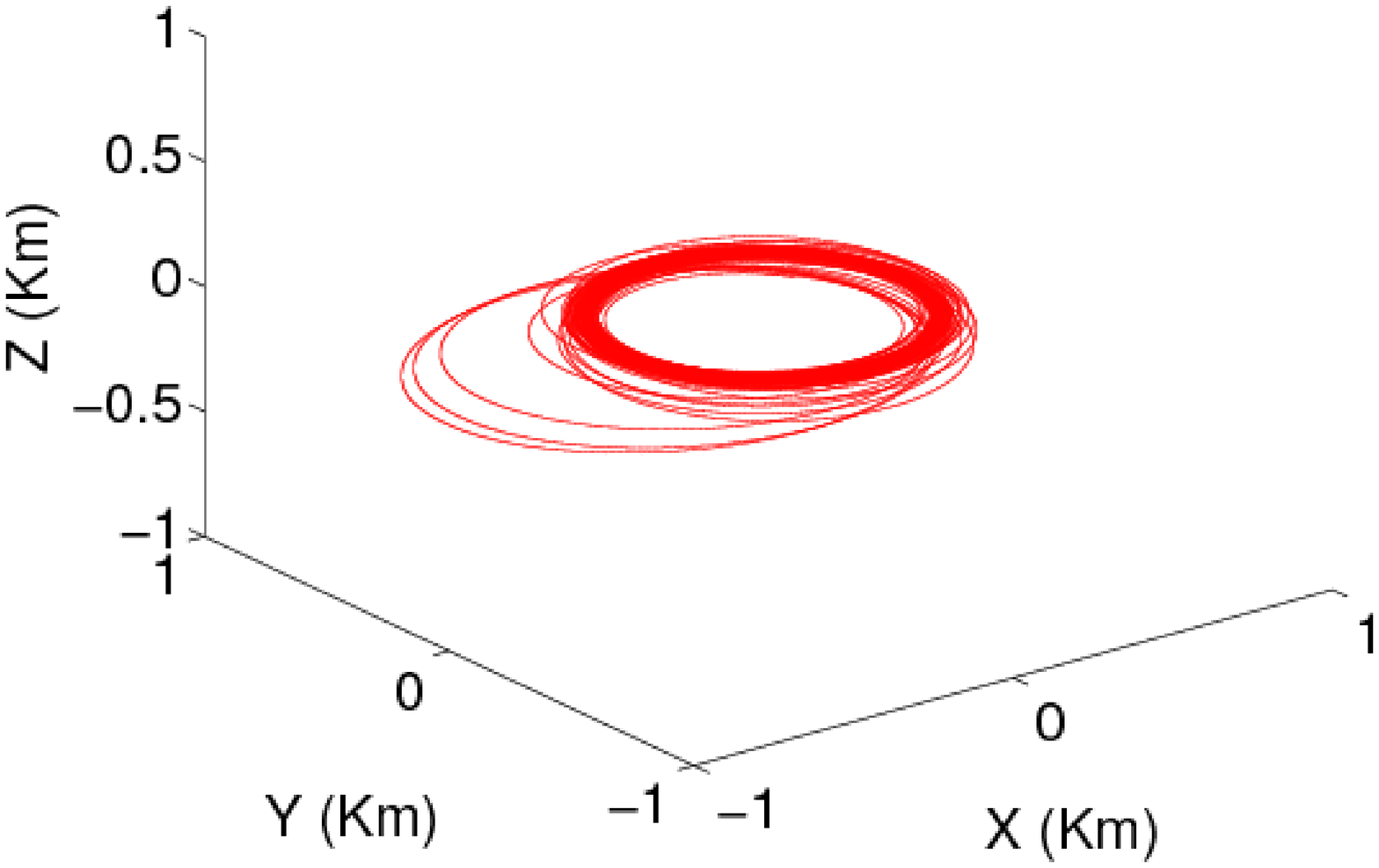}
     \includegraphics[width=0.25\linewidth]{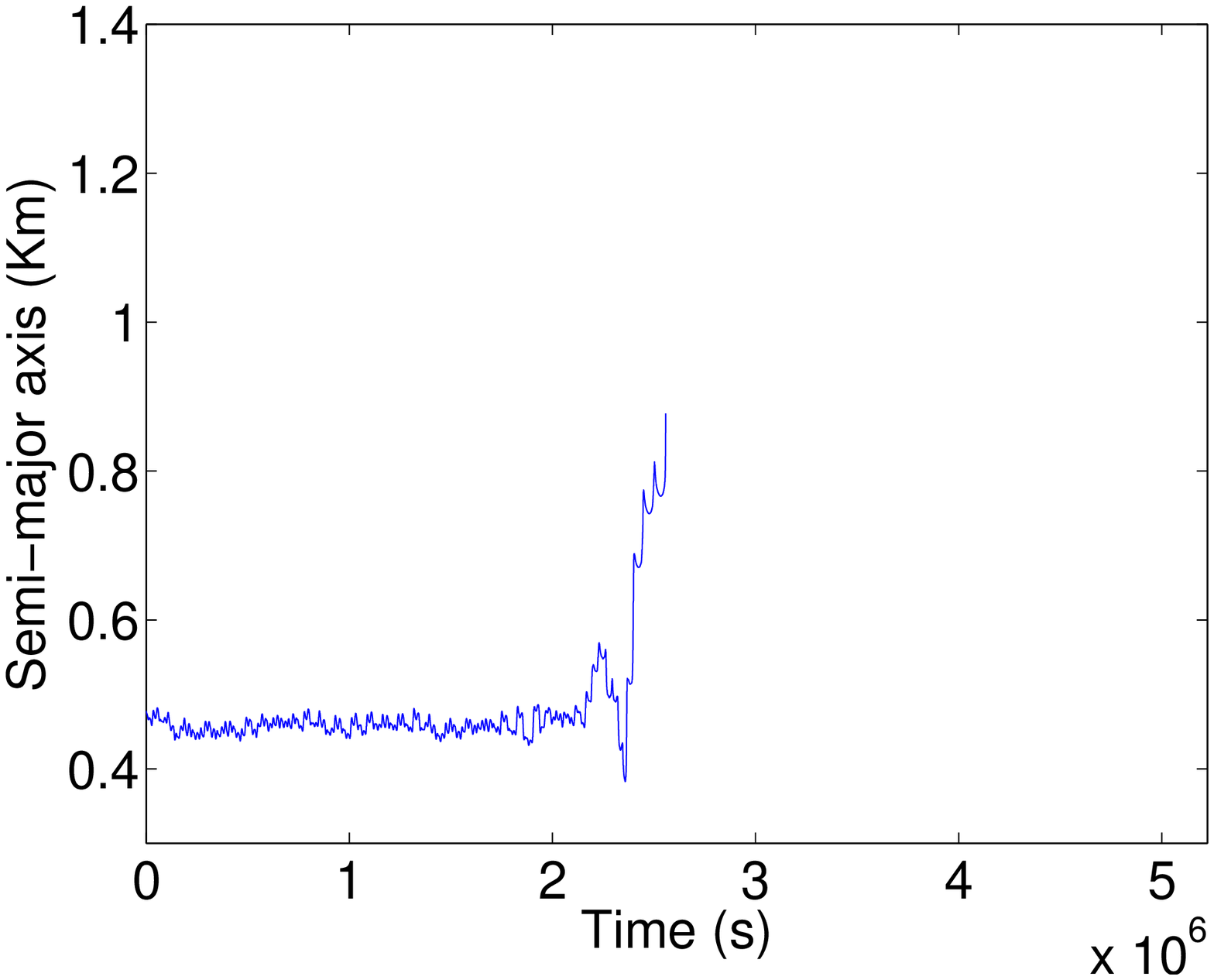}
     \includegraphics[width=0.25\linewidth]{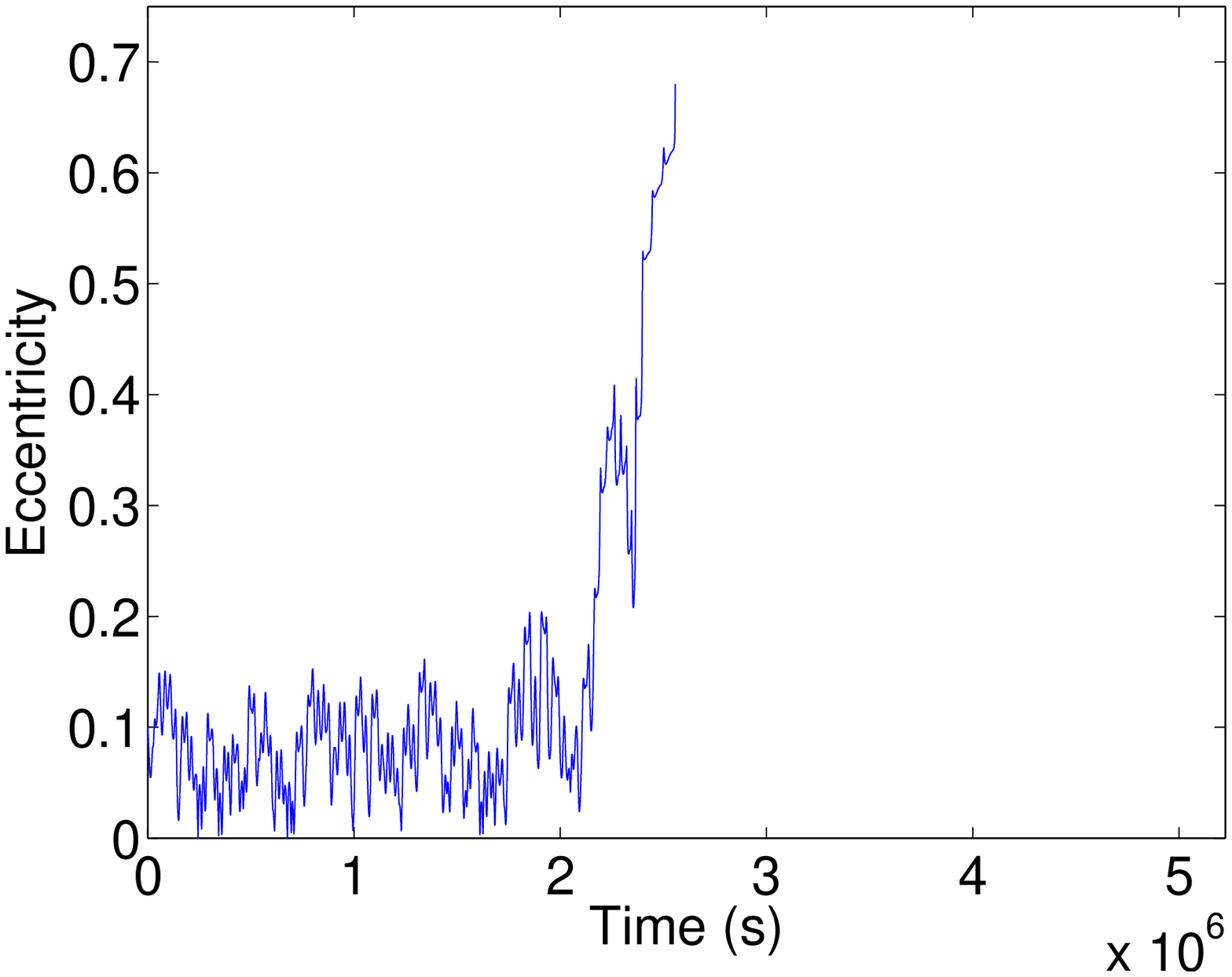}\\   
    2) \includegraphics[width=0.25\linewidth]{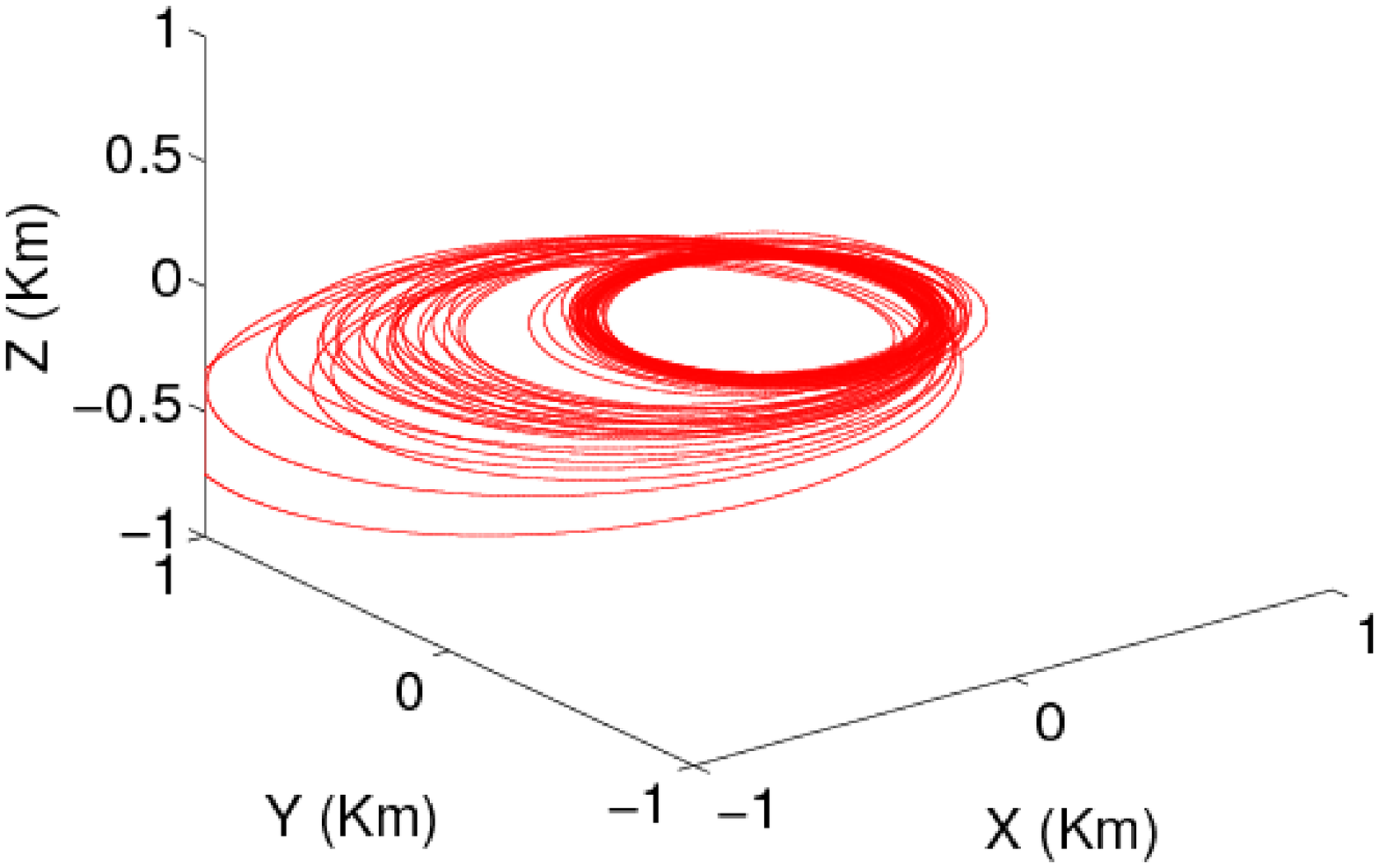}
     \includegraphics[width=0.25\linewidth]{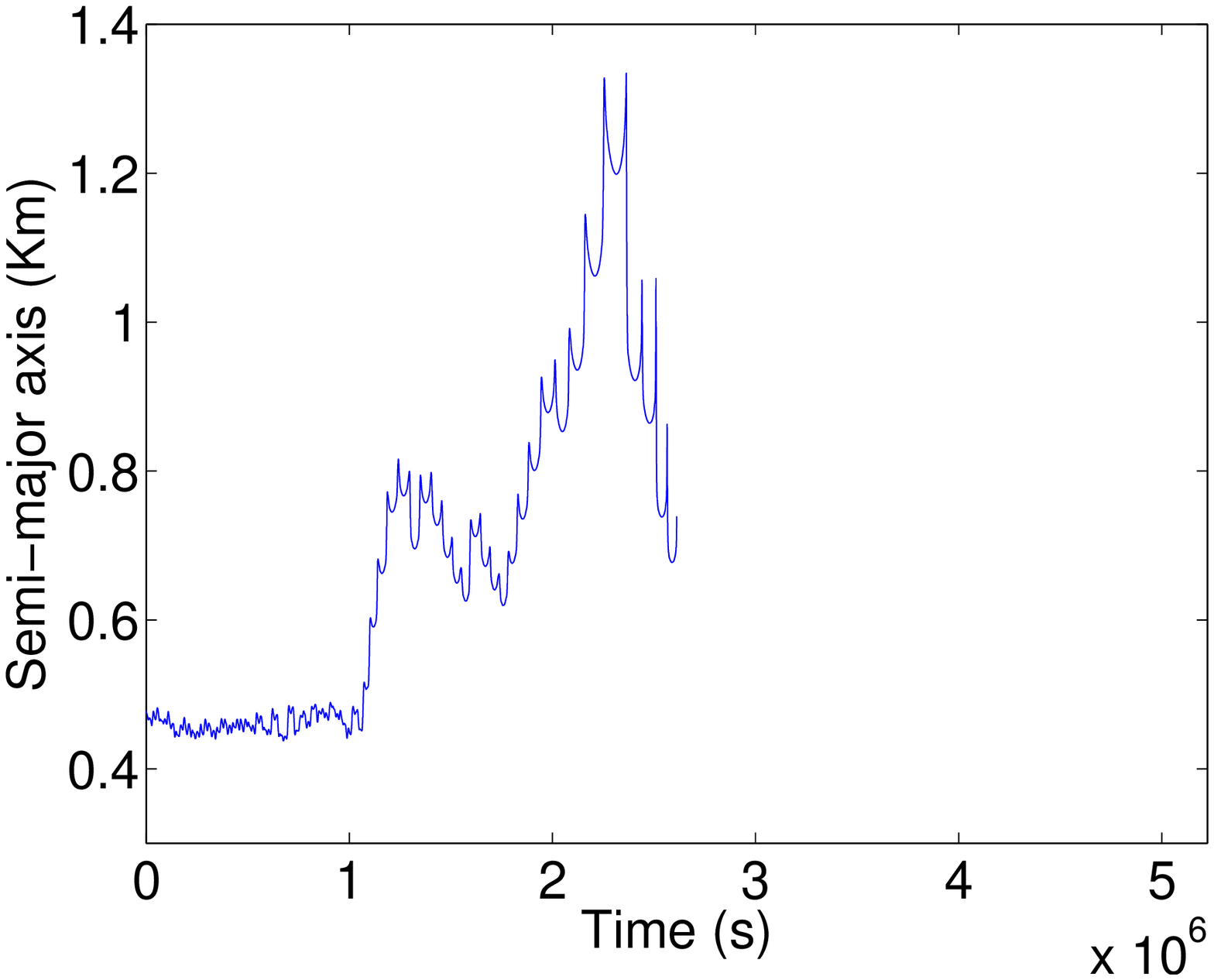}
     \includegraphics[width=0.25\linewidth]{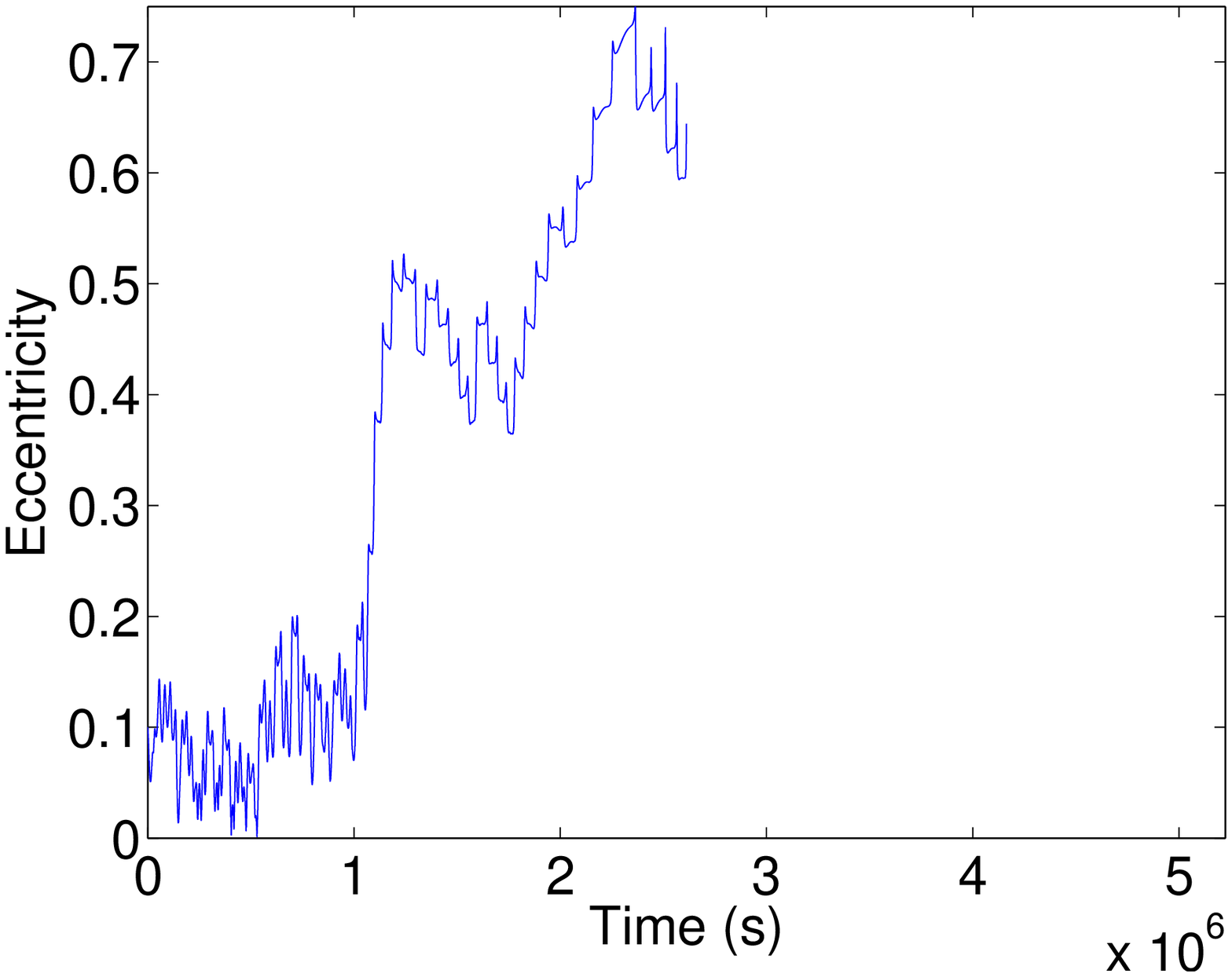}\\
     \caption{3D equatorial orbits around (101955) Bennu after $60$ days. These orbits are launched from the longitude $\lambda =90^o$. The initial eccentricity is 0.1 and its initial periapsis radius is 0.43 $km$. The solar radiation pressure is accounted for and shown like in Fig. 10.
}
         \label{Fig11} 
\end{figure*}
From the results of Fig. 8, two examples of 3D equatorial orbits around (101955) Bennu after 60 days are displayed in Figs. 10-11. We take these examples to show that the solar radiation pressure can disturb the orbit, making it precess or regress depending on the initial conditions. The initial eccentricity is 0.1 with initial periapsis radius 0.43 $km$ and $\lambda = 0^0 $ for the  first and $\lambda = 90^0 $ for the last one. In both cases, we show the effects of the passage of the spacecraft through the projected shadow of the asteroid. The behaviour of the semi-major axis and the eccentricity with respect to time are also shown. In the first case, the shadow prevents the probe from leaving its stable orbit while in the second case it increases the time of the orbit and allows the probe to visit regions far away, with passages closer to the body.
From Figs. 10 and 11, we can conclude that the shadow avoids the solar radiation pressure to be strong enough to pull the probe towards the asteroid, which also increases the orbital time before the collision. 

\section{Conclusions}
In this work, we modeled the gravity field of the asteroid (101955) Bennu using the new approach of mascon gravitation, developed by \citet{Chanut_2015a}, where the Mascon $\text{gravity}$ $\text{framework}$ \citep{Geissler_1997} was applied using the shaped $\text{polyhedral}$ source \citep{Werner_1994}. We tested this model analysing the equilibria near (101955) Bennu when the solar radiation pressure is not accounted for.  We found the same results that \citet{Wang_2016} have shown within the limits of density and rotation defined by \citet{Chesley_2014}. In this case,  the centre point $E$8 can become linearly stable by changing the topological structure from case 5 to case 1 \citep{Jiang_2014}. However, when we took the two-layered model described as a surface model in \citet{Scheeres_2016}, we can highlight that the centre point $E$8, which was linearly stable in the case of the homogeneous mass distribution, returns to be unstable when changing its topological structure from case 1 to case 5. This case already existed when we set the low density and can cause that previously stable trajectory about the centre point becomes unstable. The spacecraft or any particle may not remain in the proximity of the centre point and can briefly collide with the asteroid. This possibility will be accounted for by the $OSIRIS-REx$ mission. The space probe also may not find debris near the surface, which would greatly facilitate the approach manoeuvres.
Even though not so accurate on the surface when compared to the classical polyhedron method, we found that the model Mascon 10 is suitable regarding the $\text{computational}$ effort. We have, for example, a remarkable difference between the method Mascon 4 and Mascon 10 with only an accretion of ten minutes in execution time. Moreover, we verified that the Mascon 10 model is adequate when modelling different layers of densities, with respect to the classical polyhedron method that considerably increases the execution time, making it unviable. Unlike the Tsoulis method, the calculations of the mascon model were done only one time for the total potential, independently of each layer density.

We also modelled the solar radiation pressure with a dynamics around rotating asteroids and shadow effect \citep{Xin_2016}.
Taking into account the maximum and minimum values of the $g$ parameter of the solar radiation pressure, we considered the motion of the spacecraft close to the asteroid (101955) Bennu when it is at the perihelion distance $D=0.8969\,AU$, semi-major axis distance $D=1.1264\, AU$ and aphelion distance $D=1.3559\,AU$ from the Sun. So, we did 3D numerical simulations of initially $\text{equatorial}$ orbits of the probe OSIRIS-REx near the asteroid with $B= 96\, kg/m^2$ for the spacecraft OSIRIS-REx mass-area ratio and a reflectance of 20\%. We found that, below 0.34 km, the orbits are unstable and the limit radius for direct, initially equatorial circular orbits that will not impact with (101955) Bennu surface is 0.42 km when the solar radiation pressure is not accounted for. The results for the whole simulations with the Sun initial longitude $\psi_0 = -180^o$ and $\psi_0 = -135^o$ have shown that the solar radiation pressure greatly increases the instability of the equatorial orbits.  Moreover, starting with longitude  $\lambda = 0$, the orbits suffer fewer impacts and some (between 0.4 and 0.5 km) remain stable even if we consider the maximum solar radiation. We found the same behaviour at the longitude $\lambda = 90^o $ when the Sun initial longitude is $\psi_0 = -90^o$. The farther orbits are more unstable due to the fact that solar radiation pressure may not balance the gravitation force. When we change the initial longitude of the Sun to be $\psi_0 = -135^o$, the orbits with initial longitude $\lambda = 90^0 $ appear to be more stable than those with the Sun at initial longitude $\psi_0 = -180^o$. However, the orbits that start at the longitude $\lambda = 0^o $ are less stable than in the first case. The asteroid's shadow diminishes the effects of solar radiation in the vicinity of the body. This allows the eccentricity of the equatorial orbits closer to the body to be maintained, with the most pronounced effect when the orbits are started with relatively low initial eccentricity. We can conclude that the $OSIRIS-REx$ spacecraft may encounter regions where the solar radiation pressure counterbalances the gravity force to make its approach manoeuvres. These regions lie in a range between 400 and 500 m from the asteroid, where the manoeuvres must occur preferably in the anti-sunward direction. It follows, that in this case, the spacecraft is in the shaded region. It is important to also emphasize that the spacecraft may encounter a great instability near (101955) Bennu when the asteroid is in the closest position from the Sun.
Finally, we found a certain symmetry on the behaviour of the spacecraft around the asteroid (101955) Bennu regarding the initial position of the Sun and it is certainly due to the  small ellipticity of the shape, as shown in Fig 1.

These are preliminary results and the spacecraft can encounter other configurations since we do not take into account the retrograde equatorial orbits. Although more stable nearby the body when the solar radiation is accounted for, some tests have shown that nearly circular orbits can only serve as transition orbits to the polar orbits when asteroid (101955) Bennu will be in its aphelion position around the Sun. In its approach to the Sun, the eccentricity of these orbits should be increased. Furthermore, we do not also investigate polar orbits in this study but the shape of (101955) Bennu and the zero velocity curves may suggest that polar orbits are more unstable nearby the body. These orbits should be investigated in a future research. 

\section*{Acknowledgements}
 The authors are very grateful to the reviewer of our article for the helpful comments and suggestions that led to improve of the quality of this work.

 The authors wish to thank the S\~ao Paulo State Science Foundation (FAPESP) (Grants 13/15357-1 and 14/06762-2) and CNPq (Grants 150360/2015-0 and 312313/2014-4) for their generous support of this work.

\bsp
\label{lastpage}
\appendix
\section{Stability Maps}
\begin{figure*}
   \centering
 SRP ($g=6.941835 \times 10^{-11}km\cdot s^{-2}$, $R=0.8969AU$) \\
{\small \textcolor{white}{.} \hspace{0.75cm} $0^{\circ}$ \hspace{3.75cm} $90^{\circ}$ \hspace{3.75cm}  $180^{\circ}$ \hspace{3.75cm} $270^{\circ}$}\\
 1)\includegraphics[width=0.24\linewidth]{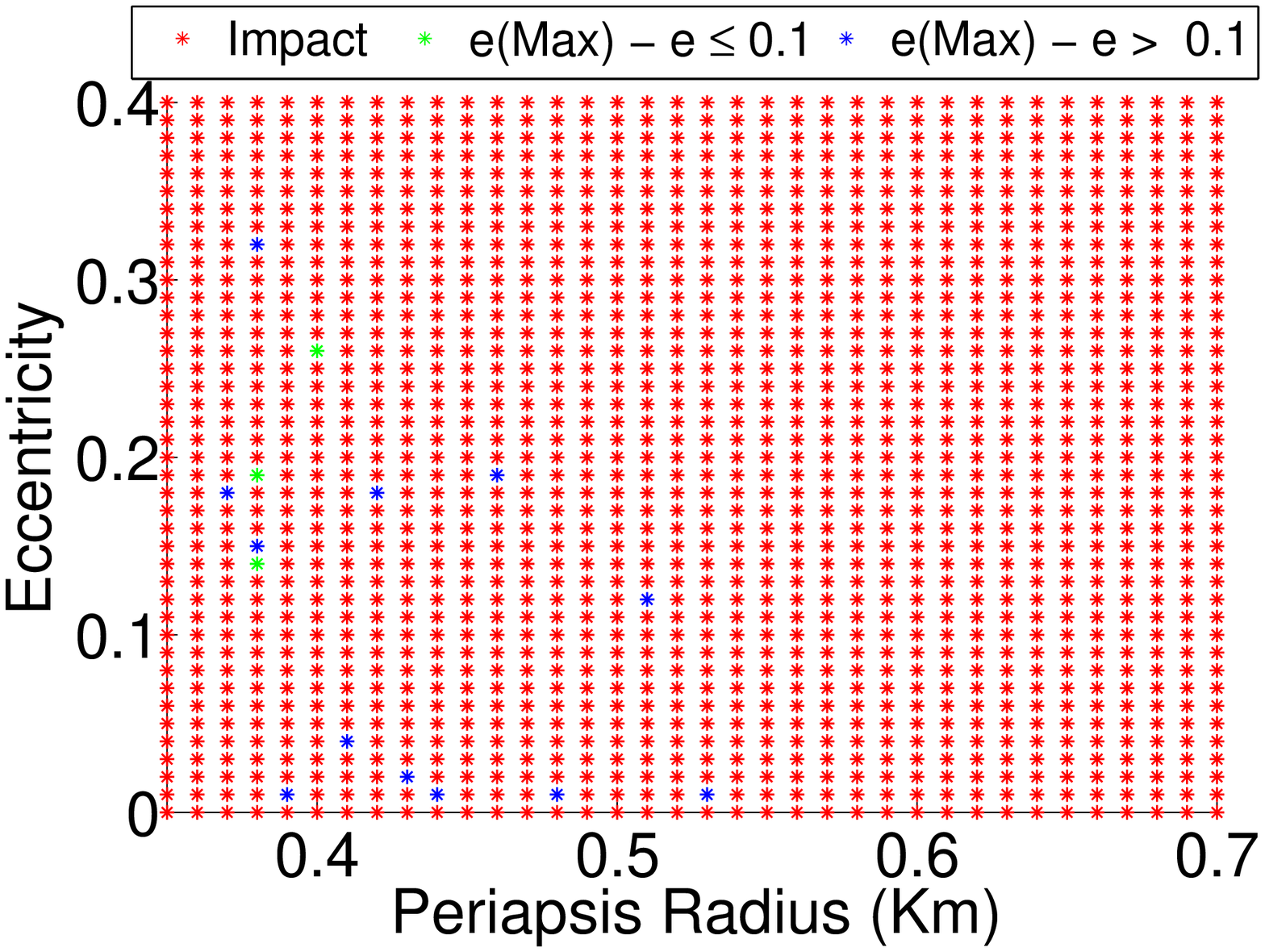}
 \includegraphics[width=0.24\linewidth]{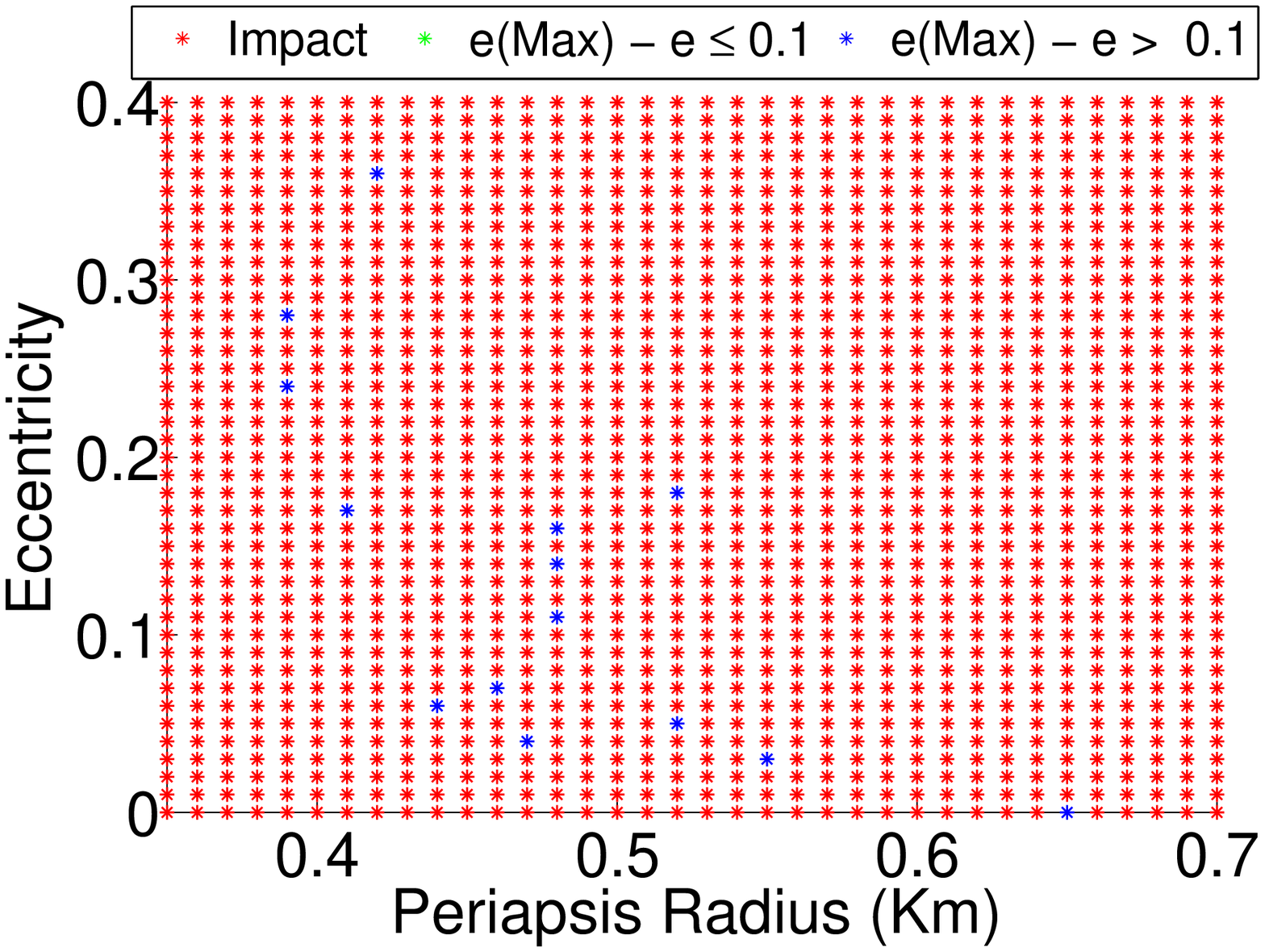}
 \includegraphics[width=0.24\linewidth]{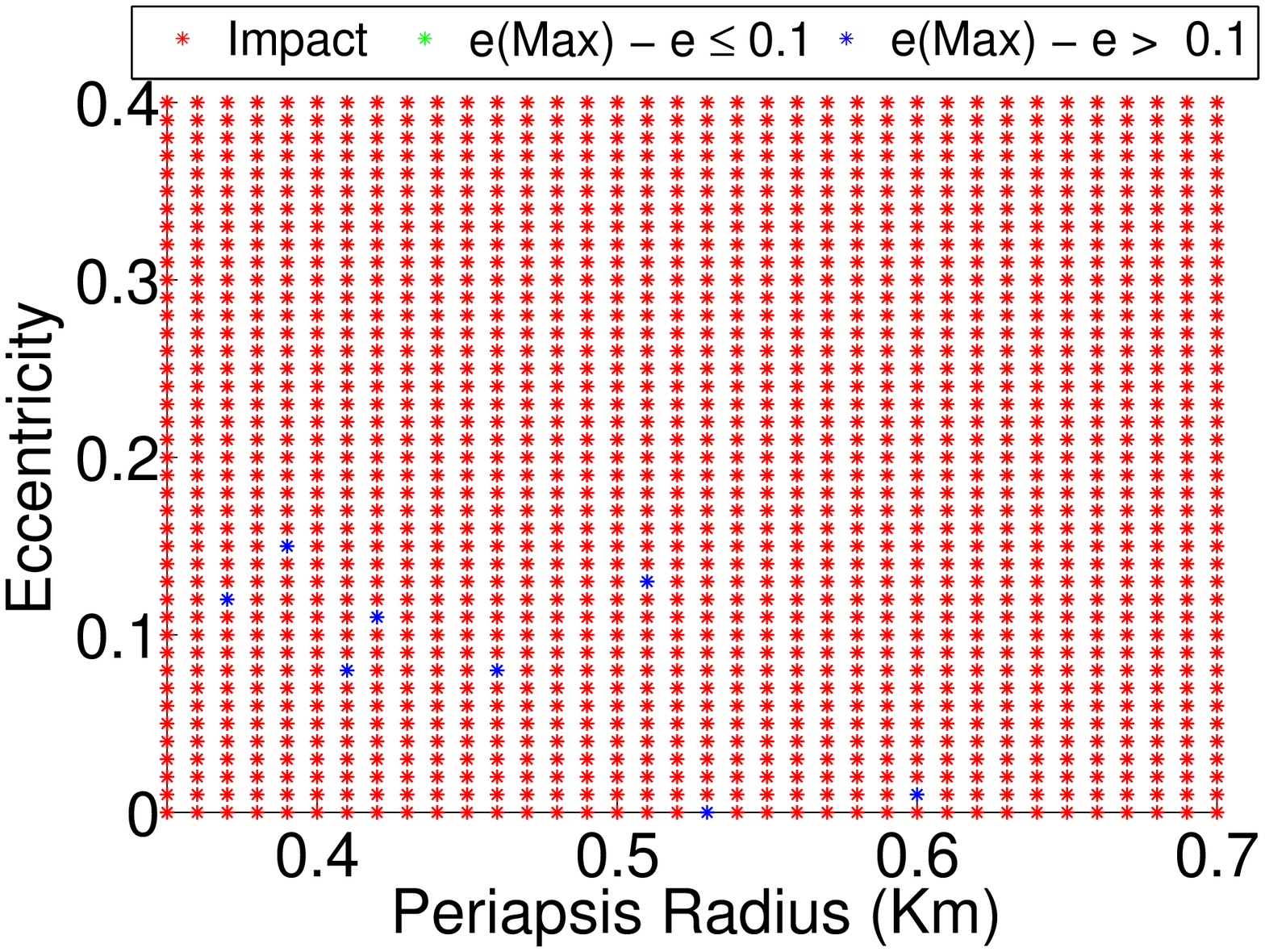}
 \includegraphics[width=0.24\linewidth]{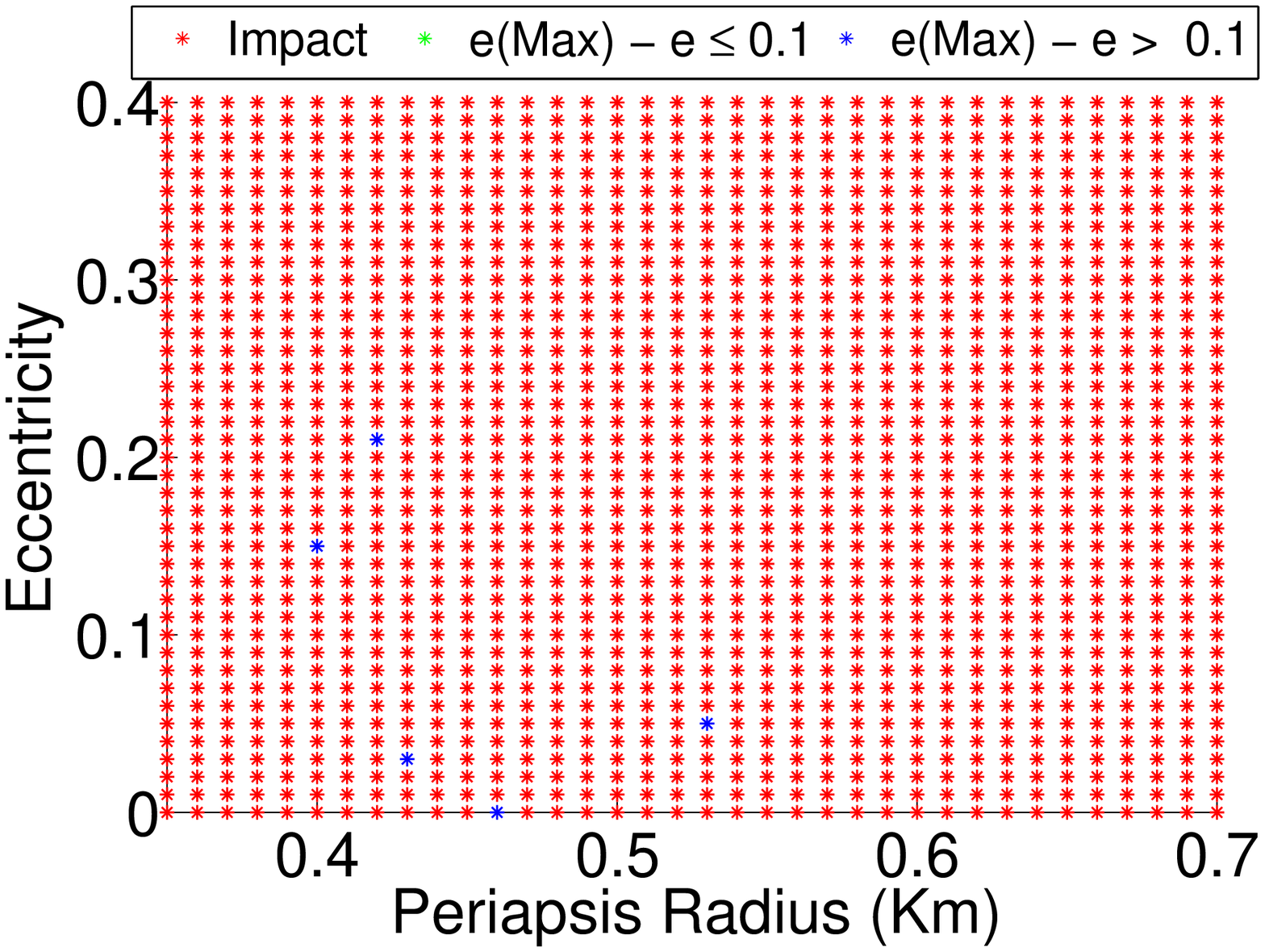}\\
{\small \textcolor{white}{.} \hspace{0.75cm} $0^{\circ}$ \hspace{3.75cm} $90^{\circ}$ \hspace{3.75cm}  $180^{\circ}$ \hspace{3.75cm} $270^{\circ}$}\\
 2)\includegraphics[width=0.24\linewidth]{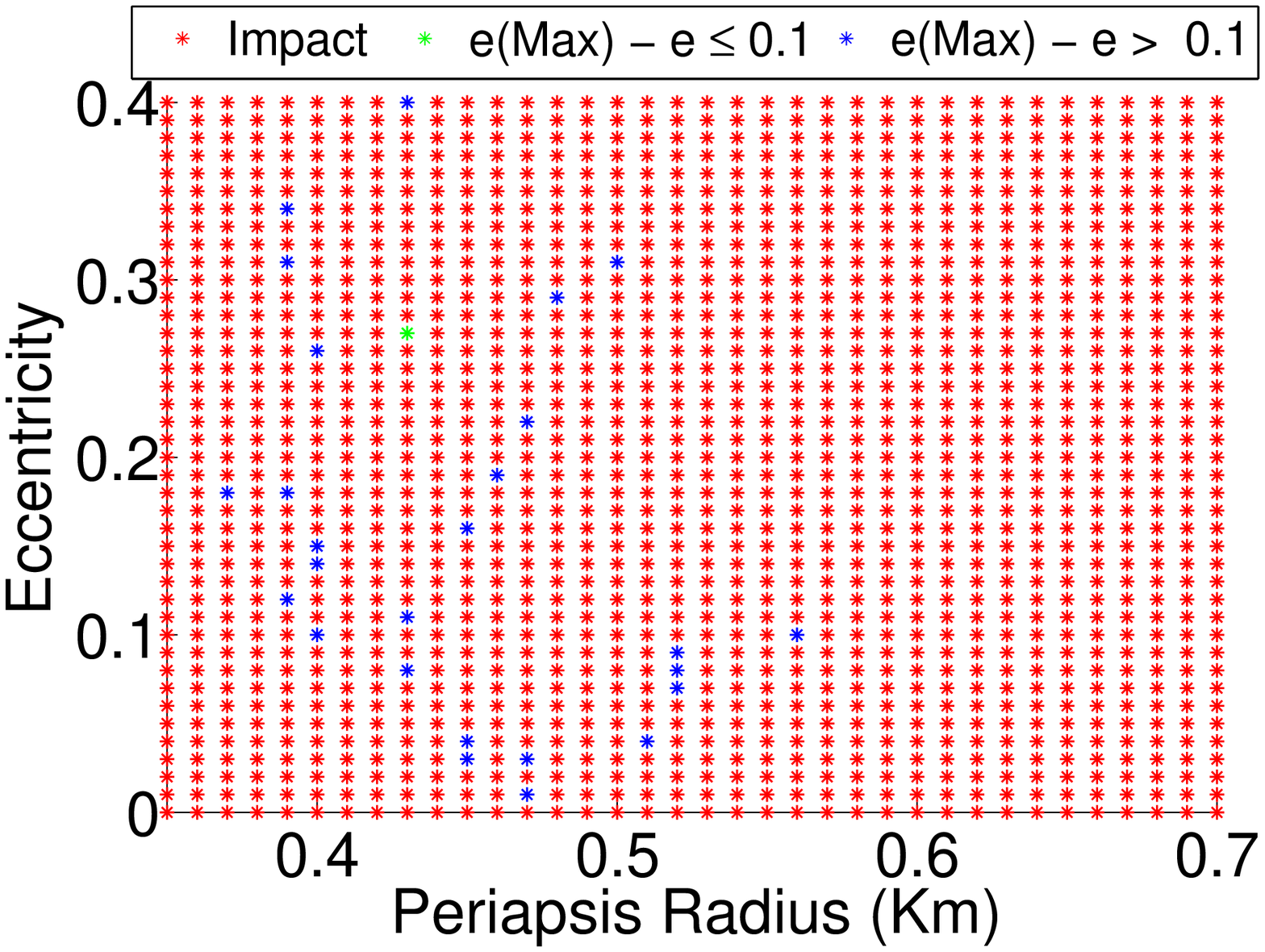}
 \includegraphics[width=0.24\linewidth]{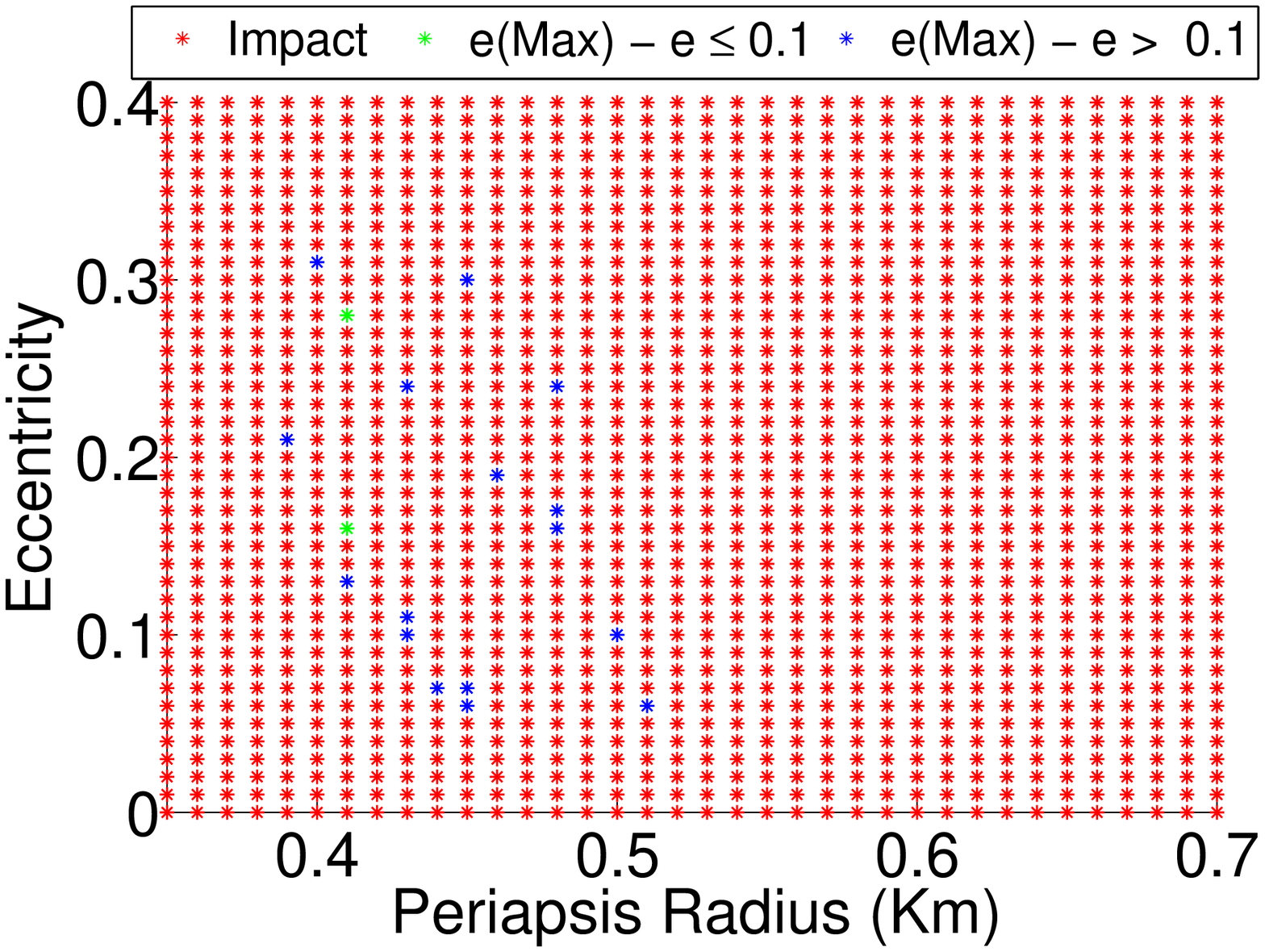}
 \includegraphics[width=0.24\linewidth]{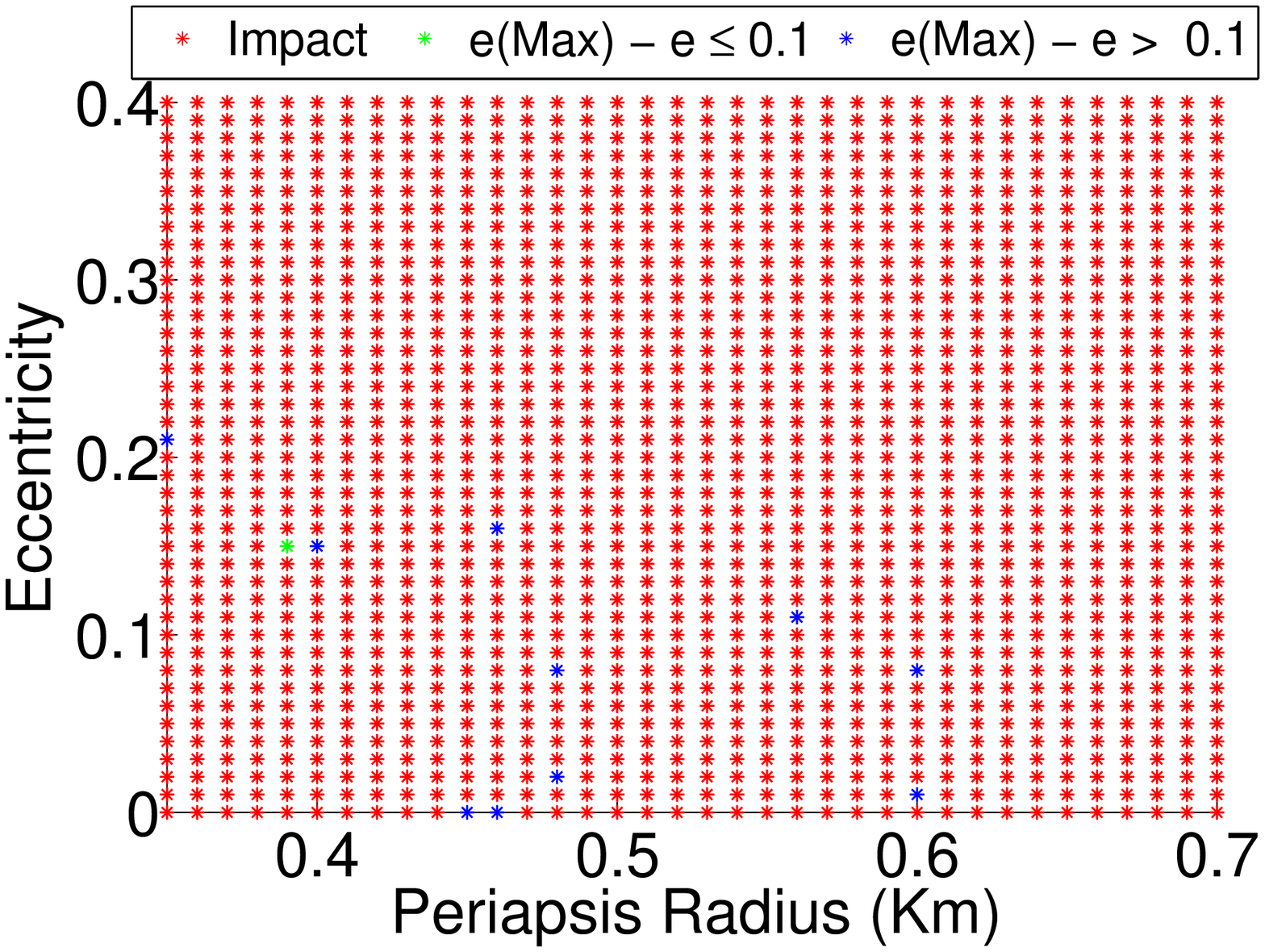}
 \includegraphics[width=0.24\linewidth]{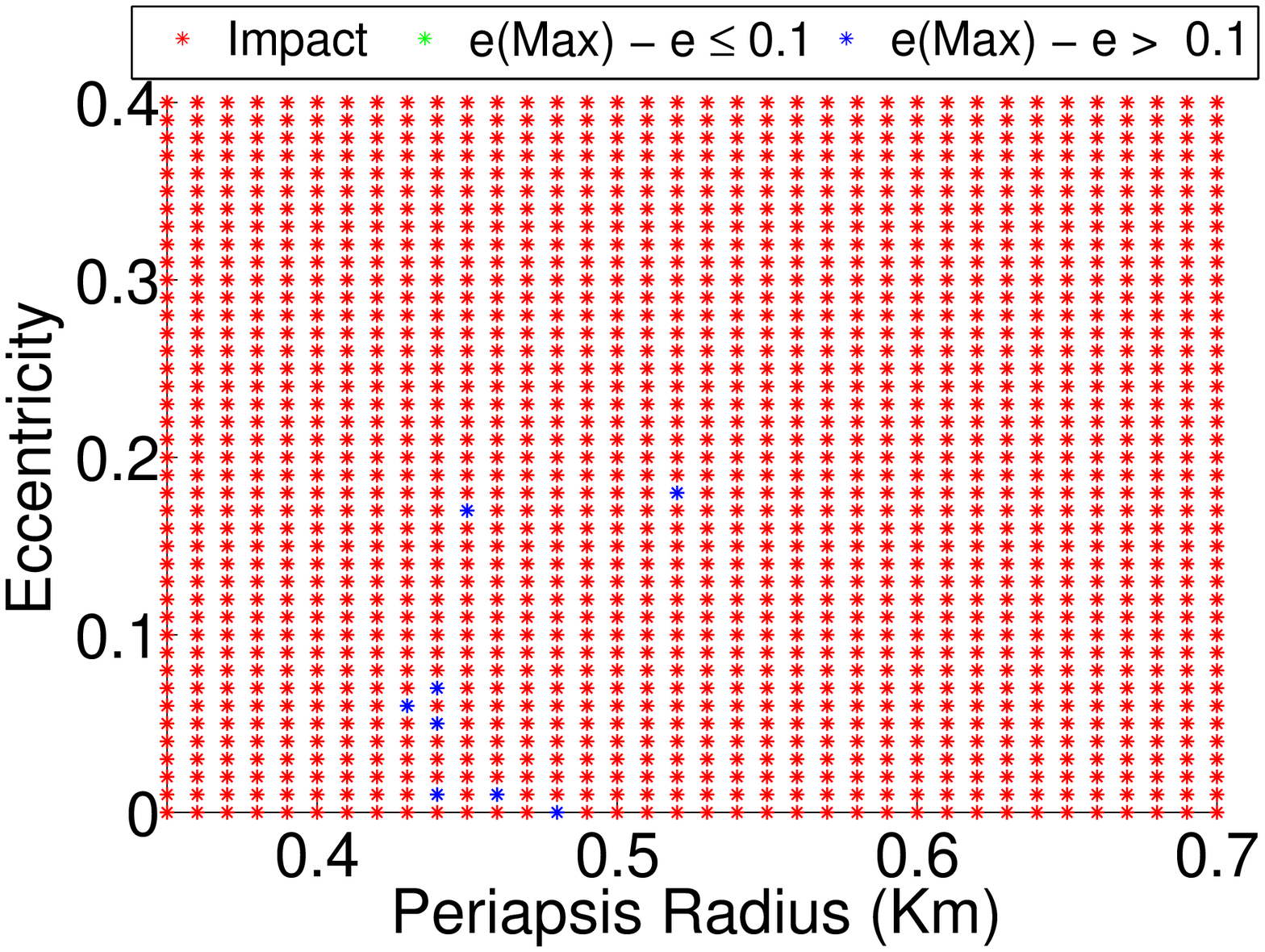}\\
      
 \caption{Stability maps of the equatorial orbits relative to (101955) Bennu where the solar radiation pressure is accounted for with the Sun initial longitude $\psi_0 = -135^o$. The (101955) Bennu's distance from the Sun $R$ is noted on the top of the related grafics and the eclipse is not taken into account in 1) and accounted for in 2). }
  \label{Fig9} 
\end{figure*}

\begin{figure}
   \centering
  SRP ($g=3.037665 \times 10^{-11}km\cdot s^{-2}$,  $R=1.3559AU$)\\
 \includegraphics[width=0.96\linewidth]{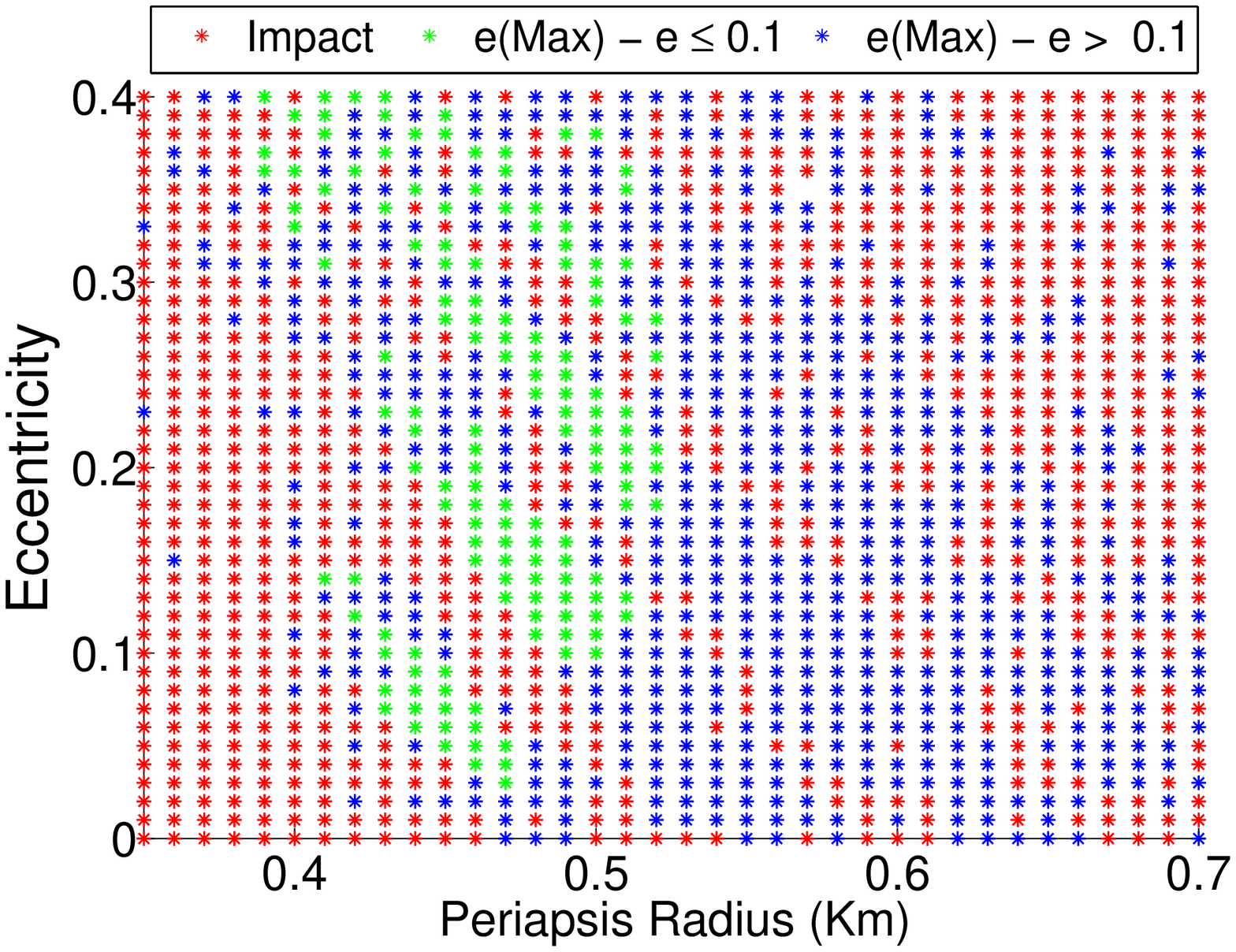} \\
  SRP ($g=4.405370 \times 10^{-11}km\cdot s^{-2}$,  $R=1.1264AU$) \\
 \includegraphics[width=0.96\linewidth]{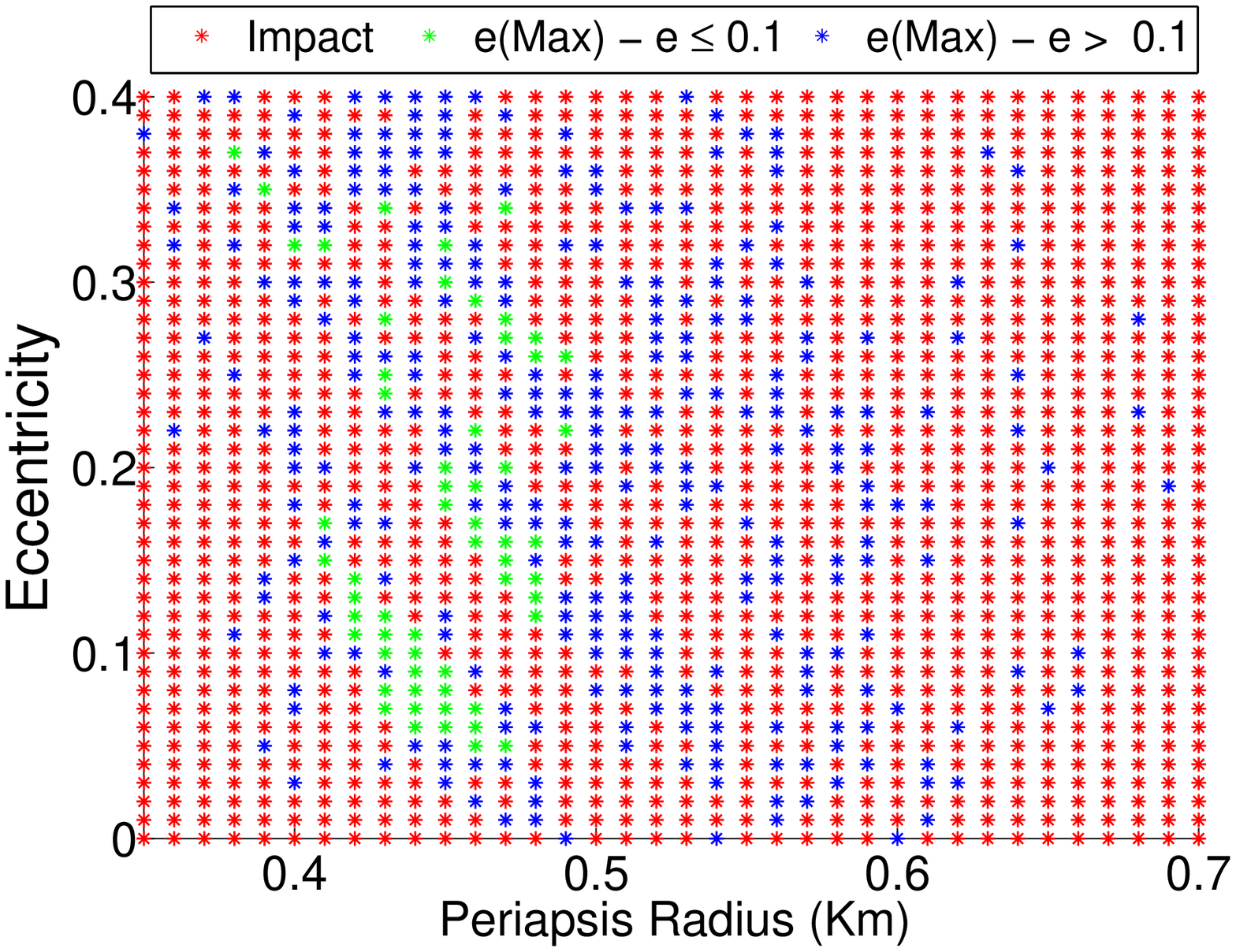}\\
  SRP ($g=6.941835 \times 10^{-11}km\cdot s^{-2}$, $R=0.8969AU$) \\
 \includegraphics[width=0.96\linewidth]{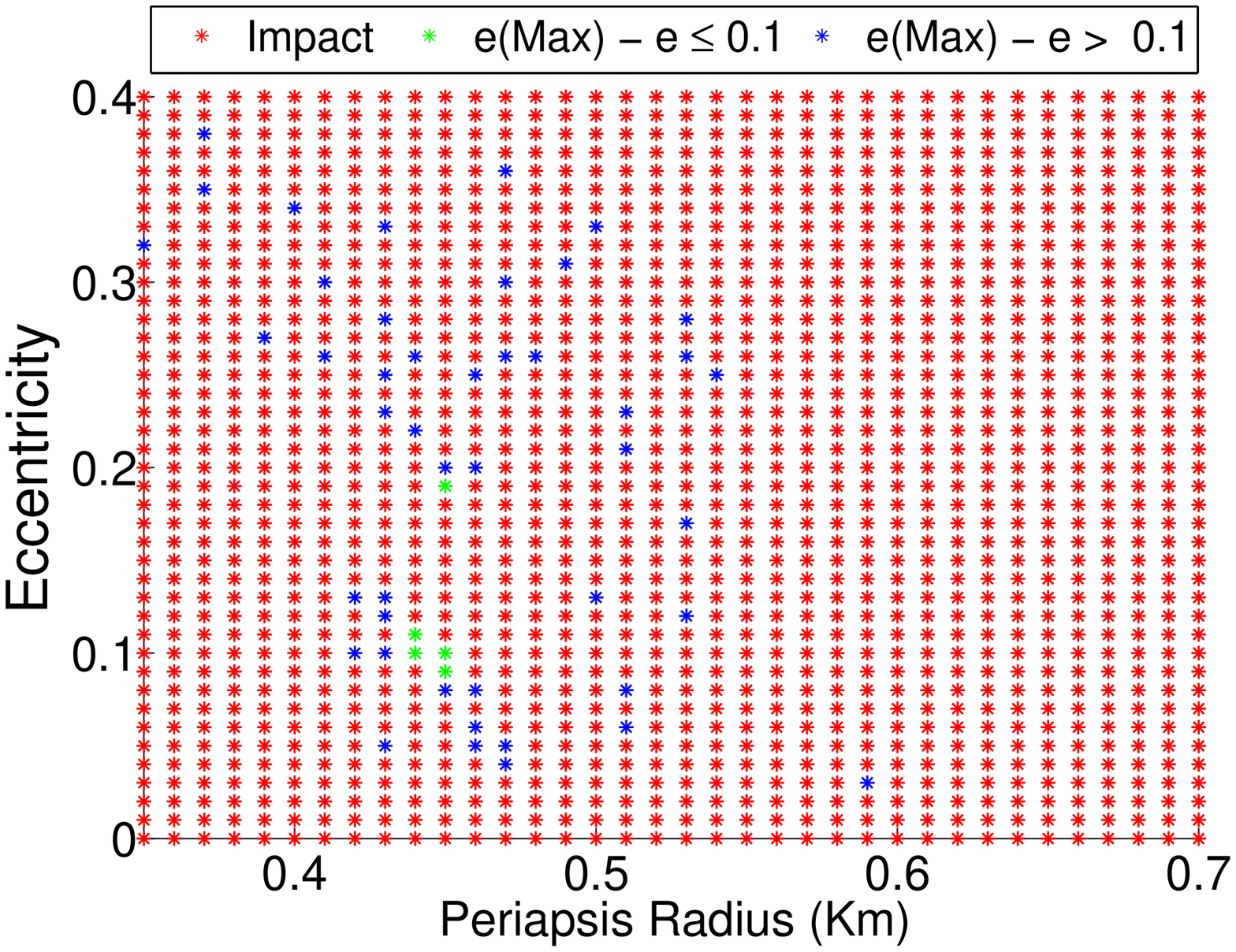}\\
       
 \caption{Stability maps of the equatorial orbits relative to (101955) Bennu with initial longitude $\lambda = 90^o $. The solar radiation pressure with the Sun initial longitude $\psi_0 = -90^o$ and the eclipse are accounted for. The (101955) Bennu's distance from the Sun $R$ is noted on the top of the related grafic. 
 }
  \label{FigA2} 
\end{figure}

\begin{figure}
   \centering
  SRP ($g=3.037665 \times 10^{-11}km\cdot s^{-2}$,  $R=1.3559AU$)\\
 \includegraphics[width=0.96\linewidth]{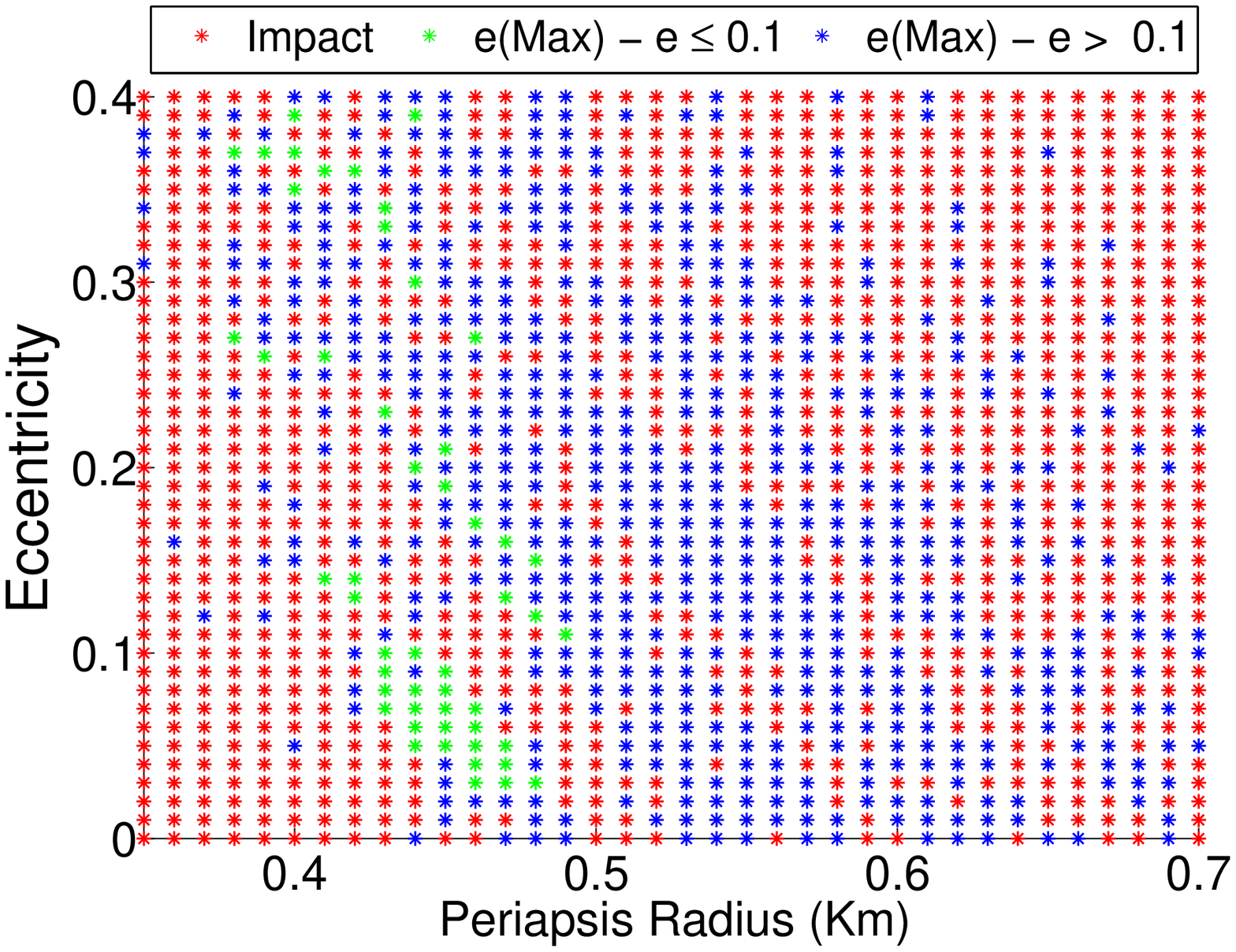} \\
  SRP ($g=4.405370 \times 10^{-11}km\cdot s^{-2}$,  $R=1.1264AU$) \\
 \includegraphics[width=0.96\linewidth]{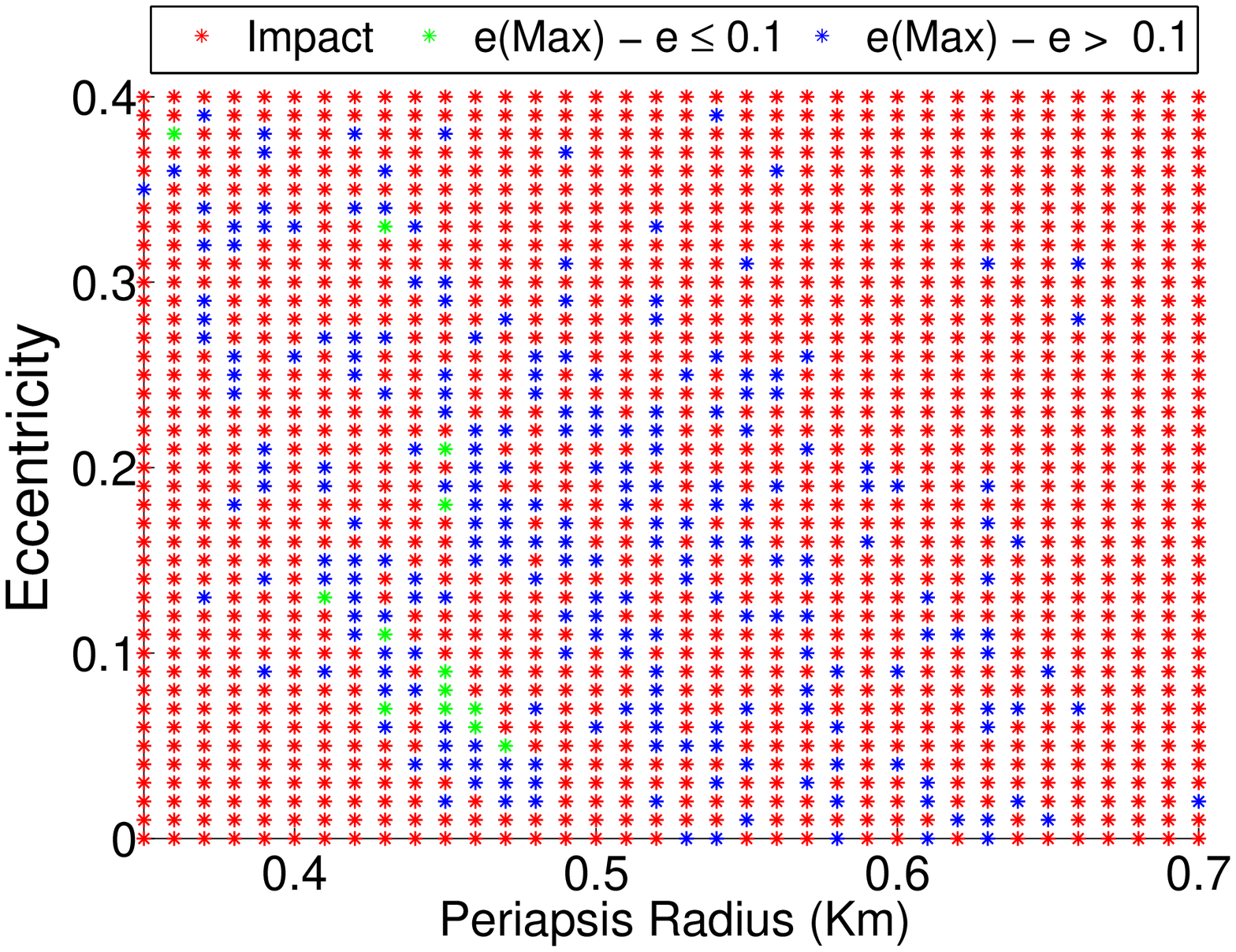}\\
  SRP ($g=6.941835 \times 10^{-11}km\cdot s^{-2}$, $R=0.8969AU$) \\
 \includegraphics[width=0.96\linewidth]{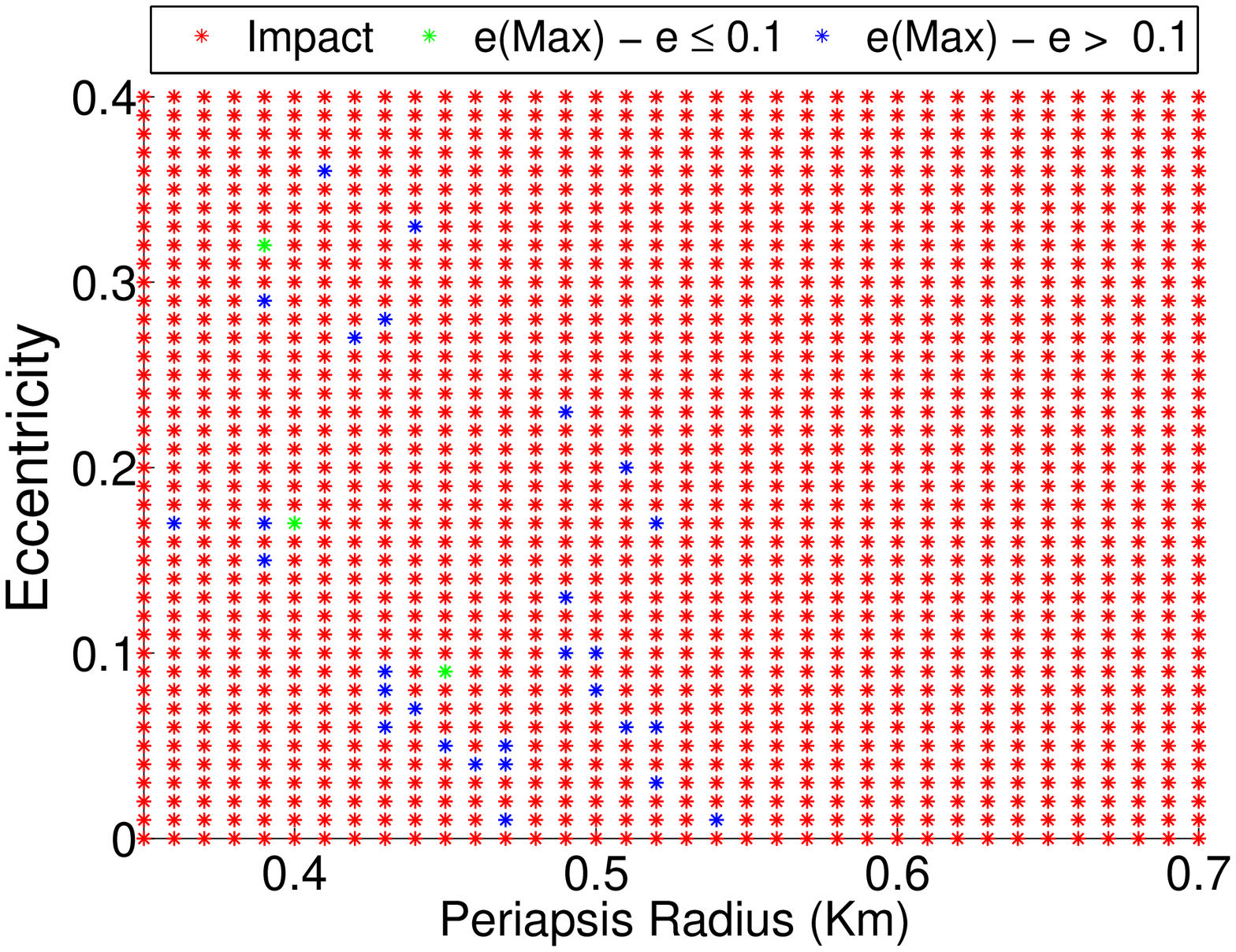}\\
       
 \caption{Stability maps of the equatorial orbits relative to (101955) Bennu with initial longitude $\lambda = 90^o $. The solar radiation pressure with the Sun initial longitude $\psi_0 = -45^o$ and the eclipse are accounted for. As Fig.A2, the (101955) Bennu's distance from the Sun $R$ is noted on the top of the related grafic.
 }
  \label{FigA3} 
\end{figure}

\end{document}